\documentclass[preprint,prd,aps,11pt,showpacs,nofootinbib,tightenlines,eqsecnum,aps,epsf,epsfig]{revtex4}

\usepackage[section]{placeins}
\usepackage{graphicx}
\usepackage{dcolumn}
\usepackage{bm}

\usepackage{epsfig}
\usepackage{amssymb}
\usepackage{amsmath}
\usepackage{flafter}
\usepackage{array}

\usepackage[dvips,usenames]{color}

\newlength{\dinwidth}
\newlength{\dinmargin}
\begin{document}

%
\title{Symmetry Principle Preserving and Infinity Free Regularization and renormalization of
  quantum field theories and the mass gap }
\author{ Yue-Liang  Wu }
\affiliation{ Institute of Theoretical Physics, Chinese Academy of sciences, \\
 P.O. Box 2735, Beijing 100080, P.R. China }

\begin{abstract}
 Through defining irreducible loop integrals (ILIs), a set of consistency conditions
 for the regularized (quadratically and logarithmically) divergent ILIs are obtained
 to maintain the generalized Ward identities of gauge invariance in non-Abelian
 gauge theories. The ILIs of arbitrary loop graphs can be evaluated from the
 corresponding Feynman loop integrals by adopting an ultraviolet (UV) divergence
 preserving parameter method. Overlapping UV divergences are explicitly shown
 to be factorizable in the ILIs and be harmless via suitable subtractions.
 A new regularization and
 renormalization method is presented in the initial space-time dimension of the theory.
 The procedure respects unitarity and causality. Of interest, the method leads to an
 infinity free renormalization and meanwhile maintains the symmetry principles
 of the original theory except the intrinsic mass scale caused
 conformal scaling symmetry breaking
 and the anomaly induced symmetry breaking. Tadpole
 graphs of Yang-Mills gauge fields are found to play an essential role for maintaining
  manifest gauge invariance via cancellations of quadratically divergent ILIs.
 Quantum field theories (QFTs) regularized through the new method
 are well defined and governed by a physically meaningful characteristic
 energy scale (CES) $M_c$ and a physically interesting sliding energy scale (SES) $\mu_s$
  which can run from $\mu_s \sim M_c$ to a dynamically generated mass gap $\mu_s=\mu_c$
  or to $\mu_s =0$ in the absence of mass gap and infrared (IR) problem.
  For $M_c\to \infty $, the initial UV divergent properties of
  QFTs are recovered and well-defined. In fact, the CES $M_c$ and SES
  at $\mu_s=\mu_c$ play the role of UV and IR cutoff energy scales respectively.
  It is strongly indicated that the conformal scaling symmetry and its
  breaking mechanism play an important role for
  understanding the mass gap and quark confinement.
  The new method is developed to be applicable for both underlying renormalizable QFTs
  and effective QFTs. It also leads to a set of conjectures on mathematically
  interesting numbers and functional limits which may provide deep insights in mathematics.
\end{abstract}

\maketitle

\newpage

\section{Introduction}

  Symmetry has plaid an important role in particle physics\cite{TDL}. All known basic forces
  of nature, i.e., gravitational, electromagnetic, weak and strong forces, are governed by the
  symmetry principles. Three of them (electromagnetic, weak and strong forces)
  have turned out to be characterized by the gauge symmetry $U(1)_Y \times SU(2)_L \times SU(3)_c $.
  Thus the three fundamental gauge interactions among the building blocks or elementary
  particles (quarks and leptons) are mediated via the Abelian and non-Abelian
  Yang-Mills gauge fields\cite{YM}. The real world of particles has been found to be
  successfully described by quantum field theories (QFTs) \cite{SW1,CL}. In the meantime,
  the gauge invariance has been viewed as a basic principle\cite{GTH}
 in addition to the well-known basic principles of Lorentz invariance and
 translational invariance. QFTs have also been applied to deal
 with effective theories for composite fields and particles at low energies
  and also critical phenomena (or phase transitions) in statistical mechanics and condensed
 matter physics\cite{JZJ}. Nevertheless, QFTs cannot be defined by a straightforward perturbative
 expansion due to the presence of ultraviolet (UV) divergences.
 Namely, the definition of Feynman diagrams in the perturbative expansion
 may be meaningless because of lack of convergence. To avoid such difficulties,
 one may modify the behavior of field theory at very large
 momentum,\footnote{If there exist infrared
 divergences, the behavior of field theory at very small momentum may also need to be
 modified} or introduce so called regulators (more general speaking, the regularization quanta),
 so that all Feynman diagrams become
 well-defined finite quantities. Such a procedure is usually called regularization.
 The most important features required for the regularization are that the regularization
 should maintain the basic symmetry principles of the theory,
 such as gauge invariance, Lorentz invariance and translational invariance (or Poincare invariance),
 and also preserve the initial but well-defined divergent properties
 of the theory. Many regularization methods have been introduced in the literature, such as:
 cut-off regularization\cite{COR}, Pauli-Villars regularization\cite{PV}, Schwinger's
 proper time regularization\cite{PR}, BPHZ regularization\cite{BPHZ},
 dimensional regularization\cite{DR}, lattice regularization\cite{LR},
 differential regularization\cite{DFR}. All the regularizations have
 their advantages and shortcomings.

The cut-off regularization is the simplest one by naively setting an upper bound to the
integrating loop momentum. This method is often used to treat QFTs
 in statistical mechanics or in certain low energy dynamical systems, where the divergent behavior of
the theory plays the crucial role, and the Lorentz or Poincare invariance beomes unimportant
and the gauge symmetry is not involved at all. In contrast, the
 method fails in applying to the QFTs for elementary particles in which Lorentz or Poincare invariance
and Yang-Mills gauge invariance play an imporatnt role. This is because
 the method destroys the principle of Lorentz and translational ( or Poincare) invariance and
also the principle of gauge invariance for gauge theories.
The spirit of Pauli-Villars's regularization is to modify the propagator.
Its prescription is simple: replacing any propagator
by a sum of propagators with large masses, and choosing appropriate coefficients so that the
 large momentum behavior becomes well controlled. More general modifications on the propagators
may be introduced in Schwinger's proper time regularization. Whereas
in Pauli-Villars's regularization, a set of regulator fields are usually introduced
to modify the action of original theory. Though Pauli-Villars's inspired regularizations
may preserve a large number of symmetries of the theory, there are still a class
of Feynman diagrams which cannot be regularized by this approach. Such field theories
include the non-Abelian Yang-Mills gauge theories\cite{YM}
and also the non-linear $\sigma$-model.
In ref.\cite{AAS}, a higher covariant derivative Pauli-Villars regularization was proposed for
maintaining the gauge symmetry in non-Abelian gauge theories. Nevertheless,
it has been shown that such a regularization violates
 unitarity\cite{LMR} and also leads to an inconsistent quantum chromodynamics (QCD)\cite{MR}.
 BPHZ prescription relies on the theorem proved by
 Weinberg\cite{SW2}, which states that the requirement for the
 actual convergence of the amplitude corresponding to any graph is
 that the power-counting should give a negative superficial degree
 of divergence for the complete multiple integral for the whole amplitude and also for
 any subintegration defined by holding any one or more linear combinations of the loop momenta fixed.
 The prescription does eliminate all superficial divergences and render the renormalized
 Feynman integrals to be convergent. The forest formula provides
 a powerful tool to construct proofs of renormalizability
 to all orders. In fact, the method is a regularization-independent subtraction
 scheme up to the finite part. To proceed the
 practical calculation so as to extract the finite part by means of the forest formula, one still has
 to adopt a regularization scheme. As such a subtraction process is based on expanding around an external momentum,
 which modifies the structure of the Feynman integral amplitudes, thus the
 gauge invariance is potentially destroyed when applying the BPHZ
 subtraction scheme to non-Abelian gauge theories. In addition,
 the unitarity, locality and causality may also not rigorously hold in
 such a subtraction scheme. In the lattice regularization, both
 space and time are made to be the discrete variables. The method can preserve gauge symmetries of the theory,
 but the principle of Lorentz and translational (or Poincare) invariance is lost in this
 method. Though there is a great advantage that lattice
 regularization may be extended to the nonperturbative calculations by a numerical method,
 it may also, at the same time, lead to a disadvantage due to a very complicated
 perturbative calculation. The principle
 of dimensional regularization is to define all Feynman diagrams by analytic continuation
 in the space-time dimension parameter, when the space-time dimension is required to
 be the initial one, the Feynman diagrams will recover the initial
 logarithmically UV divergences but the quadratically UV divergences are suppressed.
 This method provides the simplest symmetry principle preserving regularization
in the perturbative calculations. Whereas it seems to be meaningless beyond perturbative
expansion as it involves continuation of Feynman diagrams in space-time dimension
 parameter to arbitrary values. The method also fails if quantities are specific to
the initial space-time dimension, such as $\gamma_5$ in four dimensions and
 the complete antisymmetric tensor $\varepsilon_{\mu\nu\rho\sigma}$ as well as the case
 that scaling behavior becomes important. The cases include the
 chiral, topological and supersymmetric theories. For that a so-called
 dimensional reduction regularization was suggested\cite{DRED1} as a variant of dimensional regularization,
 in which the continuation from dimension d=4 to d=n  is made by
 compactification. Thus the number of field components remains unchanged and
 the momentum integrals are n-dimensional. Such a prescription may cause ambiguities in the finite parts
 of the amplitudes and also in the divergent parts of high order corrections. The scheme has been applied particularly
 in supersymmetric models. In general, it seems to hold only at one loop level and
 is actually inconsistent with analytical continuation
 concerning $\gamma_5$ \cite{DRED2}. On the other hand, the dimensional
 regularization may not be applicable to deal with the dynamics of effective QFTs at
a physically meaningful finite energy scale. The well-known example is
the derivation of gap equation in the gauged Nambu-Jona-Lasinio model\cite{NJL}.
It was shown\cite{TG} that the dimensional regularization cannot lead to a correct
gap equation. This is because the dimensional regularization destroys
the quadratic `divergent' (or quadratic `cut-off' momentum)
term in the gap equation though it maintains gauge invariance.
Instead, when adopting the simple cut-off regularization to regulate loop integrals,
one can arrive at a desired gap equation, but the gauge invariance is destroyed
by a quadratically `divergent' (or quadratic `cut-off' momentum) term and
also by ambiguities associated with arbitrary routing of loop momentum due to
lack of translational invariance. Another example is for the evaluation of the hadronic matrix
elements of four quark operators in the kaon decays $K\rightarrow \pi\pi$.
If applying the dimensional regularization to the long-distance operator evaluation
in the chiral perturbation theory, one arrives at a wrong sign for the leading order
contributions when matching to the corresponding short-distance operator evaluation in
 perturbative QCD\cite{YLWU}. The reason is the same as the previous one,
 the dimensional regularization destroys the leading quadratic cut-off momentum terms
which have the same sign as the leading order terms in perturbative QCD. Actually, the quadratic terms
 in the chiral perturbation theory were found to play a crucial role for understanding
 the $\Delta I = 1/2$ rule and for providing a consistent prediction on the
direct CP-violating parameter $\varepsilon'/\varepsilon$ in kaon
decays\cite{YLWU}. From these two realistic and interesting
examples, it is not difficult to further understand the
shortcoming of the dimensional regularization. In general, the
dimensional regularization suppresses the scaling behavior somehow
via a cancellation between UV and IR divergences. Actually, in the
dimensional regularization there is no `divergent' contribution
 that can appear in a power of cut-off energy scale. The
 differential regularization was proposed as a method working
 directly on Feynman graphs in coordinate space\cite{DFR}. Where
 one substitutes singular expressions by derivatives of
 well-behaved distributions. However, in gauge theories, the
 unsatisfactory feature arises due to the arbitrariness of the different renormalization scales in the
 Ward identities among renormalized Green functions. Thus one has
 to adjust the scales in an appropriate way so as to
 preserve the gauge invariance. Lately, a so-called constrained differential
 regularization was introduced\cite{CDFR} to fix the ambiguities of the local counterterms,
 so that the gauge invariance can be maintained in a convenient way.
 While from the calculational viewpoint, one loses the advantage of working
 directly in the momentum space.

   From the above points of view, one may arrive at the conclusion that up to now there exists no
   single satisfactory regularization that can be applied to all purposes in QFTs.
   It is then natural to ask whether one is able to
   find a regularization which can combine the advantages appearing in the above
   mentioned regularizations. To realize this purpose, the new regularization
  should match at least four criteria:

  (i) the regularization is rigorous that it can maintain the basic
 symmetry principles in the original theory, such as: gauge invariance, Lorentz invariance
 and translational invariance, except the anomaly induced symmetry
 breaking and the intrinsic  mass scale caused conformal scaling symmetry
 breaking.

  (ii) the regularization is general that it can be
  applied to both underlying renormalizable QFTs (such as QCD) and
 effective QFTs (like the gauged Nambu-Jona-Lasinio model and chiral perturbation
 theory).

  (iii) the regularization is also essential in the sense that it can lead to the well-defined
 Feynman diagrams with maintaining the initial but well-defined
 divergent properties of the theory, so that the regularized theory only needs to
 make an infinity free (finite) or infinity-controlled
 renormalization.

  (iv) the regularization must be simple that
 it can provide the practical calculations.
\\

   The first and forth criteria are clearly understood.
 The second and third ones are in principle related, for which we may recall the following
 point of view:  it has generally been assumed that the fundamental laws of nature are governed
 by a quantum theory of fields since the time when the electromagnetic and
 also the weak and strong interactions were found to be well described by the QFTs.
 However, this idea has been challenged from the
 studies of quantum gravity. As it is known that the Einstein's theory of general relativity
 is not a perturbatively renormalizable QFT in the power-counting sense.
 Also because of divergent problem in the perturbative
 expansion of QFTs, one has raised the issue that
 underlying theory might not be a quantum
 theory of fields, it could be something else, for example, string or superstring\cite{SST}.
  If this could turn out to be the case, even the quantum electrodynamics (QED) and
  QCD as well the quantum flavor dynamics (QFD) of electroweak interactions
 might be regarded as the effective QFTs. In general, effective QFTs
 are thought to be resulted from low energy approximations of a more fundamental theory.
 This may become more clear from the so-called folk theorem by Weinberg\cite{SW3,SW1}.
 The theorem states that any quantum theory that at sufficiently low energy and
 large distances looks Lorentz invariant and satisfies the cluster decomposition principle
 will also at sufficiently low energy look like a quantum field theory.
 According to such a folk theorem, there likely exists in any case a
 characteristic energy scale (CES) $M_c$ which can be either a fundamental-like energy
 scale (such as the Planck scale $M_P$ and/or the string scale $M_s$ in string theory) or
 a dynamically generated energy scale (for instance, the chiral symmetry breaking scale
 $\Lambda_{\chi}$ in chiral perturbation theory and the critical temperature for superconductivity),
  so that any effective QFTs become meaningful only at a sufficiently
 low energy scale in comparison with the CES $M_c$.

  In general, there is no restriction on the CES $M_c$. A particularly interesting case
  is to run the CES $M_c$ going to infinity. If in this case, the QFTs can remain to be
  well-defined by performing a renormalization, namely one can formally recover the
  initial divergent properties of QFTs in the perturbative expansion at $M_c\rightarrow \infty$
 and consistently eliminate all infinities via an appropriate renormalization of coupling constants
  and fields in the original theory, we may then mention such kinds of QFTs as the underlying
 renormalizable QFTs. It is not difficult to show that QFTs in which
 there are only Yang-Mills gauge fields interacting with Dirac spinor fields
 can be viewed as underlying renormalizable QFTs.
 While the Einstein's theory of general relativity may be regarded as an effective theory at
  the sufficiently low energy scale in comparison with the CES which is in the order of
 magnitude of the Planck mass, i.e., $M_c \sim M_P$.

  On the other hand, basing on the idea of renormalization group developed by either
  Wilson\cite{KGW} or Gell-Mann and Low\cite{GML}, one should be able to deal with
  physical phenomena at any interesting energy scale by integrating out the physics at
  higher energy scales. This implies that one can define the renormalized theory
  at any interesting renormalization scale or the so-called sliding energy scale (SES)
  $\mu_s$ which is not related to masses of particles or fields
  and can be chosen to be at any scale of interest,
  so that the physical effects above the SES $\mu_s$
  are integrated in the renormalized couplings and fields.

   With these considerations, it becomes reasonable to conjecture that there
  should exist an alternative new regularization scheme that can realize the above mentioned
  attractive properties. Of particular, the regularization for which we are looking
  must be governed by a physically meaningful CES $M_c$ and also a physically interesting
  SES $\mu_s$, so that the laws of nature can well be described by a quantum theory of fields
  when the energy scale in the considered phenomena becomes sufficiently lower than the CES $M_c$,
  and also the laws of nature at an interesting energy scale can well be dealt with
  by renormalizing at the SES $\mu_s$ which can be chosen to be the order of magnitude
  of the energy concerned
  in the considered process. It is this motivation that comes to the purpose of the present paper
  which may be organized as follows: in section II, by conceptually defining
  a set of so-called irreducible loop integrals (ILIs), we perform an
  explicit evaluation for the gauge field vacuum polarization
  graphs at one-loop order in the non-Abelian gauge theory with a general $R_{\xi}$ gauge.
  As a consequence, we arrive at two necessary conditions for the regularized
  quadratically and logarithmically divergent ILIs
  in order to maintain the gauge invariance of non-Abelian gauge theories (or to satisfy
  the so-called generalized Ward identities). In analogous to the quantum
  electrodynamics, the gauge fixing term $ (\partial^{\mu}A_{\mu}^a )^2$
  is found to be unaffected by the loop graphs. While it is unlike the usual case (such as the case
  in the dimensional regularization), the tadpole graphs of gauge fields
   are found in the most general case to play an important role for maintaining manifest gauge invariance.
   At the end of section, we present a set of consistency
   conditions which ensures the gauge invariance for all one loop
   graphs. We then explore in section III a new regularization method which satisfies the
   consistency conditions presented in section II for the regularized  divergent ILIs.
  To be more clear, we begin with an explicit check on the two consistency conditions
   in the well-known cut-off regularization and dimensional regularization methods.
  It is then easily seen why the cut-off regularization destroys the gauge invariance
  and the dimensional regularization does maintain the gauge invariance.
  We then present an alternative new
  regularization method to achieve the desired four criteria mentioned above.
  It is shown in section IV that the regulators (or regularization quanta) are indeed decouple
   from the regularized ILIs. Consequently, we arrive at, through a numerical check,
  three conjectures on mathematically interesting
  numbers and functional limits. We then arrive at a consistent regularization scheme.
  In section V, we study in details the overlapping Feynmay integrals
  involved in the general two loop graphs and present an explicit evaluation for the corresponding
  ILIs by adopting the usual Feynman parameter method and a newly formulated UV-divergence
  preserving parameter method. The latter method ensures that all the Feynman parameter integrations contain
  no UV divergences.  The explicit forms and remarkable features of the ILIs enable
  us to deduce a set of key theorems which contain the factorization theorem and subtraction theorem for
  overlapping divergences, and the reduction theorem for overlapping tensor type integrals
   as well as the relation theorem for the tensor and scalar type ILIs.
   In section VI, we present a general prescription for the new regularization and
  renormalization method. In particular, an explicit demonstration is provided
  for two loop diagrams. The overlapping divergences are shown to be harmless after an appropriate
  subtraction. The subtraction process can be made to be analogous to the one in the dimensional regularization.
  All the Feynman parameter integrations are convergent. The scheme is seen to respects unitarity and causality.
  We show in section VII how to evaluate the ILIs of arbitrary loop graphs.
  After a detailed evaluation for the ILIs of three loop graphs, a general procedure for
  the evaluation of any fold ILIs is presented,  which should be practically useful to make a
  realistic computation for arbitrary loop diagrams. In section VIII, we pay our special
  attention to the issues on the conformal scaling symmetry breaking, the mass gap genesis
  in Yang-Mills gauge theories and the quark confinement and deconfinement.
  Our conclusions are presented in the last section. Some useful formulae and technical
   details concerned in the text are presented in appendices.

   \section{Consistency Conditions for preserving Gauge symmetry}

   We start from the Lagrangian of gauge theory with Dirac spinor fields $\psi_n$
   ($n=1,\cdots, N_f$) interacting with Yang-Mill gauge field $A_{\mu}^a$ ($a=1, \cdots, d_G$)
   \begin{eqnarray}
  {\cal L}  =  \bar{\psi}_n (i\gamma^{\mu}D_{\mu} - m) \psi_n
   - \frac{1}{4} F^a_{\mu\nu}F_a^{\mu\nu}
   \end{eqnarray}
   with
 \begin{eqnarray}
  & & F_{\mu\nu}^a  =  \partial_{\mu} A_{\nu}^a - \partial_{\nu} A_{\mu}^a
  -g f_{abc}A_{\mu}^b A_{\nu}^b
  \\
  & & D_{\mu}\psi_n  = (\partial_{\mu} + ig T^a A_{\mu}^a)\psi_n
   \end{eqnarray}
 Here $T^a$ are the generators of gauge group and $f_{abc}$ the structure function of the gauge
 group with $ [T^a, \ T^b ] = i f_{abc} T^c $. To quantize the gauge theory,
 it is necessary to fix the gauge by adding the gauge fixing term and
  introducing the corresponding Faddeev-Popov ghost term\cite{FP} with the ghost fields $\eta_a$.
 In the covariant gauge, the additional Lagrangian ${\cal L'}$ has been found to take the
 following form
 \begin{eqnarray}
{\cal L'} = -\frac{1}{2\xi} (\partial^{\mu} A_{\mu}^a )^2 + \partial^{\mu}\eta^{\ast}_a (
\partial_{\mu} \eta^a + g f_{abc}\eta^b A_{\mu}^c )
 \end{eqnarray}
 where $\xi$ is an arbitrary parameter. Thus the modified Lagrangian is given by
 \begin{eqnarray}
 \hat{{\cal L}} & = & {\cal L}+{\cal L'} = \bar{\psi}_n (i\gamma^{\mu}D_{\mu} - m) \psi_n
   - \frac{1}{4} F^a_{\mu\nu}F_a^{\mu\nu} \nonumber \\
   & - & \frac{1}{2\xi} (\partial^{\mu} A_{\mu}^a )^2 + \partial^{\mu}\eta^{\ast}_a (
\partial_{\mu} \eta^a + g f_{abc}\eta^b A_{\mu}^c )
  \end{eqnarray}
 Based on this modified Lagrangian,  one can derive the Feynman
 rules\cite{F1,F2,FP} for propagators and vertex interactions\cite{CL}.

     For simplicity, we begin with our considerations at one-loop level.
     To find out the gauge symmetry preserving conditions, it is
     of help to introduce a set of loop integrals  which consist of the scalar
    type ones
    \begin{eqnarray}
     I_{-2\alpha} = \int \frac{d^4 k}{(2\pi)^4}\ \frac{1}{(k^2 - {\cal M}^2)^{2+\alpha}}\ ,
     \qquad \alpha = -1, 0, 1, 2, \cdots
    \end{eqnarray}
    and tensor type ones
    \begin{eqnarray}
  & & I_{-2\alpha\ \mu\nu} = \int \frac{d^4 k}{(2\pi)^4}\
    \frac{k_{\mu}k_{\nu}}{(k^2 - {\cal M}^2)^{3 + \alpha} }\ , \nonumber \\
  & & I_{-2\alpha\ \mu\nu\rho\sigma} = \int \frac{d^4 k}{(2\pi)^4}\
    \frac{k_{\mu}k_{\nu} k_{\rho}k_{\sigma} }{(k^2 - {\cal M}^2)^{4+ \alpha} }\ , \qquad \alpha =-1, 0, 1,
     2, \cdots
    \end{eqnarray}
   where the number ($-2\alpha$) in the subscript labels the power counting dimension of
   energy momentum in the integrals. Here $\alpha = -1$ and $\alpha = 0$ are
   corresponding to the quadratically divergent integrals ($I_2$, $I_{2 \mu\nu \cdots}$) and
   the logarithmically divergent integrals ($I_0$, $I_{0 \mu\nu \cdots}$).
   We will explicitly see below that all one-loop Feynman integrals
   of vacuum polarization Feynman diagrams can be expressed in terms of the above set
   of integrals by adopting the usual Feynman parameter method. In general, all Feynamn
   integrals from the one-particle irreducible (1PI) graphs can be evaluated into the
   above simple one-fold integrals. Note that the mass factor
   ${\cal M}^2$ is regarded as an effective one which depends on the Feynman
   parameters and the external momenta $p_i$, i.e.,
   ${\cal M}^2 = {\cal M}^2 (m_1^2, p_1^2, \cdots)$.

     To be conceptually helpful, we may mention the above set of loop integrals as the one-fold
     irreducible loop integrals (ILIs) which are evaluated from one loop Feynman diagrams.
     In general, n-fold ILIs are evaluated from n-loop overlapping Feynman integrals of loop
     momenta $k_i$ ($i=1,2,\cdots n$) and are generally defined as the loop integrals in which
     there are no longer the overlapping factors $(k_i-k_j + p_{ij})^2$ $(i\ne j)$
     which appear in the original overlapping Feynman integrals.
     It will be shown that any loop integrals can be evaluated into the corresponding
     ILIs by repeatedly using the Feynman parameter method (see appendix A) and
     a newly formulated UV-divergence preserving parameter method (see eq. (5.6)).

    We now evaluate the vacuum polarization diagrams of gauge fields to one-loop order.
    There are in general four non-vanishing one-loop diagrams (see Figures (1)-(4)).
    Their contributions
    to the vacuum polarization function are denoted by $\Pi_{\mu\nu}^{(i)ab}$ ($i=1,2,3,4$)
    respectively. The first three diagrams arise from the pure Yang-Mills gauge theory,
    their contributions to the vacuum polarization function may be labeled as
    $\Pi_{\mu\nu}^{(g) ab} \equiv \Pi_{\mu\nu}^{(1) ab} + \Pi_{\mu\nu}^{(2) ab} +
    \Pi_{\mu\nu}^{(3) ab} $. The last diagram
  is from the fermionic loop and its contribution to the vacuum polarization function
  is denoted by $\Pi_{\mu\nu}^{(f) ab} = \Pi_{\mu\nu}^{(4)ab}$.
  The total contributions to the gauge field vacuum polarization function are then given by
  summing over all the four diagrams
  \begin{eqnarray}
  \Pi_{\mu\nu}^{ab} = \sum_{i=1}^4 \Pi_{\mu\nu}^{(i) ab}
  \equiv \Pi_{\mu\nu}^{(g) ab} + \Pi_{\mu\nu}^{(f) ab}
  \end{eqnarray}
  Gauge invariance requires that
  $k^{\mu} \Pi_{\mu\nu}^{ab} = \Pi_{\mu\nu}^{ab} k^{\nu} = 0$ which should be true for any
  non-Abelian gauge group and valid for arbitrary fermion number $N_f$. This indicates that
  both parts $\Pi_{\mu\nu}^{(f) ab}$ and $\Pi_{\mu\nu}^{(g) ab}$ should satisfy
  the generalized Ward identities
  \begin{eqnarray}
 & &  k^{\mu} \Pi_{\mu\nu}^{(f) ab} = \Pi_{\mu\nu}^{(f) ab} k^{\nu} = 0\ , \\
  & & k^{\mu} \Pi_{\mu\nu}^{(g) ab} = \Pi_{\mu\nu}^{(g) ab} k^{\nu} = 0
   \end{eqnarray}

  In terms of the one-fold ILIs, the vacuum polarization function  $\Pi_{\mu\nu}^{(f) ab}$
  from fermionic loop is simply given by (for more details see appendix A)
  \begin{eqnarray}
    \Pi_{\mu\nu}^{(f) ab}
     =  -g^2 4N_f C_2  \delta_{ab} \  \int_{0}^{1} dx\ [\ 2 I_{2\mu\nu} (m)
     - I_2(m) g_{\mu\nu} + 2x(1-x) (p^2 g_{\mu\nu} - p_{\mu}p_{\nu} ) I_0(m) \ ]
    \end{eqnarray}
  which shows that the gauge invariance is spoiled only by the quadratic divergent ILIs.
  It then implies that the regularization for which we are
   looking should lead to the following general condition
  \begin{equation}
  I_{2\mu\nu}^R(m) = \frac{1}{2} I_{2}^R(m) \ g_{\mu\nu}
  \end{equation}
  so that the gauge non-invariant terms caused by the quadratically divergent ILIs
  cancel each other and the vacuum polarization function $\Pi_{\mu\nu}^{(f) ab}$ satisfies
  the generalized Ward identity and maintains the gauge invariance.
  Here the integrals with superscript $R$  means the regularized
  ILIs. Obviously, under this condition the regularized vacuum polarization function
  $\Pi_{R \mu\nu}^{(f) ab}$ from fermionic loop becomes gauge invariant and takes a simple form
  \begin{eqnarray}
    \Pi_{R \mu\nu}^{(f) ab} = -g^2 4N_f C_2  \delta_{ab} \
    (p^2 g_{\mu\nu} - p_{\mu}p_{\nu} ) \ \int_{0}^{1} dx\  2x(1-x) I_0^R(m)
  \end{eqnarray}

  The gauge field vacuum polarization function $\Pi_{\mu\nu}^{(g) ab}$ in the pure
  Yang-Mills gauge theory is much complicated as it receives contributions from three diagrams (for a more detailed
  evaluation see appendix A).
  Summing over the contributions from the three diagrams and making some algebra, we then obtain
  \begin{eqnarray}
  \Pi_{\mu\nu}^{(g) ab} & = & g^2 C_1 \delta_{ab}\ \int_{0}^{1} dx\ \{ \ [\
  2(\ 2I_{2\mu\nu} - I_{2}g_{\mu\nu}\ )
  - \lambda \Gamma(3) (1-2x) ( \ I_{2\mu\nu} - \frac{1}{2} I_2 g_{\mu\nu} \ ) \  ] \nonumber \\
  & + & [\ \left(5/2 - 2x(1-x) -3(1-x)(1-2x) \ \right) p^2 g_{\mu\nu}
  - \left( 1 + 4x(1-x)\ \right) p_{\mu} p_{\nu} \  ] \ I_0  \nonumber \\
  & + & \lambda \Gamma(3) \ [\ \left(\ (1-x) (1 + 2x +8x^2) p^2g_{\mu\nu}/2
  - x(2+x-2x^2) p_{\mu}p_{\nu}\ \right) \ I_0
  - 4x^3 p^{\rho}p^{\sigma}g_{\mu\nu}I_{0\rho\sigma} \nonumber \\
  & - & \left( \ (1-2x - 2(3-2x)x^2 \ ) p^2 g_{\mu}^{\rho}g_{\nu}^{\sigma}
  - (3x -4x^2 + 4x^3) p^{\rho}\ ( g_{\mu}^{\sigma} p_{\nu} + g_{\nu}^{\sigma} p_{\mu} )
  \ \right) \ I_{0\rho\sigma}\ ] \nonumber \\
  & + & \lambda \Gamma(3) \ [ \ (\ x+x^2 -6x^4 + 4x^5\ )p_{\mu}p_{\nu}
  - (\ x + 2x^2 -3x^3 -4x^4 + 4x^5) p^2 g_{\mu\nu} \ ]\ p^2\ I_{-2} \nonumber \\
  & + & \frac{1}{2}\lambda^2 \Gamma(4)\ x(1-x)\ [ \ p^4 g_{\mu}^{\rho}g_{\nu}^{\sigma}
   +  p_{\mu}p_{\nu}p^{\rho}p^{\sigma}
  - p^2 p^{\rho}\ ( g_{\mu}^{\sigma} p_{\nu} + g_{\nu}^{\sigma} p_{\mu} )\ ]\ I_{-2\rho\sigma} \ \}
  \end{eqnarray}

  To simplify the above expression, it is useful to notice the following property of the
   integrals with respect to the Feynman parameter $x$
  \begin{equation}
  \int_{0}^{1} dx\ x\ I[x(1-x)] =  \int_{0}^{1} dx\ (1-x)\ I[x(1-x)]
  = 1/2 \int_{0}^{1} dx\ I[x(1-x)]
  \end{equation}
  where the integrand $I[x(1-x)]$ is the function of the combination
  $x(1-x)$ and is invariant under the change of the integral parameter
  $x\rightarrow 1-x$. With the aid of this property, one can derive more useful
  identities (see appendix A).

   To be more useful, we may formally express the tensor type ILIs
   in terms of the scalar type ones
   \begin{eqnarray}
   & & I_{0\mu\nu} = \frac{1}{4} a_0\ I_0\ g_{\mu\nu} \\
   & & I_{-2\mu\nu} = \frac{1}{4} a_{-2}\ I_{-2}\ g_{\mu\nu}
   \end{eqnarray}
   Here $a_0$ and $a_2$ are quantities relying on the regularization and expected to be
   determined from the gauge invariant conditions.
   Adopting the above definitions and identities,
  the gauge field vacuum polarization function $\Pi_{\mu\nu}^{(g) ab} $ is simplified to be
  \begin{eqnarray}
  \Pi_{\mu\nu}^{(g) ab} & = & g^2 C_1 \delta_{ab}\ \int_{0}^{1} dx\ \{ \
   2(\ 2I_{2\mu\nu} - I_{2}g_{\mu\nu}\ ) +
  [1 + 4x(1-x)]\ I_0\ (p^2g_{\mu\nu} - p_{\mu}p_{\nu}) \nonumber \\
  & + & \frac{1}{4}\lambda \Gamma(3)\ [\ 1 + 6x(1-x)(a_0 + 2) -
  3a_0 \ ]
  I_{0}\ (p^2g_{\mu\nu} - p_{\mu}p_{\nu})  \nonumber \\
   & + & \frac{1}{8}\lambda^2 \Gamma(4)\ a_{-2}\ x(1-x)\ p^2\ I_{-2}\
     (p^2g_{\mu\nu} - p_{\mu}p_{\nu}) \nonumber \\
   & + & \frac{1}{4}\lambda \Gamma(3)\
   [\ 2x(1-x)(a_0 + 2) -1\ ]\ I_0\  p_{\mu}p_{\nu}   \\
   & + & \lambda \Gamma(3)\ x(1-x)[\ \left(\ 1 + 4x(1-x) \right) p_{\mu}p_{\nu}
   -\left(\ 1/2 + 6x(1-x) \right) p^2 g_{\mu\nu} \ ] p^2\ I_{-2} \  \} \nonumber
   \end{eqnarray}
   It is seen that not only the quadratically divergent integrals in the first and last terms,
   but also the logarimically divergent term (the fifth term) can in general destroy
   the gauge invariance. To explicitly obtain the gauge invariant conditions,
   we may further adopt the identity
  \begin{equation}
  \int_{0}^{1} dx\ [\ 6x(1-x) - 1\ ]\ I_{0}\
  = - \int_{0}^{1} dx\ 2x(1-x)[\ 1-4x(1-x)\ ]\ p^2\ I_{-2}
  \end{equation}
  which is obtained by performing partial integration with respect to $x$ and using the
  relevant integral identities presented in appendix A together with the following identity
  \begin{equation}
  \frac{\partial I_0}{\partial x} = -2(1-2x) p^2 \ I_{-2}
  \end{equation}
  Thus the integral in the fifth term of eq.(2.18) can be rewritten as
  \begin{eqnarray}
 \int_{0}^{1} dx\ [\ 2x(1-x)(a_0 + 2) -1\ ]\ I_0 & = &
  - 2\ \int_{0}^{1} dx\ x(1-x)[\ 1- 4x(1-x) ]\ p^2 \ I_{-2} \nonumber \\
  &  + & (a_0 -1)\ \int_{0}^{1} dx\ 2x(1-x)\ I_{0}
  \end{eqnarray}
  Substituting the above results into eq.(2.18), the gauge field vacuum polarization function
  $\Pi_{\mu\nu}^{(g) ab} $  is further simplified to be
  \begin{eqnarray}
  \Pi_{\mu\nu}^{(g) ab} & = & g^2 C_1 \delta_{ab}(p^2g_{\mu\nu} - p_{\mu}p_{\nu})
  \ \int_{0}^{1} dx\ \{ \
  [1 + 4x(1-x)]\ I_0\  \nonumber \\
  & + & \frac{1}{4}\lambda \Gamma(3)\ [\ \left(\ 1 + 6x(1-x)(a_0 + 2) - 3a_0 \right)
  I_{0}\ -  2x(1-x) \left(\ 1 + 12x(1-x)\ \right) p^2\ I_{-2} \ ] \nonumber \\
    & + & \frac{1}{8}\lambda^2 \Gamma(4)\ a_{-2}\ x(1-x)\ p^2\ I_{-2}\ \} \nonumber  \\
   & + &  g^2 C_1 \delta_{ab}\ \int_{0}^{1} dx\ \{\ 2(\ 2I_{2\mu\nu} - I_{2}g_{\mu\nu}\ )
  +  \lambda \Gamma(3)\frac{a_0 -1}{2}\ p_{\mu}p_{\nu}\ x(1-x)\ p^2 \ I_{-2}\ \}
   \end{eqnarray}
  Here the gauge non-invariant term is associated with the factor $(a_0 - 1)$.
  It then implies that the new regularization for which we are looking should lead
  $a_0$ to be unit (i.e., $a_0=1$). Based on the above simplified form $\Pi_{\mu\nu}^{(g) ab}$,
  it becomes clear that in order to maintain the gauge invariance for the gauge field
  vacuum polarization function $\Pi_{\mu\nu}^{(g) ab} $ in the pure Yang-Mills gauge theories,
  the new regularization must satisfy the following two conditions
  \begin{eqnarray}
   & & I_{2\mu\nu}^R = \frac{1}{2} I_{2}^R\ g_{\mu\nu} \\
   & & I_{0\mu\nu}^R = \frac{1}{4} I_{0}^R\ g_{\mu\nu}
   \end{eqnarray}
   where the superscript $R$ denotes the regularized ILIs.
  The first condition for the quadratically divergent ILIs is the same
   as the one required from the gauge invariance of the gauge field vacuum polarization
   function $\Pi_{\mu\nu}^{(f) ab} $ due to fermionic loops.  The gauge invariance
   of the vacuum polarization function $\Pi_{\mu\nu}^{(g)ab} $ from non-Abelian Yang-Mills
   gauge theories requires an additional condition concerning the logarithmically divergent
   ILIs.

   So far, we have provided at one-loop order a general proof for
   two necessary conditions that must be satisfied for any regularization to maintain the
   gauge symmetry. Under these
   two conditions, the regularized whole gauge field vacuum polarization function $\Pi_{\mu\nu}^{ab} $
   can explicitly be expressed in terms of gauge invariant form
 \begin{eqnarray}
  \Pi_{R \mu\nu}^{ab} & = & \Pi_{R \mu\nu}^{(g) ab} + \Pi_{R \mu\nu}^{(f) ab} \nonumber \\
  & = & g^2 C_1 \delta_{ab}(p^2g_{\mu\nu} - p_{\mu}p_{\nu})
  \ \int_{0}^{1} dx\ \{ \
  [1 + 4x(1-x)]\ I_0^R\  \nonumber \\
  & + & \frac{1}{2}\lambda \Gamma(3)\ [\ \left(\ 9x(1-x) - 1  \right)
  I_{0}^R\ - x(1-x) \left(\ 1 + 12x(1-x)\ \right) p^2\ I_{-2}^R \ ] \nonumber \\
   & + & \frac{1}{8}\lambda^2 \Gamma(4)\ a_{-2}\ x(1-x)\ p^2\ I_{-2}^R\ \} \nonumber \\
   & - & g^2 4N_f C_2  \delta_{ab} (p^2g_{\mu\nu} - p_{\mu}p_{\nu})\
   \int_{0}^{1} dx\ 2x(1-x)  I_0^R(m)
 \end{eqnarray}
  Once applying again the identity eq.(2.19),  the gauge field vacuum polarization function
  $\Pi_{R \mu\nu}^{ab}$ can be re-expressed into a much simpler gauge invariant form
 \begin{eqnarray}
  \Pi_{R \mu\nu}^{ab}
  & = & g^2\delta_{ab}\ (p^2g_{\mu\nu} - p_{\mu}p_{\nu}) \ \int_{0}^{1} dx\ \{ \
  C_1\ [\ 1 + 4x(1-x) + \lambda/2 \ ] \ I_0^R\  \nonumber \\
   & - & N_f C_2\ 8x(1-x)\  I_0^R(m)  - 4C_1 \lambda\ [\
   1 - 3\lambda a_{-2}/16\  ] \ x(1-x)\ p^2\ I_{-2}^R\  \  \}
 \end{eqnarray}

  Before proceeding, we would like to address the following observations implied
  from the above general evaluations:
  First,  only UV divergent ILIs destroy the generalized
  Ward identities (or gauge invariant conditions) for
  the vacuum polarization function of gauge fields, which strongly
  suggests the necessity of regularizing the divergent ILIs in order to make
   the divergent loop integrals to be meaningful.  As expected, the UV convergent ILIs
  are not constrained from gauge invariant conditions.
  Second, any consistent regularization must satisfy the necessary conditions
  resulted from the generalized Ward identities (or gauge invariant conditions)
  $k^{\mu}\Pi_{\mu\nu}^{ab}=0$.  It also becomes clear that under the two necessary conditions
  the gauge fixing term $ (\partial^{\mu}A_{\mu}^a )^2$ in non-Abelian gauge theories
  is not affected by the loop graphs, which arrives at the same conclusion
  as the one in Abelian gauge theory.
  Third, the quadratically divergent terms may not
  necessarily be a harmful source for the gauge invariance once a
  rigorous regularization can be found to satisfy the necessary conditions
  for the regularized quadratically divergent ILIs, so that all the regularized quadratically
  divergent ILIs cancel each other.
  Last but not least, in contrast to the usual case in the dimensional regularization,
   in the most general case, the tadpole graphs of Yang-Mills gauge fields actually play
   an essential role for maintaining the gauge invariance, which can explicitly be seen from
  the above general evaluation on the vacuum polarization function.
  In fact, it is the tadpole graph that leads to
  the manifest gauge invariant form of the vacuum polarization function even
  without carrying out any explicit integrations over the Feynman parameter $x$ and
  the loop momentum $k$.

      Here we have only presented the explicit evaluation for the vacuum polarization function of gauge bosons.
      For all other two-point, three-point and four-point one loop Feynman graphs of fermions, gauge
      and ghost bosons arising from the gauge theory described by the Lagrangian eqs.(2.1-2.5), it can be shown
      that they all maintain the gauge invariance and satisfy the generalized Ward
      identities at one-loop level as long as the following
      conditions for the regularized ILIs hold
     \begin{eqnarray}
     & & I_{2\mu\nu}^R = \frac{1}{2} g_{\mu\nu}\ I_2^R, \\
    & &  I_{2\mu\nu\rho\sigma }^R = \frac{1}{8} g_{ \{\mu\nu} g_{\rho\sigma\} }\ I_2^R
     \end{eqnarray}
     for quadratically divergent ILIs, and
     \begin{eqnarray}
     & & I_{0\mu\nu}^R = \frac{1}{4} g_{\mu\nu} \ I_2^R, \\
     & & I_{0\mu\nu\rho\sigma }^R = \frac{1}{24} g_{ \{\mu\nu} g_{\rho\sigma\} }\ I_0^R
     \end{eqnarray}
     for logarithmically divergent ILIs, as well as
     \begin{eqnarray}
     & & I_{-2\mu\nu}^R = \frac{1}{6} g_{\mu\nu}\ I_{-2}^R, \\
     & & I_{-2\mu\nu\rho\sigma }^R = \frac{1}{48} g_{ \{\mu\nu} g_{\rho\sigma\} }\ I_{-2}^R
     \end{eqnarray}
     for convergent ILIs.   Where we have used the notation
     \begin{equation}
      g_{ \{\mu\nu} g_{\rho\sigma\} } \equiv g_{\mu\nu}g_{\rho\sigma} + g_{\mu\rho}g_{\nu\sigma}
     + g_{\mu\sigma}g_{\rho\nu}
      \end{equation}

       We may mention the above conditions as the general regularization-independent
       {\it consistency conditions} for maintaining the gauge invariance of
       theories. It is easily shown that the dimensional
       regularization does satisfy the above consistency
       conditions.

        It is of interest to see that once the above consistency conditions for the regularized ILIs hold, the
        divergent structure of the theories can be well characterized
        by only two regularized scalar type ILIs $I_0^R$ and $I_2^R$.
        In general, one may not need to evaluate those ILIs by any
        specific regularization scheme. As it has been turned out that in the underlying gauge theories
        the quadratic divergences arising from the vacuum polarization functions cancel each other,
        while the logarithmic divergences can be fully absorbed into the redefinitions of the
        coupling constants and the relevant fields. In this sense, one
        arrives at a regulator free scheme for the underlying
        theories. Mathematically, one only needs to prove that
        such a regularization exists and can result in the
        above consistency conditions among the ILIs. Practically, we shall explicitly show in the
        following sections that there does exist, in addition to the dimensional regularization,
        an alternative regularization scheme that can lead to the above consistency conditions and
        in the meantime match to the four criteria stated in the introduction.

        Before ending this section, we would like to point out that
        the above consistency conditions hold exactly
        without any ambiguities, which differs from the so-called consistency
       relations imposed in the implicit regularization\cite{IR}.
       Where some local arbitrary counterterms parameterized by
       (finite) differences of divergent integrals arise, they may (or not)
        be determined by the physical conditions. The reason
       is that the consistency relations in the implicit
       regularization are imposed among the basic divergent
       integrals which are independent of the external momenta.
       Namely the mass factor in those basic divergent integrals remains the original mass
       of fields. This is explicitly seen from the following identity and truncation which have been
       used in the implicit regularization

       \[ \frac{1}{[(k+p_i)^2 -m^2]} = \sum_{j=0}^{N}
       \frac{(-1)^j (p_i^2 + 2p_i \cdot k )^j}{(k^2 -m^2)^{j+1} }
        +  \frac{(-1)^{N+1} (p_i^2 + 2p_i \cdot k )^{N+1}}{(k^2 -m^2)^{N+1} [(k+p_i)^2 -m^2 ] }   \]

       where $p_i$ are the external momenta and $N$ is chosen so
       that the last term is finite under integration over $k$. It
       must be this truncation that causes the local arbitrary
       counterterms.

  \section{Symmetry principle preserving and infinity free regularization}

    We now proceed to explore a possible new regularization which can preserve
    the basic symmetry principles of the theory and lead to an
    infinity free renormalization. Before doing so, it is helpful
   to briefly look at the well-known two simple regaularizations, i.e.,
    cut-off regularization and dimensional regularization, through checking
   the consistency conditions resulted from the generalized Ward identities (or
   gauge invariant conditions).

    The regularized ILIs in the cut-off regularization are presented in appendix B.
    From the explicit forms overthere, one easily sees that
    \begin{eqnarray}
     I_{2\ \mu\nu}^R |_{cutoff}
     & = & \frac{1}{2}g_{\mu\nu}\ I_2^R |_{cutoff} - \frac{1}{4}g_{\mu\nu}\ \frac{-i}{16\pi^2}
           \ [\ \Lambda^2 + \frac{1}{2} {\cal M}^2 \ ]
           \neq \frac{1}{2} g_{\mu\nu} I_2^R |_{cutoff}\ ,    \\
  I_{0\ \mu\nu}^R |_{cutoff}
      & = &  \frac{1}{4}g_{\mu\nu}\ I_0  |_{cutoff} +   \frac{i}{64\pi^2}g_{\mu\nu}
     \left( \frac{{\cal M}^2}{\Lambda^2 + {\cal M}^2}
     - \frac{{\cal M}^4}{2(\Lambda^2 + {\cal M}^2)^2}  - \frac{1}{2} \right)
     \neq \frac{1}{4} g_{\mu\nu} I_0^R |_{cutoff}
    \end{eqnarray}
  where $\Lambda^2$ is the cut-off momentum. Obviously, the regularized ILIs
  by the simple cut-off regularization do not satisfy the consistency conditions. In fact,
  not only the quadratically divergent terms can destroy the gauge invariance,
  even the finite terms may spoil the gauge conditions in the simple cut-off regularization.
  Whereas one may notice that the involved logarithmically divergent terms in the regularized ILIs
  do satisfy the consistency conditions. This is actually the case realized by the
  dimensional regularization. From the explicit results given in appendix B for
  the regularized ILIs in the dimensional
  regularization, it is not difficult to yield, by using the
  identity $\Gamma(1 + z) = z\ \Gamma(z)$,
  the desired relations
  \begin{eqnarray}
  I_{2\ \mu\nu}^R |_{Dim.} = \frac{1}{2}g_{\mu\nu}\ I_2^R |_{Dim.}\ , \qquad I_{0\ \mu\nu}^R |_{Dim.}
  = \frac{1}{4}g_{\mu\nu}\ I_0^R |_{Dim.}
  \end{eqnarray}
  It then becomes manifest why the dimensional regularization preserves gauge symmetry.

   To see the difference between the cut-off regularization and dimensional regularization,
    one may make an expansion for $\varepsilon\rightarrow 0$. Using the properties of the
    $\Gamma$-function
    \begin{equation}
    \Gamma(\frac{\varepsilon}{2}) = \frac{2}{\varepsilon} - \gamma_E + O(\varepsilon), \qquad
    \Gamma (-1 + \frac{\varepsilon}{2} ) =
     - \Gamma(\frac{\varepsilon}{2})/(1- \frac{\varepsilon}{2})
    \end{equation}
    with the Euler number $\gamma_E=0.5772 \cdots $,  one has
   \begin{eqnarray}
     & & I_2^R |_{Dim.} = \frac{i}{16\pi^2} \ {\cal M}^2 \  [ \ \frac{2}{\varepsilon}
      - \gamma_E + 1 - \ln (4\pi {\cal M}^2 ) + O(\varepsilon) \ ] \\
     & & I_0^R |_{Dim.} = \frac{i}{16\pi^2}\  [ \ \frac{2}{\varepsilon}  - \gamma_E
     - \ln (4\pi {\cal M}^2 ) + O(\varepsilon) \ ]
    \end{eqnarray}
    Comparing them to the results in the cut-off regularization, the divergences
    given by $1/\varepsilon$ must be logarithmic, i.e.,
    $1/\varepsilon \rightarrow \ln \Lambda $.
    This may become more clear from the vanishing integral in the dimensional regularization
    \begin{eqnarray}
    \int \frac{d^D k}{k^2} = 0\
    \end{eqnarray}
    It is this property that suppresses the quadratic divergence in the dimensional regularization.
    Actually, the dimensional regularization destroys
    all the divergences in a power of cut-off momentum because of the vanishing integrals
    \begin{eqnarray}
    \int \ d^D k\ ( k^2 )^l = 0\  \qquad l=0,\ 1,\ \cdots
    \end{eqnarray}
    which implies that divergences appearing in any loop integrals are
    expressed only in terms of the expansion by powers of $1/\varepsilon$ in the dimensional
    regularization.

      We now turn to investigate the possible new regularization.
      It has been seen from the previous section that the divergences actually arise
      from loop integrals. It is the UV divergent loop integrals which destroy
      the generalized Ward identities of gauge invariance. This observation naturally
      motivates us to propose a new regularization scheme which is supposed to regularize
      the ILIs of loop graphs, i.e., the whole Feynman diagrams, instead of only the
      propagators carrying out in the Pauli-Villars inspired methods. Unlike the Pauli-Villars
      regularization, we are not going to introduce any regulator fields to modify
      the action of original theory. Alternative to the dimensional regularization, we shall not
      change the space-time dimension of initial theories.
      Our motivation is only to modify the very high energy behavior of loop integrals and/or
      the very low energy behavior of loop integrals if the integrals concern IR divergent.
      In fact, we do not well understand the physics at very short distances,
      especially the physics above the Planck scale. Thus it must not be surprised
      that at infinitely large loop momentum, the Feynman diagrams and the corresponding
      Feynman integrals become ill-defined due to the lack of convergence.
      They are actually meaningless because of the UV divergent
      integrals which destroy the basic gauge symmetry principle in gauge theories.
      Moreover, it is also a hard task in certain QFTs to deal with the physics
      at a very low energy scale when the nonperturbative effects of strong interactions
      become important at that scale. In particular, the same serious problem may be faced when the
      Feynman integrals also become ill-defined at zero momentum due to
      the existence of IR divergences. All of these considerations strongly
      suggest the necessity of regularization for QFTs.

      Here we are going to present a new regularization
      without introducing any additional Lagrangian formulation to modify the original theory
      as usually done in the Pauli-Villars inspried regularization.
      What we are doing is only to require the new regularization possessing
      the most important and necessary properties, namely: the regularization shall
      make the meaningless divergent
      ILIs to be well-defined and physically meaningful, and also to be applicable
      to both underlying (renormalizable) QFTs and effective QFTs. Furthermore the new
      regularization shall preserve the initial divergent behavior and
      meanwhile maintain the important symmetry principles in the original theory.
      More specifically we are going to work out the regularization
     which will ensure the regularized divergent ILIs satisfying the
     consistency conditions of gauge invariance even when keeping the
      quadratic `cut-off' terms in the loop integrals. In general, the quadratic `cut-off' terms
      will be found in the new regularization to be harmless for the underlying (renormalizable) QFTs
      but very important for the effective QFTs.
      Of important, we shall finally verify that the physical quantities are independent of
      any regulators (or regularization quanta) introduced in the regularization
      scheme, namely the regulators (or regularization quanta)
      will eventually be decouple from the well-defined regularized theory.

      The prescription for the new regularization is simple: supposing that the ILIs of loop graphs
      have been rotated into the four dimensional Euclidean space of momentum, we then replace
      in the ILIs the loop integrating variable $k^2$ and the loop integrating
      measure $\int d^4 k$ by the corresponding regularized ones $[k^2]_l$ and $\int [d^4 k]_l$
    \begin{eqnarray}
       k^2 \rightarrow [k^2]_l \equiv k^2 + M^2_l\ , \qquad
      \int d^4 k \rightarrow
      \int [d^4 k]_l \equiv \lim_{N, M_l^2} \sum_{l=0}^{N} c_l^N \int d^4 k
      \end{eqnarray}
     In here $M_l^2$ ($ l= 1,\ \cdots $) may be regarded as the mass factors of regulators
     (or regularization quanta). For $l=0$,
     we shall have $M_0^2 = 0$ to maintain the original integrals. If the original integrals
     are IR divergent, one can set $M_0^2 = \mu_s^2$ to avoid IR divergent
     problem. In general, the mass factors $M_l^2$ ($ l= 1,\ \cdots $) of the regulators
     (or regularization quanta) are introduced as unknown parameters and their values
     can be infinitely large. It is seen that the initial integral is recovered by letting
     the regulator mass factors $M_l^2$ go to infinity, i.e., $M_l^2\rightarrow \infty$ ($ l= 1,\ \cdots $).
      Where $c_l^N$ ($l=1,2, \cdots N $)
      are the corresponding coefficients and will be judiciously chosen to
      modify the short-distance behavior of loop integrals and remove the divergences.
      Specifically, in order to suppress the divergences at high orders, the coefficients
      $c_l^N$ with the initial condition $ c_0^N = 1$ may be chosen to satisfy the conditions
     \begin{equation}
     \lim_{N, M_l^2}\sum_{l=0}^{N} c_l^N\ (M_l^2)^n = 0 \  \quad (n= 0, 1, \cdots ) \ ,
     \qquad  c_0^N = 1
      \end{equation}
      so that the regularized integrals of $\int d^4 k (k^2)^n $ vanish
      \begin{eqnarray}
       \int [d^4 k]_l \  (k^2 + M_l^2)^n \equiv
       \lim_{N, M_l^2} \sum_{l=0}^{N} c_l^N \int d^4 k \  (k^2 + M_l^2)^n = 0 \
      \qquad (n= 0, 1, \cdots )
       \end{eqnarray}
      which coincide with the ones in the dimensional regularization.
      Where the integer $N$ counts the corresponding numbers of regulators (or regularization quanta)
      and is in principle a free parameter. Its value can be infinitely large. Practically,
      the regulator number $N$ may also only need to reach a sufficiently large value
      around which the regularized ILIs become insensitive to the choice for its specific values.
      Finally, we shall prove that the regularized ILIs
      must not be sensitive to the choices
      for the various mass factors $M_l^2$ of regulators (or regularization quanta) and
      the corresponding coefficients $c_l^N$ as well as the regulator number $N$.

       We would like to point out that the order of operations for performing the integration
       over the loop momentum and making the limitation to the regulator mass factors
       is very important for divergent integrals. One should first carry out
       the integration over the loop momentum, then take the
       infinite limit for the regulator mass factors, ($M_l^2\rightarrow \infty$ ($ l= 1,\ \cdots
       $). Physically, this order of operations implies that
     \begin{equation}
      \frac{M_l^2}{k^2} \rightarrow 0 \qquad \mbox{at} \quad k^2 \rightarrow \infty
      \quad \mbox{and} \quad M_l^2 \rightarrow \infty \   \quad (l =1,2, \cdots)
      \end{equation}
      Namely, the loop momentum in the Euclidean space goes to infinity in a much faster way
      than the mass factors of regulators ( or regularization quanta).

       Before a detailed discussion, one may notice that
       the present regularization scheme will result
       in a quite different property for the quadratically divergent integral
       in comparison with the dimensional regularization.
       This is because in the dimensional regularization,
       the quadratically divergent integral vanishes for the massless case (see eq.(3.7)).
       While in the present regularization scheme, only when the loop momentum $k$ becomes
       sufficiently large with $k^2 \gg M^2_l$, one then approximately has
     \begin{eqnarray}
      \int_{k^2 \gg M_l^2}^{\infty} [d^4 k]_l \ \frac{1}{k^2 + M_l^2}
     \sim \int_{k^2 \gg M_l^2} [d^4 k]_l \ \frac{1}{k^2}
     \sum_{n=0}^{\infty} \left(\frac{M_l^2}{k^2} \right)^n =0
     \end{eqnarray}
     which will lead to a distinguishable property in the present regularization scheme.
     We shall come to a more detailed discussion below.

      Let us first apply the new regularization scheme to the scalar type divergent ILIs
     \begin{eqnarray}
     I_2^R & = &  i \int \frac{[d^4 k]_l}{(2\pi)^4}\ \frac{1}{-k^2 - \hat{M}_l^2 }  \equiv
     -i\lim_{N, M_l^2}\sum_{l=0}^{N} c_l^N \int
     \frac{d^4 k}{(2\pi)^4}\ \frac{1}{k^2 + \hat{M}_l^2 } \nonumber \\
     & = & \frac{-i}{16\pi^2} \  \lim_{N, M_l^2}\sum_{l=0}^{N} c_l^N\  \hat{M}_l^2 \ln \hat{M}_l^2 \\
     I_0^R & = &  i\int \frac{[d^4 k]_l}{(2\pi)^4}\
     \frac{1}{(-k^2 - \hat{M}_l^2)^2 } \equiv
     i\lim_{N, M_l^2}\sum_{l=0}^{N} c_l^N \int \frac{d^4 k}{(2\pi)^4}\
     \frac{1}{(k^2 + \hat{M}_l^2)^2 } \nonumber \\
     & = & - \frac{i}{16\pi^2} \  \lim_{N, M_l^2}\sum_{l=0}^{N} c_l^N\  \ln \hat{M}_l^2
     \end{eqnarray}
     with
     \[ \hat{M}_l^2 \equiv M_l^2 + {\cal M}^2 \]

      To check whether the above new method
     satisfies the consistency conditions for maintaining gauge invariance,
     we further evaluate the corresponding tensor type ILIs by the same regularization scheme
    \begin{eqnarray}
     I_{2\mu\nu}^R & = & i\int \frac{[d^4 k]_l}{(2\pi)^4}\
     \frac{-k_{\mu}k_{\nu}}{(-k^2 - \hat{M}_l^2 )^2 }  \equiv
     -i\lim_{N, M_l^2}\sum_{l=0}^{N} c_l^N \int
     \frac{d^4 k}{(2\pi)^4}\ \frac{k_{\mu}k_{\nu}}{(k^2 + \hat{M}_l^2 )^2 } \nonumber \\
     & = & \frac{1}{4}g_{\mu\nu}\ \frac{-i}{16\pi^2} \ \lim_{N, M_l^2}\sum_{l=0}^{N} c_l^N  \int d^2 k
         \  [\ 1 + \frac{-2\hat{M}_l^2 }{k^2 + \hat{M}_l^2 }
      + \frac{\hat{M}_l^4 }{(k^2 + \hat{M}_l^2)^2 } \  ] \nonumber \\
    & = & \frac{1}{2} g_{\mu\nu}\ \frac{-i}{16\pi^2}
    \  \lim_{N, M_l^2}\sum_{l=0}^{N} c_l^N\  \hat{M}_l^2 \ln \hat{M}_l^2 \nonumber \\
    & = & \frac{1}{2} g_{\mu\nu}\ I_2^R \\
     I_{0\mu\nu}^R & = & i \int \frac{[d^4 k]_l}{(2\pi)^4}\
     \frac{-k_{\mu}k_{\nu}}{(-k^2 - \hat{M}_l^2)^3 } \equiv
     i\lim_{N, M_l^2}\sum_{l=0}^{N} c_l^N \int \frac{d^4 k}{(2\pi)^4}\
     \frac{k_{\mu}k_{\nu}}{(k^2 + \hat{M}_l^2)^3 } \nonumber \\
     & = & \frac{1}{4}g_{\mu\nu}\ \frac{i}{16\pi^2} \ \lim_{N, M_l^2}\sum_{l=0}^{N} c_l^N  \int d^2 k
         \ [\ \frac{1}{k^2 + \hat{M}_l^2 } + \frac{-2\hat{M}_l^2 }{ (k^2 + \hat{M}_l^2)^2 }
      + \frac{\hat{M}_l^6 }{(k^2 + \hat{M}_l^2)^3 } \  ] \nonumber \\
     & = & \frac{1}{4}g_{\mu\nu}\ \frac{-i}{16\pi^2}
     \  \lim_{N, M_l^2}\sum_{l=0}^{N} c_l^N\  \ln \hat{M}_l^2
     \nonumber  \\
     & = & \frac{1}{4}g_{\mu\nu}\  I_0^R
     \end{eqnarray}
    which shows that the method does preserve gauge invariance.
    Note that in obtaining the above results, only two conditions have actually been used, i.e.,
     \begin{equation}
     \lim_{N, M_l^2}\sum_{l=0}^{N} c_l^N\ (M_l^2)^n = 0 \  \quad \mbox{with} \quad n= 0, 1
      \end{equation}

     To be able to directly compare with the cut-off regularization and dimensional regularization,
     and also to explicitly check whether the present new regularization maintains
     the harmless and important quadratic `cut-off' terms, it is of help to find out the
     manifest expressions for the mass factors $M_l^2$ of regulators (or regularization quanta) and the
     corresponding coefficients $c_l^N$. For a self consistent reason,
     the explicit forms of $M_l^2$ and $c_l^N$ are assumed to be uniquely determined through the
     conditions given in eq.(3.11). This requires that for a given $N$ the number $n$ of
     conditions should be sufficient in comparison with the unknown variables $2N$ which consist of
     N's mass parameters $M_l^2$ and N's coefficient functions $c_l^N$.
     Also for an economic reason, it is supposed that the regularization scheme only involves a minimal
     set of parameters. With these considerations, we come to the following simple
     choice for the mass factors $M_l^2$ of regulators (or regularization quanta)
     \begin{equation}
      M_l^2 = \mu_s^2 + l M_R^2 \qquad l= 0,1, \cdots
     \end{equation}
      Here $M_R$ may be regarded as a basic mass scale for all regulators
      (or regularization quanta). The energy scale $\mu_s$ may be viewed as an intrinsic
      mass scale and will be seen to provide an IR cut-off energy scale.
      In the absence of IR divergences,
      the scale $\mu_s$ can in principle run to zero $\mu_s = 0$.
      In this case, there is only one free parameter $M_R$.
      In general, even without IR problem the scale $\mu_s$ can still be introduced to play
      the role of sliding energy scale (SES). A particularly interesting case is that
      a mass gap around the scale $\mu_s = \mu_c$ may be generated due to strong interactions,
      if this case happens, the SES $\mu_s$ is naturally set to be at the mass gap
      $\mu_s = \mu_c$.

         Substituting the above explicit form of $M_l^2$ to the
      mass-coefficient conditions given in eq.(3.11) for the regulators (or regularization quanta),
      we arrive at the interesting constraints purely for the coefficients $c_l^N$
     \begin{equation}
     \lim_{N\rightarrow \infty}\sum_{l=0}^{N} c_l^N\ l^n = 0 \  \quad \mbox{with} \quad
      n= 0, 1, \cdots N-1\,
      \end{equation}
     For a given $N$, the coefficients $c_l^N$ are completely determined by solving
     $N$'s linear equations with the initial condition $c_0^N = 1$.
     The result is explicitly given by
     \begin{equation}
     c_l^N = (-1)^l \frac{N!}{(N-l)!\ l!}
     \end{equation}
    which may be regarded as the sign-associated Combinations. More precisely speaking,
    $(-1)^l c_l^N$ is the number of
    combinations of $N$ regulators (or regularization quanta), taken $l$ at a time.
    The initial condition $c_0^N = 1$ is required so that
    we shall recover the original ILIs when no any regulator is introduced.

     Up to now, we have completely presented an explicit new regularization scheme in which
     there exists no free parameters when both the basic mass scale $M_R$ and the number $N$
     of regulators (or regularization quanta) are taken to be infinitely large, i.e.,
     $M_R ,\ N \rightarrow \infty$, and the SES $\mu_s$ is set to be zero $\mu_s = 0$.
     What we shall do next step is to prove that the regularized ILIs are insensitive to the
     regularization scheme. More specifically
     the regularized ILIs should eventually be independent of the basic mass scale
     $M_R$ and the number $N$ of regulators (or regularization quanta)
     when they are approaching to the infinitely large limits or
     practically when they are taken to be sufficiently large.
     To see that, let us apply the above explicit forms for the mass factors
     $M_l^2$ of regulators (or regularization quanta) and
     the coefficients $c_l^N$ to the regularized ILIs. After some algebra, we obtain
     the following explicit and simple forms for the regularized ILIs
     \begin{eqnarray}
     I_2^R & = & - \frac{i}{16\pi^2} \ \{\ M_c^2
      - \mu^2 [ \ \ln \frac{M_c^2}{\mu^2} - \gamma_w + 1
      +   y_2(\frac{\mu^2}{M_c^2})  \ ] \ \} \\
     I_0^R & = & \frac{i}{16\pi^2} \
      [ \ \ln \frac{M_c^2}{\mu^2 } - \gamma_w + y_0(\frac{\mu^2}{M_c^2}) \ ] \\
      I_{-2}^R & = & - \frac{i}{16\pi^2} \ \frac{1}{2\mu^2}[\ 1
      - y_{-2}(\frac{\mu^2}{M_c^2})  \ ]
       \end{eqnarray}
     where we have introduced the following definitions and notations
     \begin{eqnarray}
     & &  M_c^2 =   \lim_{N,M_R}\sum_{l=1}^{N} c_l^N\ (l \ln l) M_R^2  \\
     & & y_0 (\frac{\mu^2}{M_c^2}) = - \lim_{N,M_R}\sum_{l=1}^{N} c_l^N\
     \ln (1 + \frac{\mu^2}{lM_R^2} )
     = \sum_{n=1}^{\infty} \frac{(-)^{n-1} L_n }{n\ n!}
      \left(\frac{\mu^2}{M_c^2}\right)^n \\
     & & y_1 (\frac{\mu^2}{M_c^2}) - 1 =  \lim_{N,M_R}\sum_{l=1}^{N} c_l^N\
     \frac{lM_R^2}{\mu^2} \ln (1 + \frac{\mu^2}{lM_R^2} )
      = \sum_{n=1}^{\infty} \frac{(-)^{n-1} L_n }{(n+1) n!}
      \left(\frac{\mu^2}{M_c^2}\right)^n -1  \\
      & & y_2(\frac{\mu^2}{M_c^2})  = y_0(\frac{\mu^2}{M_c^2}) - y_1(\frac{\mu^2}{M_c^2})
      = \sum_{n=1}^{\infty} \frac{(-)^{n-1} L_n }{n(n+1) n!}
      \left(\frac{\mu^2}{M_c^2}\right)^n \\
      & & y_{-2} (\frac{\mu^2}{M_c^2}) =  - \lim_{N,M_R}\sum_{l=1}^{N} c_l^N\
     \frac{\mu^2}{lM_R^2} (1 + \frac{\mu^2}{lM_R^2} )^{-1} =
     \sum_{n=1}^{\infty} \frac{(-)^{n-1} L_n }{n!}
      \left(\frac{\mu^2}{M_c^2}\right)^n \\
       & & \mu^2 = \mu_s^2 + {\cal M}^2
     \end{eqnarray}
     Here $\gamma_w$ and $L_n$ ($n=1,2 \cdots$) are the numbers of functional limits
     \begin{eqnarray}
     & & \gamma_w = \lim_{N} \gamma_w(N) \equiv \lim_{N}\{ \ \sum_{l=1}^{N} c_l^N \ln l +
     \ln [\ \sum_{l=1}^{N} c_l^N\ l \ln l \ ] \}  \\
     & & L_n = \lim_{N}L_n(N) \equiv - \lim_{N}\sum_{l=1}^{N} c_l^N\frac{n!}{l^n}\
    [\ \sum_{l'=1}^{N} c_{l'}^N\ l' \ln l' \ ]^n
     \end{eqnarray}
     Note that the following notations are understood
     \begin{eqnarray}
      \lim_N = \lim_{N\rightarrow \infty} \  \ ,\qquad
      \lim_{N, M_R^2} = \lim_{N\rightarrow \infty, M_R^2 \rightarrow \infty}
     \end{eqnarray}

     It is manifest that the energy scale $\mu_s $ sets an IR `cut-off' for the case
     ${\cal M}^2 =0$ which could occur in a massless theory with imposing
     on mass-shell conditions, and the energy scale $M_c$ provides an UV `cut-off'.
     In fact, $M_c$ plays the role of characteristic energy scale (CES)
     and its magnitude is supposed to be known
     for the considered quantum theories, and $\mu_s$ plays the role of SES and it can
     in principle run from $\mu_s = M_c$ to $\mu_s =0$ if there is no IR divergent problem
     and no mass gap generated in the considered theory.
     Of particular, the method does maintain the quadratic `cut-off' term $M_c^2$,
     which makes it different from the dimensional
     regularization. Though this feature is analogous to the naive cut-off regularization,
     while it is quite unlike the naive cut-off regularization as the new regularization
     ensures the consistency conditions for the regularized scalar and tensor type ILIs.
     So that the new method preserves the basic symmetry principles of the theory, such as
      gauge invariance and  Lorentz invariance as well translational invariance,
     which is compatible to the dimensional regularization.
      It then becomes clear that the UV `cut-off' scale $M_c$ and the IR `cut-off' scale
      $\mu_s=\mu_c$ in the new regularization are completely distinguishable from the ones
      in the naive cut-off regularization. In the latter case, the cut-off momentum scales
      are introduced by imposing kinematically the upper and lower bound of the integrating
      loop momentum, which explicitly breaks the translational and Lorentz invariance and
      also destroys the gauge invariance in gauge theories. In the former case, which is
      in contrast to the latter one, the energy scales $M_c$ and $\mu_s$ are entered intrinsically
      as two basic mass scales from the regulators (or regularization quanta).
      For convenience and also to be distinguishable from the naive momentum cut-off,
      the energy scales $M_c$ and $\mu_s$ may be mentioned henceforth as intrinsic mass scales.

      It is particularly noted that when taking the CES
      $M_c$ to be infinite, i.e., $M_c \rightarrow \infty$, we then recover
      the initial but well-defined UV divergences of the ILIs
      including the quadratically divergent terms. For non-abelian gauge theories
      of Yang-Mills fields interacting with Dirac spinor fields,
      as all the quadratic divergent terms cancel each other in the new regularization scheme
      (which has explicitly been shown in the previous section for the gauge field vacuum
      polarization function at one loop order), thus the resulting divergent behavior
      in the regularized gauge theories at $M_c \rightarrow \infty$ is actually
      equivalent to the one by the dimensional regularization once one makes
      a simple replacement $\ln (M_c/\mu) \leftrightarrow 1/(4-D)$ with taking
      $D\rightarrow 4$ and $M_c \rightarrow \infty$. Therefore,
      such non-Abelian gauge theories may be viewed,
      according to our previous convention, as underlying renormalizable QFTs.

       Before ending this section, we would like to emphasize that
       the new regularization described above
       differs in principle from the dimensional regularization.
       The differences arise not only from the quadratic `cut-off' term,
       but actually from all the divergences and their origins.
       This is because in the dimensional regularization
      the divergences are intimately correlated to the space-time dimensions.
      Therefore it is inevitable to meet divergences in the dimensional regularization
      when one is forced to take the space-time dimension $D$ to be four $(D=4)$
      of the real world or to be the one in the original theories.
      Unlike in the dimensional regularization,
      the space-time dimension in the new
      regularization is the same as the one in the
      original theory, the initial divergences of ILIs are recovered and well-defined  by
      taking the CES $M_c$ to be infinite, $M_c \rightarrow \infty$.
      In principle, $M_c$ is a free but well-controlled parameter and its value relies
      in general on the considered phenomena described by the quantum theories.
      In this sense, the CES $M_c$ is physically meaningful, so do
      the quadratic terms or other possible high order terms
      of $M_c$. On the other hand, the finite terms in the new regularization are given
      by the polynomial of $\mu^2 / M_c^2$  rather than the one of $1/\ln \frac{M_c}{\mu} \sim
      (4-D)$ in the dimensional regularization. Of particular, in the new regularization,
      one can always make on mass-shell renormalization even for a massless
      theory, like QCD with massless quarks. This is because one can always choose the SES $\mu_s$ to
      be at an interesting finite energy scale.  Nevertheless, the new regularization is formally compatible with the
      dimensional regularization as they satisfy the same consistency conditions. Finally, we want to point out
      that the prescription of the new regularization formally appears to be similar to the one of Pauli-Villars, but they
      are different in the following two aspects: First, the new regularization is applied to the
      ILIs of Feynman loop graphs rather than the propagators imposed in the Pauli-Villars regularization which
      manifestly spoils gauge invariance. This is a conceptual extension to the Pauli-Villars regularization.
      Second, in the new regularization, the coefficients $c_l^N$ ($l=1,2, \cdots$)
      are well determined and independent of the regulator mass factors $M_l^2$ for $M_l^2 = \mu_s^2 + l M_R^2$
      ($l=1,2, \dots$). Here the basic regulator mass scale $M_R^2$ and the number $N$ of regulators are
      taken to be infinitely large in the new regularization, which is basically a crucial generalization
      to the usual Pauli-Villars regularization in which the regulator mass scale $M_R^2$ is given $M_R = \Lambda$
      and the number of regulators is normally chosen to be only a few limited one, say $N=2$.
      As we will show below that for a finite $N$ and given $M_R^2$, the integrals
      are regulator dependent (one can explicitly see from table 1 in appendix C). Only when taking the number $N$ and
      the mass scale $M_R^2$ of the regulators to be infinite (or sufficiently large), then the resulting
      regularized theory becomes independent of (or insensitive to) the regulators.

   \section{Decouple of regulators and Mathematically interesting numbers and functional limits}

      One may notice that the basic mass scale $M_R$ of the regulators (or regularization quanta)
      does not appear in the final expressions
      of the regularized ILIs, while it is intrinsically correlated to the regulator
     number $N$ in such a way that: taking $M_R$ and $N$ to be infinite,
     $M_R \rightarrow \infty$ and $N \rightarrow \infty$, but keeping
     the CES $M_c$ remain held fixed. This implies that $M_R$ must approach to infinity in terms
     of an appropriate function of $N$. To obtain its explicit form, we begin with the conjecture
     that
     \begin{eqnarray}
     M_R^2 = M_o^2 \ln N
     \end{eqnarray}
     Here $M_o$ is supposed to be a constant mass scale. Substituting this form into the definition of
     the CES $M_c$, we arrive at the relation
     \begin{eqnarray}
     M_c^2 = \lim_{N,M_R}\sum_{l=1}^{N} c_l^N\ ( l \ln l ) M_R^2 =
      M_o^2 \lim_{N} (\ln N ) \sum_{l=1}^{N} c_l^N\ ( l \ln l )  = M_o^2/h_w
     \end{eqnarray}
    with $h_w$ being given by
      \begin{eqnarray}
      h_w = \lim_{N\rightarrow \infty} h_w(N) =
      \lim_{N\rightarrow \infty} \frac{1}{ \sum_{l=1}^{N} c_l^N\ ( l \ln l ) \ln N }
     \end{eqnarray}
     Here the function $h_w(N)$ is purely determined by the regulator number $N$ and supposed to
     have a finite limit at $N\to \infty$.

     So far the regularized ILIs are completely characterized by the three type of functional quantities
       $h_w(N)$, $\gamma_w(N)$ and $L_n(N)$ ($n=1, 2, \cdots$) ,
       they are solely determined by the regulator number $N$.
     It then becomes clear that in order to prove the regularized ILIs to be independent of the
      regulators (or regularization quanta), one actually only needs to show that the regularized ILIs
      are not sensitive to the regulator number $N$ at sufficiently large $N$. This is then
      equivalent to check whether the functional quantities  $h_w(N)$, $\gamma_w(N)$ and $L_n(N)$
      approach to certain finite limits at $N \rightarrow \infty$, namely whether $h_w$,
      $\gamma_w$ and $L_n$ are eventually reached to the constant numbers.
      To be manifest, we perform, instead of an analytic proof, a
      numerical evaluation. The numerical analysis is presented in appendix C .
      The numerical results given in table 1 of appendix C truly show that
      for a sufficiently large $N$ the functions $h_w(N)$, $\gamma_w(N)$ and $L_n(N)$
      do approach to certain values which become less sensitive to
      the changes of the number $N$ once $N$ is going to be sufficiently large.
      In general, we are led to the following three conjectures which have been
      checked by an appropriate functional fitting method up to the precision of $10^{-3}$
      (for more details, see Appendix C).
      \\

      Conjecture 1.
      \begin{eqnarray}
      h_w = \lim_{N\rightarrow \infty} h_w(N) = 1
     \end{eqnarray}
     which implies the following interesting functional limit
      \begin{eqnarray}
       W_N \equiv \sum_{l=1}^{N} (-1)^l \frac{N!}{(N-l)!\ l!}
       \ ( l \ln l )  \stackrel{\rm N\rightarrow \infty}{=} 1/ \ln N
      \end{eqnarray}

      Conjecture 2.
       \begin{eqnarray}
     \gamma_w & = & \lim_{N\to \infty} \gamma_w(N) =
     \lim_{N\to \infty} \left( \sum_{l=1}^{N} (-1)^l \frac{N!}{(N-l)!\ l!}\  \ln l
     - \ln [ h_w(N) \ln N\ ]  \right) \nonumber \\
      & = & 0.5772\cdots = \gamma_E
     \end{eqnarray}
     Namely $\gamma_w$ is equal to the Euler number
     $\gamma_E =  0.577215664901533 \cdots$. It then implies another
     interesting functional limit
     \begin{eqnarray}
      \Gamma_N \equiv \sum_{l=1}^{N} (-1)^l \frac{N!}{(N-l)!\ l!}\  \ln l
     \stackrel{\rm N\rightarrow \infty}{=} \ln \ln N + \gamma_E
    \end{eqnarray}

     Conjecture 3.
     \begin{eqnarray}
     L_n = \lim_{N \to \infty}L_n(N) = 1 \ , \qquad n = 1, 2 \cdots
     \end{eqnarray}
     which indicate the functional limits
     \begin{eqnarray}
     E_N^{(n)} \equiv \sum_{l=1}^{N} (-1)^{l-1} \frac{N!}{(N-l)!\ l!}\  \frac{1}{l^n}
    \stackrel{\rm N\rightarrow \infty}{=} \frac{(\ln N)^n }{n!}\ ,
    \ \qquad  n = 1, 2 \cdots
     \end{eqnarray}

     When the above conjectures hold, the functions $y_i (x)$ defined in the
     previous section via the polynomials can now be expressed by
     the following simple functions
      \begin{eqnarray}
      & & y_{-2} (x) = 1 - e^{-x} \\
      & & y_0 (x) = \int_0^x d \sigma\ \frac{1 - e^{-\sigma} }{\sigma} \\
      & & y_1 (x) = \frac{1}{x} \left(e^{-x} - 1 + x\right) \\
      & & y_2 (x) = y_0 (x) - y_1(x)
      \end{eqnarray}

     It is then clear that the regulators (or regularization quanta) are
    indeed decouple from the regularized ILIs. Furthermore, the consistency conditions eqs.(2.27-2.33) do hold
    in the new regularization (one can easily check that by using the useful integrals given in appendix D).
     Therefore it can be concluded that we do arrive at a consistent description for the new regularization
     scheme.

     Obviously, an analytical proof for the above conjectures must be very
     helpful and important. It may also provide deep insights in
     mathematics.

  \section{Factorization of overlapping divergences and \\
  Reduction of Overlapping Tensor Integrals}

     The gauge symmetry preserving and infinity free regularization described above
     is well applied to regularize one loop graphs and the relevant ILIs.
     In order to obtain a consistently regularized theory, such a regularization should be
     applicable to all loop graphs. Furthermore, though the regularized loop graphs become well-defined and
     infinity free, the coupling constants and the fields in the quantum theories still need
     to be renormalized. The only difference is that the subtraction terms become
     finite quantities instead of infinite ones.
     The procedure of such subtractions (or the so-called renormalization)
     must be consistently carried out order by order in perturbative expansion.
     In fact, whether the subtraction terms in the renormalization are finite or infinite,
     in a given order any new subtraction terms must be finite polynomials in external momenta
     in order to maintain unitarity of the theory. Such a statement should in general be
     independent of any regularization. Therefore, one shall be able to prove this
     statement in a more general way without involving any regularization.

      To be specific and also for an explicit demonstration, we will treat in this section the
     Feynman integrals involved in two loop diagrams. The generalization to more loop graphs
     is straightforward. It is different from one loop graphs,
     two loop diagrams require us to make a careful treatment for
     overlapping divergences. Eventually, we must show that after
     appropriate subtractions of one loop divergences, the coefficient functions of
     all the divergent terms are finite polynomials in the external momenta.
     Namely one has to prove the theorem that all eventual divergences must be
     harmless divergences. To arrive at this purpose, we shall first prove that
     overlapping divergences can completely be factorized. More specifically we shall show that
     the overall divergences of two loop graphs and the sub-integral divergences corresponding
     to one loop graphs can be well separated. It will be seen below that all of these
     properties will become manifest after performing the evaluation of two-fold ILIs involved in
     two loop diagrams.

     The explicit demonstration in the previous section shows that all divergences
     in one loop diagrams are harmless ones. In fact, as all one loop integrals can be
     expressed in terms of the one-fold ILIs by using the Feynman parameter method,
     therefore the divergences are completely characterized by
     the one-fold ILIs. As the coefficient functions of all divergent one-fold ILIs
      are finite polynomials of external momenta and masses,
     therefore all the divergences at one loop level are truly harmless
     ones.

      We now come to study two loop graphs. Consider first the scalar type
      overlapping Feynman integrals involved in the general two loop diagrams (i.e., the
     so-called general $\alpha\beta\gamma$ diagrams \cite{DR})
     \begin{eqnarray}
     I_{\alpha\beta\gamma}^{(2)} = \int d^4 k_1 \int d^4 k_2
     \frac{1}{(k_1^2 +  {\cal M}_1^2 )^{\alpha}\left(k_2^2 + {\cal M}_2^2\right)^{\beta}
     \left((k_1-k_2 +p )^2 + {\cal M}_{12}^2\right)^{\gamma} }
     \end{eqnarray}
     The above integral has been given in the Euclidean space of momentum
     through an analytic Wick rotation.
     Here $\alpha,\ \beta,\ \gamma > 0 $ since if one of them is zero,
     the integral is no longer an overlapping one and it actually becomes a factorizable one.
     In the gauge theories, the usual power counting rule shows
      that the divergent overlapping integrals of two loop diagrams
      are at most quadratically divergent, i.e., $\alpha,\ \beta,\ \gamma \ge 1 $.
     The above form of integrals can be
     regarded as the most general one for two loop graphs. Here we have supposed that
     one can always apply the Feynman parameter method to
     all momentum factors $k_1$, so that
     all propagators concerning $k_1$
     and external momenta $p_i^2$ ($i=1,2, \cdots$) are combined into a single one.
     Similarly for the propagators concerning $(k_1-k_2)$ and $k_2$. Therefore,
     the three mass factors ${\cal M}_1^2$, ${\cal M}_2^2$ and ${\cal M}_{12}^2$
     are in general the functions of masses $m_i^2$ and external momenta $p_i^2$ ($i=1,2, \cdots$)
     \begin{eqnarray}
      {\cal M}_{12}^2 \equiv {\cal M}_{12}^2 (m_1^2, p_1^2, m_2^2, p_2^2, \cdots), \quad
      {\cal M}_i^2 \equiv {\cal M}_i^2 (m_1^2, p_1^2, m_2^2, p_2^2, \cdots) \quad (i=1,2)
     \end{eqnarray}

      Based on the usual power counting rule, we may first recall some useful definitions
      following the ref.\cite{DR}:
     (i) the sub-integral over $k_1$ is said to be
     convergent or divergent according to $\alpha + \gamma > 2$ or $\alpha + \gamma \le 2$,
     similarly, the sub-integral over $k_2$ is said to be
     convergent or divergent according to $\beta + \gamma > 2$ or $\beta + \gamma \le 2$;
     (ii) the overall integral of  $\alpha\beta\gamma $ diagram is
     said to be overall convergent or overall divergent according to
     $\alpha + \beta + \gamma > 4$ or $\alpha + \beta + \gamma \le 4$;
     (iii) a harmless divergence
     is a divergence with its coefficient functions a polynomial of finite order in the
     external momenta. In the gauge theories, the overall divergence of a nontrivial
     overlapping integral is at most quadratic, thus $\alpha + \beta + \gamma \ge 3$.

      In two loop graphs, we shall also involve the general tensor type Feynman integrals
     \begin{eqnarray}
     I_{\alpha\beta\gamma\ (\mu\nu\ \cdots)}^{(2)} = \int d^4 k_1 \int d^4 k_2
     \frac{(k_{1\mu}k_{1\nu}\ , \ k_{1\mu}k_{2\nu}\ , \ k_{2\mu}k_{2\nu}\ , \ \cdots )}{
     (k_1^2 +{\cal M}_1^2 )^{\alpha}\left(k_2^2 + {\cal M}_2^2\right)^{\beta}
     \left((k_1-k_2 +p )^2 + {\cal M}_{12}^2 \right)^{\gamma} }
     \end{eqnarray}
     Similar to the one loop case, by adopting appropriate
     parameter methods, the tensor type Feynman integrals can eventually be expressed
     in terms of the scalar type integrals through the aid of
     the metric tensor $g_{\mu\nu}$ and the external momentum vector $p_{\mu}$.
     We shall first focus on the general scalar
     type Feynman integrals $ I_{\alpha\beta\gamma}^{(2)} $ and come back to the
     tensor type Feynman integrals at the end of this section.

     Though the above general forms of Feynman integrals may be more concise than
     the original ones due to the uses of Feynman parameter method,
     they are not yet the desired ILIs because of the overlapping momentum factor
     $(k_1-k_2 + p)^2$. To obtain the ILIs of two loop graphs, according to the definition of ILIs,
     we shall further adopt appropriate parameter methods to treat the factor
     $(k_1-k_2 + p)^2$. As the first step, we still apply the usual Feynman parameter method to
     the sub-integral over $k_1$ (one can also consider first the sub-integral over $k_2$
     as the two sub-integrals are symmetric with the exchanges $k_1 \leftrightarrow k_2$,
     $\alpha \leftrightarrow \beta $, $ {\cal M}_1^2 \leftrightarrow {\cal M}_2^2 $ and
     $p \leftrightarrow -p$, for convention, we will henceforth make the treatments according to
     the order $k_1$, $k_2$, $\cdots$).

     Taking the Feynman parameter variables $a_1=k_1^2 + {\cal M}_1^2$ and
     $a_2=(k_1-k_2 + p)^2 + {\cal M}_{12}^2$, the sub-integral over $k_1$ becomes a quadratic
     (see Appendix A)
      \begin{eqnarray}
     I_{\alpha\beta\gamma}^{(2)} = \frac{\Gamma(\alpha + \gamma)}{\Gamma(\alpha)\Gamma(\gamma)}
     \int_0^1 dx \int d^4 k_1 \int d^4 k_2
     \frac{x^{\gamma -1}(1-x)^{\alpha -1}}{[k_1^2 + x(1-x)(k_2-p)^2
     + {\cal M}_x^2 \ ]^{\alpha + \gamma}
     \left(k_2^2 + {\cal M}_2^2\right)^{\beta} }
     \end{eqnarray}
     with definition
     \begin{equation}
     {\cal M}_x^2 \equiv {\cal M}_1^2 + x ({\cal M}_{12}^2 -{\cal M}_1^2)
     \end{equation}
     Where we have made a momentum shift $k_{1\mu} \rightarrow k_{1\mu} + x (k_{2\mu}-p_{\mu})$.

     To further combine the denominators into a single quadratic,
     it is found to be of help to adopt an alternative identity
      \begin{equation}
    \frac{1}{a^{\alpha}b^{\beta}} = \frac{\Gamma(\alpha + \beta)}{\Gamma(\alpha)\Gamma(\beta)}
    \int_{0}^{\infty}\ du \frac{u^{\beta -1} }{[a + b u]^{\alpha +\beta } }
    \end{equation}
    The advantage of this identity is that once $u$ approaches to infinite in a
    way similar to $a$, i.e., $u \sim a \rightarrow \infty$, the divergent structure
    of the integral coincides with the one of the term $a$. For convenience, we may mention this
    parameter method as an UV-divergence preserving parameter method, and the parameter $u$ as
    the corresponding UV-divergence preserving integrating
    variable. Its important consequence is that by appropriately applying this UV-preserving
    parameter method to the combined propagators obtained by carefully using the usual Feynman parameter method,
    one can factor out all the UV divergences, so that the usual Feynman parameter integrations contain no
    UV divergences. Only IR divergences could be hidden in the Feynman parameter integrations.
     This is crucial point for analyzing the overlapping
     divergences. In fact, it will be shown next section that both UV and IR divergences can be
     simultaneously regularized in the new regularization. Thus
     all the Feynman parameter integrals are convergent.

     Taking $a = x(1-x)(k_2-p)^2 + k_1^2 + {\cal M}_x^2 $ and $b= k_2^2 + {\cal M}_2^2 $
     and making a momentum shift $k_{2\mu} \rightarrow k_{2\mu} + x(1-x)p_{\mu}/(u + x(1-x))$,
     we then obtain from eqs.(5.4) and (5.6) that
   \begin{eqnarray}
     I_{\alpha\beta\gamma}^{(2)}& = &
     \frac{\Gamma(\alpha + \beta + \gamma)}{\Gamma(\alpha)\Gamma(\beta)\Gamma(\gamma)}
     \int_0^1 dx  \int d^4 k_1 \int d^4 k_2  \int_0^{\infty} du  \nonumber \\
     & & \frac{x^{\gamma -1}(1-x)^{\alpha -1} u^{\beta - 1} }{ [\ k_1^2 + {\cal M}_x^2 +
     \left( u + x(1-x) \right) k_2^2 + u
     ( {\cal M}_2^2 + \frac{x(1-x)}{u + x(1-x)} p^2 )\ ]^{\alpha + \beta + \gamma} }
     \end{eqnarray}
     It is clear that once $u$ approaches to infinite in a way similar to $k_1^2$, i.e.,
     $u \sim k_1^2 \rightarrow \infty$, the usual power counting rule implies
     that the resulting divergent structure does coincide with the one in the
     original sub-integral over $k_1$. In other word, the UV divergent
     property of the sub-integral over $k_1$ will be characterized by the one of
     the integral over $u$. It will become more clear below from the
     explicit form of ILIs.

     In general, $\alpha + \beta + \gamma > 2$, the sub-integral over $k_1$
     becomes convergent and can be carried out safely. It is straightforward to perform
     the integration over $k_1$, which leads to the result
     \begin{eqnarray}
     I_{\alpha\beta\gamma}^{(2)} & = &
     \Gamma_{\alpha \beta \gamma }\int_0^1 dx\ x^{\gamma -1}(1-x)^{\alpha -1}
     \int_0^{\infty} d u \frac{\pi^2 \ u^{\beta - 1}}{
     (u + x(1-x))^{\alpha + \beta + \gamma - 2} } \nonumber  \\
     & &  \int d^4 k_2 \frac{1}{(\ k_2^2 + {\cal M}_2^2
     + \mu_{u}^2 \ )^{\alpha + \beta + \gamma -2 } }
     \end{eqnarray}
     where we have factored out the coefficient factor $[u + x(1-x)]$ of the momentum square $k_2^2$
     and introduced the definitions
     \begin{eqnarray}
     & & \Gamma_{\alpha \beta \gamma }= \frac{\Gamma(\alpha + \beta
     + \gamma)}{\Gamma(\alpha)\Gamma(\beta)\Gamma(\gamma)} \
     \frac{1}{(\alpha + \beta + \gamma -1)(\alpha + \beta + \gamma -2)} =
     \frac{\Gamma(\alpha + \beta
     + \gamma - 2)}{\Gamma(\alpha)\Gamma(\beta)\Gamma(\gamma)}  \\
    & &  \mu_{u}^2 = \frac{1}{u + x(1-x) } \ \left( {\cal M}_x^2
     -x(1-x) (\ {\cal M}_2^2 - p^2\ ) - \frac{x^2 (1-x)^2 }{u + x(1-x)}\  p^2 \ \right)
     \end{eqnarray}

        It is useful to change the dimensionless integrating
       variable $u$ into the dimensionful one $\hat{k}_1^2 \equiv q_o^2u$.
       Here $q_o^2$ with dimension of mass square is introduced as an
       universal energy scale. In general, it only requires $ -q_o^2 <
       4\mu_s^2$ in order to respect unitarity. Typically, one may take
       $q_o^2 \sim \mu_s^2$ to reduce unnecessary parameters. The variable $\hat{k}_1^2$ may be regarded
       as a momentum-like one in connecting with the initial loop momentum $k_1$.
       Though $\hat{k}_1^2$ cannot wholly be compatible with the
       initial momentum $k_1^2$, whereas the integral over $\hat{k}_1^2$ does maintain
       the one-loop divergent behavior of the initial loop momentum $k_1$ .
       Using the new integrating variable and noticing the formal relation of the integrating
       measure
       \begin{equation}
        \int_0^{\infty} du = \int \frac{d^4 \hat{k}_1}{\hat{k}_1^2} \frac{1}{q_o^2 \pi^2} \ ,
       \end{equation}
      We can rewrite the above integral eq.(5.8) into the following form
     \begin{eqnarray}
     I_{\alpha\beta\gamma}^{(2)} & = & \Gamma_{\alpha \beta \gamma }\
     \int_0^1 dx\ x^{\gamma -1}(1-x)^{\alpha -1}\  (q_o^2)^{\alpha + \gamma -2 }
     \sum_{i=0}^{\beta -1} c_i^{\beta -1}(x(1-x)q_o^2)^{i} \nonumber  \\
     & & \int \frac{d^4 \hat{k}_1}{\hat{k}_1^2} \frac{1}{(\ \hat{k}_1^2
     + x(1-x)q_o^2\ )^{\alpha + \gamma -1 + i} } \int d^4 k_2 \frac{1}{
     (\ k_2^2 + {\cal M}_2^2  + \mu^2_{\hat{k}_1^2} \ )^{\alpha + \beta + \gamma -2 } } \nonumber \\
     &\equiv & \Gamma_{\alpha \beta \gamma }\
     \int_0^1 dx\ x^{\gamma -1}(1-x)^{\alpha -1}\  (q_o^2)^{\alpha + \gamma -2 }
     \sum_{i=0}^{\beta -1} c_i^{\beta -1}(x(1-x)q_o^2)^{i} \nonumber  \\
     & & \int \frac{d^4 \hat{k}_1}{\hat{k}_1^2} \frac{1}{(\ \hat{k}_1^2
     + x(1-x)q_o^2\ )^{\alpha + \gamma -1 + i} }
     I_{\alpha\beta\gamma}^{(1)}({\cal M}_2^2  + \mu^2_{\hat{k}_1^2} )
     \end{eqnarray}
     where $I_{\alpha\beta\gamma}^{(1)}({\cal M}_2^2  + \mu^2_{\hat{k}_1^2} )$ is an overall
     one-fold ILI involved in two loop graphs
     \begin{eqnarray}
     I_{\alpha\beta\gamma}^{(1)}({\cal M}_2^2  + \mu^2_{\hat{k}_1^2} ) =
    \int d^4 k_2 \frac{1}{(\ k_2^2 + {\cal M}_2^2  + \mu^2_{\hat{k}_1^2}
    \ )^{\alpha + \beta + \gamma -2 } }
     \end{eqnarray}
     with
     \begin{eqnarray}
      & & c_i^{\beta -1} = (-1)^i\ \frac{(\beta - 1)!}{(\beta - 1 -i)!\ i !} \\
      & & \mu^2_{\hat{k}_1^2} = \frac{q_o^2}{\hat{k}_1^2 + x(1-x)q_o^2 } \ \left( {\cal M}_x^2
     -x(1-x) (\ {\cal M}_2^2 - p^2\ ) - \frac{x^2 (1-x)^2 q_o^2 }{\hat{k}_1^2
     + x(1-x)q_o^2}\ p^2 \ \right)
     \end{eqnarray}

     According to the definition of ILIs, the above resulting loop integrals may be called
     as the scalar type two-fold ILIs.
     Basing on such two-fold ILIs and noticing the interesting
     property that $\mu_{\hat{k}_1^2}^2$ vanishes in the infinite limit of $\hat{k}_1^2$,
     i.e.,
     \begin{equation}
      \mu_{\hat{k}_1^2}^2 \rightarrow 0 \ \qquad \mbox{at} \qquad \hat{k}_1^2 \rightarrow \infty
     \end{equation}
     we arrive at the following important observations:

     (i) For the simple case that $\alpha + \gamma = 2$ and $\beta = 1 $, it becomes manifest that
     the UV divergent behavior of the two-fold ILIs at
     $\hat{k}_1^2 \rightarrow \infty $ and $k_2^2 \rightarrow \infty$
     coincides with the one of the original integral. This explicitly
     shows that the one-loop UV-divergence preserving sub-integral over the momentum-like variable
     $\hat{k}_1^2$ does characterize the high energy
     behavior of the sub-integral over the loop momentum $k_1$ in the initial integral.

     (ii) The sub-integral over the loop momentum $k_2$ turns out to be an overall
      one-fold ILI which is correlated to the sub-integral over the momentum-like variable
      only via the $\hat{k}_1^2$-dependent mass factor
      ${\cal M}_2^2  + \mu^2_{\hat{k}_1^2}$.

     (iii)  When $\hat{k}_1^2 \rightarrow \infty $, the two sub-integrals over $k_2$
     and $\hat{k}_1$ become factorized ones due to the property
      $\mu_{\hat{k}_1^2}^2 \rightarrow 0$ at $\hat{k}_1^2 \rightarrow \infty $.
     Thus the divergent properties of the initial two loop integral can completely be described
     by the ones of the divergence factorizable two-fold ILIs.

     (iv) The most divergent behavior of the sub-integral over $\hat{k}_1$
     is governed by the power counting of $\alpha +\gamma$ which reflects the
     divergence of one-loop sub-diagram, and the divergent property
     of the sub-integral over $k_2$ in the ILIs is solely
     characterized by the power counting of $\alpha + \beta + \gamma$ which
     actually describes the overall divergence of the general $\alpha\beta\gamma$
     two loop diagrams.

     (v) The external momentum dependence is contained only in the
     mass factor ${\cal M}_2^2  + \mu^2_{\hat{k}_1^2}$ of the sub-integral over $k_2$.
     At the divergent point of the sub-integral over $\hat{k}_1$ ($\hat{k}_1 \rightarrow \infty$),
      the possible external momentum dependence is solely given by the mass function
      ${\cal M}_2^2$ which appears only in the sub-integral over $k_2$, therefore
     such a harmful divergence is expected to be eliminated by one subtraction term.

     From the above observations, it becomes clear that in the two-fold ILIs
     the overall divergences of two loop graphs are completely factorized from
     the sub-integral divergences of one loop graphs. Consequently,
     we are able to deduce the following theorems without involving any regularization.

     {\bf Theorem I} (Factorization Theorem for Overlapping divergences).
      Overlapping divergences which contain divergences of sub-integrals
     and overall divergences in the general Feynman loop integrals
     become completely factorizable in the corresponding ILIs.

     This theorem is the crucial one to treat the problem of overlapping divergences
     involved in two and more loop diagrams. With the observations (iv), (iii) and (ii) in
     the ILIs, its proof becomes manifest. It is actually a direct consequence of
     the ILIs. Therefore, the whole demonstration of the theorem is equivalent to the evaluation
     of ILIs from the general overlapping Feynman integrals of loop graphs. More explicit
     demonstrations will be presented in next section from the regularized ILIs.

     {\bf Theorem II} (Subtraction Theorem for Overlapping divergences).
        The difference of the general Feynman loop integral with the subtraction
        term corresponding to the divergent sub-integral contains only harmless divergences.

  This theorem is the central one to obtain a consist theory. Obviously, the
  factorization theorem (theorem I) becomes crucial  to yield this theorem. Its proof is
  straightforward from the observations (i)-(v) in the corresponding ILIs of the general
  Feynman loop integrals. Let us present the simple demonstration based on
  the observations from the two-fold ILIs. Note that as the external momentum
  dependence in the two-fold ILIs only appears in the mass factor
  ${\cal M}_2^2  + \mu^2_{\hat{k}_1^2}$
  of the overall one-fold ILI over the loop momentum $k_2$, and also as the sub-integral over
  the momentum-like variable $\hat{k}_1^2$ has the same one-loop UV divergent structure
  as the one over the loop momentum $k_1$, one only needs to introduce
  one subtraction term of the sub-integral over the loop momentum $k_1$
  \begin{eqnarray}
     I_{\alpha\beta\gamma}^{(2)S} = \int d^4 k_1
     \frac{1}{(k_1^2 + m_o^2 )^{\alpha + \gamma} }
     \int d^4 k_2 \frac{1}{\left(k_2^2 + {\cal M}_2^2\right)^{\beta} } \ ,
  \end{eqnarray}
     where the superscript $S$ denotes the subtraction term.
     With particularly noticing the property
     $\mu_{\hat{k}_1^2}^2 \sim O( \frac{1}{\hat{k}_1^2} ) \rightarrow 0$
     at $\hat{k}_1^2 \rightarrow \infty $, it then becomes clear that
     the difference of the integrals, i.e., ( $I_{\alpha\beta\gamma}^{(2)}
   - I_{\alpha\beta\gamma}^{(2)S}$ ), contains only harmless divergences. Its manifest
   demonstration will be given in next section for the regularized ILIs.

   {\bf Theorem III} (Harmless Divergence Theorem).
   If the general loop integral contains no divergent sub-integrals,
   then it contains only a harmless single divergence arising from the overall divergence.

   This theorem may be deduced from the Theorems I and II.
   Its proof is also obvious
   with the observations (iv), (iii) and (ii) in the ILIs.
   Specifically, as $\alpha + \gamma > 2 $ and $\beta + \gamma > 2 $
   the integral over the momentum-like variable $\hat{k}_1^2$ is
   convergent in the two-fold ILIs. Only the integral over the loop momentum $k_2$ may contain
   divergence. As the integral over the loop momentum $k_2$
   characterizes the overall divergence of the initial loop integral and it is actually
   an overall one-fold ILI, also as the sub-integral over the momentum-like variable
   $\hat{k}_1$ contains no external momentum-dependence, we then come to the statement
   in the theorem.

   {\bf Theorem IV} (Trivial Convergence Theorem).
   If the general loop integral contains no overall divergence and also no
   divergent sub-integrals, then it is convergent.

   This theorem is really trivial and it is presented only for completeness.

    We now turn to the tensor type Feynman integrals. It is not difficult to arrive at
   the following theorems:

   {\bf Theorem V} (Reduction Theorem for Overlapping Tensor Type Integrals). The general
   overlapping tensor type Feynman integrals of arbitrary loop graphs
   are eventually characterized by the overall one-fold tensor type ILIs
    of the corresponding loop graphs.

    This theorem is the key theorem for the generalization of treatments and also for the
     prescriptions from one loop graphs to arbitrary loop graphs. From this theorem,
     it is not difficult to deduce the following theorem

    {\bf Theorem VI}  (Relation Theorem for tensor and scalar type ILIs).
     For any fold tensor and scalar type ILIs, as long as their power counting dimension of
    the integrating loop momentum are the same, then the relations between the tensor and
    scalar type ILIs are also the same and independent of the fold number of ILIs.

    This theorem is crucial to extend the consistency conditions of
    gauge invariance from divergent one-fold ILIs (or
    one loop graphs ) to more fold ILIs (or more closed loop graphs).

    In here we provide an explicit proof for the two loop case.
    The extension of the proof to more closed loops is obvious.
   Repeating the evaluation similar to the one for the scalar type two-fold ILIs,
   one can easily observe that all the tensor type Feynman integrals with structures
   such as $k_{1\mu}k_{1\nu}\cdots$, $k_{1\mu}k_{2\nu}\cdots $ and $k_{2\mu}k_{2\nu}\cdots$
   will be evaluated into the non-trivial tensor type ILIs with tensor structures
   given only by $k_{2\mu}k_{2\nu}\cdots$ plus the tensor type integrals which
   are constituted from the scalar type ILIs and the metric tensor $g_{\mu\nu}$ as well as
   the external momentum vector $p_{\mu}$. This can easily be seen from the fact
   that in evaluating the ILIs all involving operations only concern the simple loop momentum shifts
  \begin{eqnarray*}
  k_{1\mu} \rightarrow k_{1\mu} + x(k_{2\mu} - p_{\mu}) \, \qquad k_{2\mu}
   \rightarrow k_{2\mu} + x(1-x)p_{\mu}/(u+x(1-x))
  \end{eqnarray*}
   together with the integration over the loop momentum $k_1$. Here the integration is well-defined
   as the sub-integral over $k_1$ becomes convergent after applying for the UV-divergence preserving
   parameter method.

   We finally find that the non-trivial tensor type ILIs in two loop graphs are in general
   given by
   \begin{eqnarray}
     I_{\alpha\beta\gamma\ \mu\nu \cdots}^{(2)} & = & \Gamma_{\alpha \beta \gamma }\
     \int_0^1 dx\ x^{\gamma -1}(1-x)^{\alpha -1}\ f(x)\ (q_o^2)^{\alpha + \gamma -2 }
     \sum_{i=0}^{\beta -1} c_i^{\beta -1}(x(1-x)q_o^2)^{i} \nonumber  \\
     & & \int \frac{d^4 \hat{k}_1}{\hat{k}_1^2} \frac{1}{(\ \hat{k}_1^2
     + x(1-x)q_o^2\ )^{\alpha + \gamma -1 + i} } \int d^4 k_2 \frac{k_{2\mu}k_{2\nu}\cdots}{
     (\ k_2^2 + {\cal M}_2^2  + \mu^2_{\hat{k}_1^2} \ )^{\alpha + \beta + \gamma -2 } } \nonumber \\
     &\equiv & \Gamma_{\alpha \beta \gamma }\
     \int_0^1 dx\ x^{\gamma -1}(1-x)^{\alpha -1}\ f(x)\  (q_o^2)^{\alpha + \gamma -2 }
     \sum_{i=0}^{\beta -1} c_i^{\beta -1}(x(1-x)q_o^2)^{i} \nonumber  \\
     & & \int \frac{d^4 \hat{k}_1}{\hat{k}_1^2} \frac{1}{(\ \hat{k}_1^2
     + x(1-x)q_o^2\ )^{\alpha + \gamma -1 + i} }
     I_{\alpha\beta\gamma\  \mu\nu\cdots}^{(1)}({\cal M}_2^2  + \mu^2_{\hat{k}_1^2} )
     \end{eqnarray}
     where $f(x)$ is a polynomial function of Feynman parameter $x$ and its form only relies on the
     considered tensor structure. Here the tensor type ILIs
     $I_{\alpha\beta\gamma\ \mu\nu\cdots}^{(1)}({\cal M}_2^2  + \mu^2_{\hat{k}_1^2} ) $
     are defined as a non-trivial overall one-fold tensor type ILIs with an effective
     mass factor ${\cal M}_2^2  + \mu^2_{\hat{k}_1^2}$
     \begin{eqnarray}
     I_{\alpha\beta\gamma\ \mu\nu\cdots}^{(1)}({\cal M}_2^2  + \mu^2_{\hat{k}_1^2} ) =
    \int d^4 k_2 \frac{k_{2\mu}k_{2\nu}\cdots}{(\ k_2^2 + {\cal M}_2^2  + \mu^2_{\hat{k}_1^2}
    \ )^{\alpha + \beta + \gamma -2 } }
     \end{eqnarray}
      The superscript $(1)$ means one-fold. It is obvious that
      \begin{eqnarray}
     \frac{I_{ [\delta]\ \mu\nu \cdots}^{(2)} }{ I_{ [\delta] }^{(2)} }
     = \frac{I_{[\delta]\ \mu\nu \cdots}^{(1)} }{I_{[\delta]}^{(1)} }
     \end{eqnarray}
     where $[\delta]$ labels the power counting dimension of momentum. It can be generalized to any fold
     ILIs
     \begin{eqnarray}
     \frac{I_{[\delta]\ \mu\nu \cdots}^{(n)} }{ I_{[\delta]}^{(n)} }
     = \frac{I_{[\delta]\ \mu\nu \cdots}^{(1)} }{I_{[\delta]}^{(1)} }
     \end{eqnarray}
     Here the superscript $(n)$ labels the fold number of ILIs and its value is arbitrary.

      The above explicit forms of the tensor type ILIs provide the desired
      results and complete the proof for the theorem V and theorem VI.
      It is clearly seen that we eventually only need to consider the non-trivial overall
      one-fold  tensor type ILIs.

      All of the above observations and the deduced theorems imply
      the importance of evaluating the ILIs of loop graphs.

    \section{Regularization, Renormalization and Unitarity}

     It has been seen in the previous section that the problem of overlapping divergences is
    not relevant to any regularization. In the section II, it has been shown at one-loop level
    that the generalized Ward identities (or gauge invariant conditions) are in general
    spoiled only by the divergent integrals. This implies meaningless of divergent
    integrals and suggests the necessity of regularizing the divergent integrals. Consequently,
    consistency conditions between the regularized scalar and
    tensor type ILIs have been resulted from the generalized Ward identities at one-loop level.
    In section III, it has been demonstrated that the consistency conditions can
    provide stringent constraints on the regularization methods.
    In this section, we shall apply, as a practical computation and also an explicit check,
    those theorems obtained in the previous section to the regularized ILIs. In fact, by
    explicitly carrying out the relevant integrations in the regularized ILIs,
    we shall arrive at an independent verification on those theorems.

     We shall first generalize the regularization described in the section III for the one-fold
     ILIs to the two- and n-fold ILIs with $n$ being arbitrary.
     The general prescription for the new regularization
     method is simple:

      (i)  Analytically rotate the four dimensional Minkowski space into the four
      dimensional Euclidean space of momentum by using Wick rotation.

      (ii) Appropriately evaluate  Feynman integrals of
      loop graphs into the corresponding ILIs by adopting the usual Feynman
      parameter method and the newly formulated UV-divergence preserving parameter method.

      (iii) Universally replace in the ILIs the loop momentum square $k^2$ and
      the corresponding loop integrating measure $\int d^4 k$ as well as the
      UV-divergence preserving momentum-like variable $\hat{k}^2$
      (or the UV-divergence preserving integrating parameter $u$ )
      and the corresponding integrating measure $\int d^4 \hat{k}/\hat{k}^2$
      (or $\int d u$)
      by the regularizing ones $[k^2]_l$ and $\int [d^4 k]_l$
      as well as $[\hat{k}^2]_l$ (or $[u]_l$ ) and $\int [d^4 \hat{k}/\hat{k}^2]_l$
      (or $\int [du]_l$ ), i.e.,
    \begin{eqnarray}
      & &  k^2 \rightarrow [k^2]_l \equiv k^2 + M^2_l= k^2 + \mu_s^2 + l M_R^2\ , \qquad
      \int d^4 k \rightarrow
      \int [d^4 k]_l \equiv \lim_{N,M_R} \sum_{l=0}^{N} c_l^N \int d^4 k \ , \\
      & &  \hat{k}^2 \rightarrow  [\hat{k}^2]_l \equiv \hat{k}^2 + M_l^2 =
      \hat{k}^2 + \mu_s^2 + l M_R^2\ , \qquad
       \int \frac{d^4 \hat{k}}{\hat{k}^2} \rightarrow
      \int [\frac{d^4 \hat{k}}{\hat{k}^2}]_l \equiv \lim_{N,M_R}
      \sum_{l=0}^{N} c_l^N \int \frac{d^4 \hat{k}}{\hat{k}^2}  \\
      & & or \nonumber \\
      & & \quad u \rightarrow [u]_l \equiv u + M_l^2/q_o^2 =
      u + ( \mu_s^2 + l M_R^2)/q_o^2  \ , \qquad
      \int du \rightarrow \int [du]_l \equiv \lim_{N,M_R} \sum_{l=0}^{N} c_l^N
       \int du  \nonumber
     \end{eqnarray}
     where the coefficient function $c_l^N$ is given by eq. (3.21)
     and the mass scale $M_R^2$ is characterized via eqs.(4.1-4.4).
     The above prescription should be applicable to any fold ILIs.
     Here one needs to distinguish the difference of the integrating measures
     between the loop momentum $k$ and the momentum-like variable $\hat{k}$. In general,
     the SES $\mu_s$ is taken to be a finite energy scale in order to avoid possible IR problem, so that
     one can always make on mass-shell renormalization even for a massless theory.

     With the above prescription, the two-fold ILIs in eq. (5.18) are simply regularized
     into the following integrals
     \begin{eqnarray}
     I_{\alpha\beta\gamma}^{(2)R} & = & \Gamma_{\alpha \beta \gamma }\
     \int_0^1 dx\ x^{\gamma -1}(1-x)^{\alpha -1}\  (q_o^2)^{\alpha + \gamma -2 }
     \nonumber  \\
     & &  \int [\frac{d^4 \hat{k}_1}{\hat{k}_1^2} ]_l \
      \sum_{i=0}^{\beta -1} c_i^{\beta -1}
     \frac{(x(1-x)q_o^2)^{i}}{(\ \hat{k}_1^2 + x(1-x)q_o^2 + M_l^2 \ )^{\alpha + \gamma -1 +
     i} } \nonumber \\
     & & \int [d^4 k_2]_{l'} \frac{1}{
     (\ k_2^2 + {\cal M}_2^2  + M_{l'}^2 + \mu^2_{\hat{k}_1^2 + M_{l}^2}
     \ )^{\alpha + \beta + \gamma -2 } }
     \end{eqnarray}
     where the superscript $R$ means regularized ILIs.

     Based on the factorization theorem (Theorem I) and the subtraction theorem (Theorem II)
     for overlapping divergences as well as the harmless divergence theorem (theorem III)
     and the trivial convergence theorem (theorem IV) presented in the previous section,
     we shall be able to decompose the regularized ILIs into the following general form
     \begin{eqnarray}
     I_{\alpha\beta\gamma}^{(2)R} = I_{\alpha\gamma}^{(1)RD}I_{\alpha\beta\gamma}^{(1)RD}
     + I_{\alpha\gamma}^{(1)RC}I_{\alpha\beta\gamma}^{(1)RD}
     + I_{\alpha\beta\gamma}^{(2)RC}
     \end{eqnarray}
     where the superscripts `$RD$' and `$RC$' represent the regularized divergent and convergent
     ILIs respectively, and the numbers (1) and (2) in the superscripts label the one-fold and
     two-fold ILIs respectively. The possible harmful divergences must appear only
     in the first term with double divergences in the two-fold ILIs.

     To explicitly check the above decomposition, we may first carry out the integration
     over the loop momentum $k_2$ as it is actually an overall one-fold ILIs.
     One can directly read off the results from Appendix D for the regularized one-fold
     ILIs.

     We consider first the case in which the overall integral is logarithmically
      divergent, i.e., $\alpha + \beta + \gamma = 4$. The integration over $k_2$ results in
      the following explicit form
     \begin{eqnarray}
     I_{0}^{(2)R} & = & \Gamma_{\alpha \beta \gamma }\
     \pi^2\ \int_0^1 dx\ x^{\gamma -1}(1-x)^{\alpha -1}\  (q_o^2)^{\alpha + \gamma -2 }
     \nonumber  \\
     & & \int [\frac{d^4 \hat{k}_1}{\hat{k}_1^2} ]_l \
      \sum_{i=0}^{\beta -1} c_i^{\beta -1}
     \frac{(x(1-x)q_o^2)^{i}}{(\ \hat{k}_1^2 + x(1-x)q_o^2 + M_l^2 \ )^{\alpha + \gamma -1 +
     i} } \nonumber \\
     & & \{\ [ \ \ln \frac{M_c^2}{\mu_M^2 }
     - \gamma_w + y_0(\frac{\mu_M^2}{M_c^2}) \ ]
     - [ \ \ln(\ 1 +  \mu^2_{\hat{k}_1^2 + M_l^2}/\mu_M^2\ )
     - z_0(\hat{k}_1^2 + M_l^2) \ ] \  \} \ ,
     \end{eqnarray}

     We consider next the case in which the overall integral is quadratically divergent,
     i.e., $\alpha + \beta + \gamma = 3$. The integration over $k_2$ leads to the result
     \begin{eqnarray}
     I_{2}^{(2)R} & = & \Gamma_{\alpha \beta \gamma }\
     \pi^2\ \int_0^1 dx\ x^{\gamma -1}(1-x)^{\alpha -1}\  (q_o^2)^{\alpha + \gamma -2 }
     \nonumber  \\
     & & \int [\frac{d^4 \hat{k}_1}{\hat{k}_1^2} ]_l \
      \sum_{i=0}^{\beta -1} c_i^{\beta -1}
     \frac{(x(1-x)q_o^2)^{i}}{(\ \hat{k}_1^2 + x(1-x)q_o^2 + M_l^2 \ )^{\alpha + \gamma -1 +
     i} } \nonumber \\
     & & \{\ M_c^2 - \mu_M^2\ [ \ \ln \frac{M_c^2}{\mu_M^2 }
     - \gamma_w + 1 +  y_2(\frac{\mu_M^2}{M_c^2}) \ ] \nonumber \\
      & &  - \mu^2_{\hat{k}_1^2 + M_l^2}\ [ \ \ln \frac{M_c^2}{\mu_M^2 }
     - \gamma_w + 1 +  y_2(\frac{\mu_M^2}{M_c^2}) \ ] \nonumber \\
     & & + [\ \mu_M^2 + \mu^2_{\hat{k}_1^2 + M_l^2}\ ][ \ \ln(\ 1 +
     \mu^2_{\hat{k}_1^2 + M_l^2}/\mu_M^2\ )
     - z_2(\hat{k}_1^2 + M_l^2) \ ] \  \}
     \end{eqnarray}
     Here the functions $z_0$ and $z_2$
    are defined as
    \begin{eqnarray}
    & & z_0(\hat{k}_1^2 + M_l^2) =  y_0(\frac{\mu_M^2 + \mu^2_{\hat{k}_1^2 + M_l^2}  }{M_c^2})
    - y_0(\frac{\mu_M^2}{M_c^2}) \\
    & & z_2(\hat{k}_1^2 + M_l^2) =  y_2(\frac{\mu_M^2 + \mu^2_{\hat{k}_1^2 + M_l^2}  }{M_c^2})
    - y_2(\frac{\mu_M^2}{M_c^2}) \\
    & & \mu_M^2 = \mu_s^2 +  {\cal M}_2^2
    \end{eqnarray}
   From the explicit forms of the functions $y_0(x)$ and $y_2(x)$ (see eqs.(3.26) and (3.28)),
   it is easily seen that in the high energy limit the functions $z_0$ and $z_2$ are approaching to
   zero in terms of the inverse powers of  $\hat{k}_1^2$, i.e.,
    \begin{eqnarray}
    & & z_0(\hat{k}_1^2 + M_l^2) \sim O(\frac{\mu^2_{\hat{k}_1^2 + M_l^2}}{\mu_M^2})
     \sim O(\frac{q_o^2}{\hat{k}_1^2 + M_l^2} )
     \rightarrow 0 \qquad \mbox{at} \qquad \hat{k}_1^2 \rightarrow \infty \\
    & & z_2(\hat{k}_1^2 + M_l^2) \sim
    O(\frac{\mu^2_{\hat{k}_1^2 + M_l^2}}{\mu_M^2} ) \sim O(\frac{q_o^2}{\hat{k}_1^2 + M_l^2} )
    \rightarrow 0 \qquad \mbox{at} \qquad \hat{k}_1^2 \rightarrow \infty
    \end{eqnarray}
    It will be seen that the above properties are crucial for the treatment of overlapping divergences.

    To be more clear, consider first the simple case
    with $\alpha + \gamma = 2$ and $\beta = 2$. Applying the usual power counting rule
    to the general overlapping loop integrals, one sees that the overall integral in this case is
    logarithmically divergent ($\alpha +\beta +\gamma =4$), the sub-integral over $k_1$ is
    also logarithmically divergent ($\alpha +\gamma =2$), and the  sub-integral over $k_2$ is
    superficially convergent ($\beta +\gamma =3$). Turning to the corresponding
    ILIs, the usual power counting rule shows that the most divergent part of the sub-integral
    over the momentum-like variable $\hat{k}_1$ remains logarithmically one
    ($\alpha + \gamma =2$), whereas the sub-integral over $k_2$ in the corresponding ILIs
    becomes logarithmically divergent one ($\alpha +\beta +\gamma - 2 = 2$) as it
    characterizes the overall divergence of the overlapping loop integrals.
    One can easily carry out the integration for the regularized divergent part
    in the integral eq.(6.5). Indeed, we find that the regularized ILIs in this case
    can be written into the following form
    \begin{eqnarray}
     I_{0}^{(2)R} = \hat{I}_{0}^{(1)RD} I_{0}^{(1)RD} + \hat{I}_{0}^{(1)RC}I_{0}^{(1)RD}
     + I_{0}^{(2)RC}
     \end{eqnarray}
    where the involving regularized one-fold ILIs are given by
   \begin{eqnarray}
    \hat{I}_{0}^{(1)RD} & = & \pi^2\int_0^1 dx\ [\ \ln \frac{M_c^2}{\mu_{q_o}^2 }
     - \gamma_w + y_0(\frac{\mu_{q_o}^2}{M_c^2})  \ ]  \\
     I_{0}^{(1)RD} & = & \pi^2 [\ \ln \frac{M_c^2}{\mu_M^2 }
     - \gamma_w + y_0(\frac{\mu_M^2}{M_c^2}) \ ]  \\
     \hat{I}_{0}^{(1)RC} & = & -\pi^2\int_0^1 dx\  \frac{x(1-x)q_o^2}{ \mu_{q_o}^2 }
     [\ 1 - y_{-2} (\frac{\mu_{q_o}^2}{M_c^2}) \ ]
  \end{eqnarray}
   with
  \begin{eqnarray}
   \mu_{q_o}^2 = \mu_s^2 + x(1-x)q_o^2 \ .
  \end{eqnarray}
  The regularized convergent part of the two-fold ILIs is evaluated by
    \begin{eqnarray}
    I_{0}^{(2)RC} & = & - \pi^4 \int_0^1 dx\  \int  [ d \hat{k}_1^2]_l \
     \frac{\hat{k}_1^2 + M_l^2}{(\ \hat{k}_1^2 + x(1-x)q_o^2 + M_l^2 \ )^2 } \nonumber \\
    & &  [ \ \ln(\ 1 +  \mu^2_{\hat{k}_1^2 + M_l^2}/\mu_M^2\ )
     - z_0(\hat{k}_1^2 + M_l^2) \ ] \
     \end{eqnarray}

      Consider next the case with $\alpha + \gamma = 2$ and $\beta = 1$. In this case,
      the general overlapping loop integral contains two logarithmically
      divergent sub-integrals over $k_1$ ($\alpha\gamma$) and $k_2$ ($\beta\gamma$). The overall
      integral is quadratically divergent. In the corresponding ILIs, the sub-integral over
      the momentum-like variable $\hat{k}_1$ ($\alpha+\gamma = 2$) is logarithmically
      divergent, whereas the sub-integral over $k_2$ in ILIs becomes quadratically divergent
      ($\alpha + \beta +\gamma - 2 = 1$) as it characterizes the overall divergences. Similar to
      the first case, it is not difficult to carry out the integration concerning the regularized
      divergent parts and yield the result with the following form
      \begin{eqnarray}
     I_{2}^{(2)R} = \hat{I}_{0}^{(1)RD} I_{2}^{(1)RD} + \hat{I}_{2}^{(1)RC} I_{2}^{(1)RD} +
     I_{2}^{(2)RC}
     \end{eqnarray}
 where the regularized quadratically divergent one-fold ILI and the regularized convergent
 one-fold ILI are given by
 \begin{eqnarray}
 I_{2}^{(1)RD} & = & \pi^2\ \{\ M_c^2 - \mu_M^2\ [ \ \ln \frac{M_c^2}{\mu_M^2 }
     - \gamma_w + 1 +  y_2(\frac{\mu_M^2}{M_c^2}) \ ] \} \\
   \hat{I}_{2}^{(1)RC} & = & -\pi^2\  \int_0^1 dx\  \int [ d \hat{k}_1^2 ]_l \
     \frac{ \mu^2_{\hat{k}_1^2 + M_l^2}}{(\ \hat{k}_1^2 + x(1-x)q_o^2 + M_l^2 \ ) }
 \end{eqnarray}
 The regularized convergent two-fold ILI is evaluated by
   \begin{eqnarray}
    I_{2}^{(2)RC} & = &  \pi^4 \int_0^1 dx\ \int [ d \hat{k}_1^2 ]_l \
     \frac{\mu_M^2 + \mu^2_{\hat{k}_1^2 + M_l^2}}{(\ \hat{k}_1^2
     + x(1-x)q_o^2 + M_l^2 \ ) } \nonumber \\
     & & [ \ \ln(\ 1 + \mu^2_{\hat{k}_1^2 + M_l^2}/\mu_M^2\ )
     - z_2(\hat{k}_1^2 + M_l^2) \ ]
   \end{eqnarray}

    From the above explicit forms, it is easily seen that when taking the CES
    $M_c$ to be infinite, i.e., $M_c \rightarrow \infty $, we then recover
    the initial divergent behaviors of the ILIs (or the ones of the corresponding
    general overlapping Feynman integrals). We shall be more interested in the possible
    harmful divergences contained in the above results
    at $M_c \rightarrow \infty $. One notices that the divergences associated with
    possible non-polynomial external momentum dependent coefficient functions could occur
    only in the double divergent terms $\hat{I}_{0}^{(1)RD} I_{0}^{(1)RD}$ and
    $\hat{I}_{0}^{(1)RD} I_{2}^{(1)RD}$. This is because only the regularized one-fold ILIs
    $I_{0}^{(1)RD}$ and $I_{2}^{(1)RD}$ contain the logarithmic term
    $\ln {\cal M}_2^2 $, in which the mass factor ${\cal M}_2^2$ is in general
    the function of external momenta. It is remarkable to note that
    the double divergent terms in the two cases contain a common
    one-fold ILI, i.e., $\hat{I}_{0}^{(1)RD}$. This implies that
    the two harmful divergences at $M_c \rightarrow \infty $ are actually the same. Therefore,
    we only need introduce one subtraction term to make them becoming harmless.

      Consider now the regularization to the subtraction term for the sub-integral over
   the loop momentum $k_1$ (see eq.(5.17))
  \begin{eqnarray}
     I_{\alpha\beta\gamma}^{(2)RS} & = & \int [ d^4 k_1 ]_l
     \frac{1}{(k_1^2 + m_o^2 + M_l^2 )^{\alpha + \gamma} }
     \int [ d^4 k_2 ]_{l'} \frac{1}{\left(k_2^2 + {\cal M}_2^2 + M_{l'}^2 \right)^{\beta} }
     = I_{\alpha\gamma}^{(1)R} I_{\beta}^{(1)R}
  \end{eqnarray}
  which is actually factorized and can easily be carried out. For the case with $\alpha + \gamma = 2$
  and $\beta = 2$, we have
  \begin{eqnarray}
     I_{0}^{(2)RS} = \hat{I}_{0}^{(1)RD}(m_o) I_{0}^{(1)RD}
  \end{eqnarray}
  and  for the case with $\alpha + \gamma = 2$ and $\beta = 1$, we
  yield
  \begin{eqnarray}
     I_{2}^{(2)RS} = \hat{I}_{0}^{(1)RD}(m_o) I_{2}^{(1)RD}
  \end{eqnarray}
 where the explicit subtraction point $m_o$ is chosen.

  It then becomes obvious that the differences
  \begin{eqnarray}
    & & I_{0}^{(2)R}- I_{0}^{(2)RS} =  \tilde{I}_0^{(1)RC} I_{0}^{(1)RD}
    + \hat{I}_{0}^{(1)RC}I_{0}^{(1)RD} + I_{0}^{(2)RC}  \\
    & &  I_{2}^{(2)R} -I_{2}^{(2)RS} = \tilde{I}_0^{(1)RC} I_{2}^{(1)RD}
    + \hat{I}_{2}^{(1)RC} I_{2}^{(1)RD} + I_{2}^{(2)RC}
    \end{eqnarray}
  contain only harmless divegencies at $M_c \rightarrow \infty $. Here $\tilde{I}_0^{(1)RC}$
  is the additional convergent function arising from the subtraction
  \begin{eqnarray}
  \tilde{I}_0^{(1)RC} = \pi^2\int_0^1 dx\ [\ \ln \frac{m_o^2}{\mu_{q_o}^2 } +
     y_0(\frac{\mu_{q_o}^2}{M_c^2}) - y_0(\frac{m_o^2}{M_c^2}) \ ]
   \end{eqnarray}
   which is independent of the external momenta and vanishes if choosing $m_o^2
   =\mu_{q_o}^2$,
  \begin{eqnarray}
  \tilde{I}_0^{(1)RC} = 0 \ , \quad \mbox{for} \quad m_o^2=\mu_{q_o}^2
   \end{eqnarray}

   The above demonstrations for the two interesting cases have
   provided an explicit verification on the factorization theorem (theorem I) and
   the subtraction theorem (theorem II) for overlapping divergences, as well as on the
   harmless divergence theorem (theorem III) and the trivial convergence theorem (theorem IV).
   Practically speaking, it provides an explicit demonstration on the regularization and
   renormalization prescriptions for two loop graphs. Obviously, such prescriptions
   can easily be generalized to more closed loops.

   To establish the consistency of the new regularization,
   we shall further verify the generalized Ward identities (or gauge invariant conditions)
   at two and more loop level. As we have shown in section II that
   only the divergent integrals could spoil the generalized Ward identities. On the other hand,
   the generalized Ward identities only require the regularized tensor and scalar type
   divergent ILIs to satisfy a set of consistency conditions.
   Therefore the verification is equivalent to check whether the
   consistency conditions between the regularized tensor and scalar
   type divergent ILIs are still preserved at two and more loop level. For this purpose,
   one only needs to apply the theorems V and VI (i.e., the reduction theorem for overlapping
   tensor type integrals and the relation theorem  for the tensor and scalar type ILIs)
   to the regularized tensor and scalar type divergent ILIs. Obviously,
   the desired result can easily be achieved from those two
   theorems.

    Last but not least, it must be emphasized that the procedure respects unitarity and
    causality. This is because: (i) the usual Feynman parameter method and
    the newly formulated UV-divergence preserving parameter method do not
    violate unitarity and causality, this process is independent
    of regularization and can be made in any regularization
    schemes; (ii) the evaluation of ILIs involves just the shifts of integrating loop momenta
   and the convergent integrals over the loop momenta. This procedure is
   well defined and ensured by the translational invariance and the safety of convergent
   integrals; (iii) the regulators act on the whole Feynman integrals of loop
   graphs rather than on the propagators, which distinguishes to some Pauli-Villars
   inspired regularization schemes. The Feynman integrals are well defined
   in the Euclidean momentum space;
   (iv) unlike the usual Pauli-Villars inspired regularization schemes,
   the mass factors and numbers of the regulators (or regularization quanta) in the new regularization
   are eventually taken to be infinitely large and decouple from the theory. No additional
   singularities appear via the Feynman propagators; (v) also unlike the BPHZ subtraction scheme which is based on
   expanding around an external momentum, the new regularization and renormalization scheme
   introduces two intrinsic mass scales, i.e., the CES $M_c$ and SES $\mu_s$, to characterize the
   UV and IR behaviors of the Feynman amplitudes, so that the structure of
   the corresponding amplitudes involving the external momenta is not changed. This can be
   seen explicitly from the evaluation of the vacuum polarization function presented in section II and appendix A;
   (vi) it has been shown that in a given order any new subtraction terms are finite polynomials
    in external momenta, which is also necessary for maintaining unitarity of the theory.
   In fact, the subtraction process in the new regularization scheme can be made
   in a similar way as the one in the dimensional regularization.

 \section{Evaluation of ILIs for arbitrary loop graphs}

   It has been seen that evaluating ILIs for loop graphs is a crucial step in the new regularization
   method. We shall present in this section a general description on the evaluation of ILIs
   for arbitrary loop graphs. Though the demonstration may concern some tedious formulae, while it
   must be very useful for a practical computation of more loop diagrams.

    We begin with the scalar type overlapping loop integrals involved in the general n-loop diagrams
    with $n$ an arbitrary number
   \begin{eqnarray}
   I^{(n)}_{\alpha_i \alpha_{ij}} = \int d^4 k_n \frac{1}{(k_n^2 + {\cal M}_n^2 )^{\alpha_n} }
   \prod_{j>i} \prod_{i=1}^{n-1} \int d^4 k_i \frac{1}{(k_i^2 + {\cal M}_i^2 )^{\alpha_i} }
   \frac{1}{ [(k_i-k_j + p_{ij} )^2 + {\cal M}_{ij}^2 ]^{\alpha_{ij} } }
   \end{eqnarray}
   with $\alpha_i > 0$ and $\alpha_{ij} \ge 0$.
   For the loop diagrams in which the internal lines are topologically not crossed over each other,
   the loop integrals are corresponding to the case with $j= i + 1$
   \begin{eqnarray}
   I^{(n)}_{\alpha_i \alpha_{i i+1}} = \int d^4 k_n \frac{1}{(k_n^2 + {\cal M}_n^2 )^{\alpha_n} }
   \prod_{i=1}^{n-1} \int d^4 k_i \frac{1}{(k_i^2 + {\cal M}_i^2 )^{\alpha_i} }
   \frac{1}{ [(k_i-k_{i+1} + p_{i i+1} )^2 + {\cal M}_{i i+1}^2 ]^{\alpha_{i i+1} } }
   \end{eqnarray}
   Here $\alpha_i > 0$ and $\alpha_{i i+1} > 0$ as if one of them vanishes,
   the corresponding sub-integral becomes no longer an overlapping one.
   The most divergence of the above overlapping integrals is corresponding to the
   case with $\alpha_i =1$ ($i=1, \cdots n$) and $\alpha_{i i+1} = 1$ ($i=1, \cdots n-1$), which
   is quadratically divergent.  This
   can easily be shown from the power counting rule as the momentum dimension of the overlapping
   integral in this case is two ($4n-2n -(2n-1) = 2$).

    The above overlapping integrals can be regarded as the most general
    ones for n-loop graphs. Here we have also supposed that
    one can always apply the usual Feynman parameter method to
     all the momentum factors $k_i$, so that
     all propagators concerning $k_i$ and the external momenta $p_{l}$  are combined into a single one.
     Similarly for the propagators concerning $(k_i-k_j)$ ($i\neq j$). Therefore,
     the mass factors ${\cal M}_i^2$ and ${\cal M}_{ij}^2$
     are in general the functions of masses $m_l$ and external momenta $p_l$ ($l=1,2, \cdots$)
     \begin{eqnarray}
      {\cal M}_i^2 \equiv {\cal M}_i^2 (m_1^2, p_1^2, m_2^2, p_2^2, p_1\cdot p_2, \cdots), \qquad
       {\cal M}_{ij}^2 \equiv {\cal M}_{ij}^2 (m_1^2, p_1^2, m_2^2, p_2^2, p_1\cdot p_2, \cdots)
     \end{eqnarray}

  For simplicity, it is good enough to begin with the demonstration on the evaluation
  of ILIs for three loop graphs. Its generalization to more loop graphs is obvious. The
  procedure and prescription on the evaluation of ILIs for arbitrary loop diagrams will be
  presented after an explicit demonstration. For three
  loop diagrams, the general overlapping loop integral can be expressed as
   \begin{eqnarray}
  & &  I^{(3)}_{\alpha_i \alpha_{ij}}  =  \int d^4 k_3 \int d^4 k_2 \int d^4 k_1
   \frac{1}{(k_1^2 + {\cal M}_1^2 )^{\alpha_1} }\  \frac{1}{(k_2^2 + {\cal M}_2^2 )^{\alpha_2} }
   \ \frac{1}{(k_3^2 + {\cal M}_3^2 )^{\alpha_3} }  \\
   & &    \frac{1}{ [(k_1-k_2 + p_{12} )^2 + {\cal M}_{12}^2 ]^{\alpha_{12} } }
   \frac{1}{ [(k_1-k_3 + p_{13} )^2 + {\cal M}_{13}^2 ]^{\alpha_{13} } }
   \frac{1}{ [(k_2-k_3 + p_{23} )^2 + {\cal M}_{23}^2 ]^{\alpha_{23} } } \nonumber
   \end{eqnarray}

    As the first step, we apply the usual Feynman parameter method (see Appendix A)
    to the denominators containing
    the loop momentum $k_1$. Taking the Feynman parameter variables $a_1 = k_1^2 + {\cal M}_1^2 $,
    $a_2 =  (k_1-k_2 + p_{12} )^2 + {\cal M}_{12}^2 $ and $a_3 = (k_1-k_3 + p_{13} )^2
    + {\cal M}_{13}^2$, the sub-integral over $k_1$ becomes quadratic after
    making a momentum shift
    \[ k_{1 \mu} \rightarrow k_{1\mu} + (x_1 - x_2) (k_{2\mu} - p_{12 \mu})
     + x_2 (k_{3\mu} - p_{13\mu})\ . \]
   The resulting explicit form is given by
  \begin{eqnarray}
   & & I^{(3)}_{\alpha_i \alpha_{ij}}  =  \frac{ \Gamma (\alpha_1 + \alpha_{12} + \alpha_{13} ) }{
    \Gamma(\alpha_1) \Gamma(\alpha_{12} ) \Gamma(\alpha_{13}) } \int d^4 k_3 \int d^4 k_2 \int d^4 k_1
    \int_0^1 dx_1 \int_0^{x_1} dx_2 \nonumber \\
    & & \frac{(1-x_1)^{\alpha_1 - 1} (x_1 - x_2 )^{\alpha_{12} - 1}
    x_2^{\alpha_{13} -1 } }{ [\ k_1^2 + (x_1-x_2)(1-x_1 + x_2) (k_2 -p_{12})^2
    + x_2(1-x_2) (k_3 - p_{13})^2 + {\cal M}_{x_1 x_2}^2\ ]^{\alpha_1 + \alpha_{12} + \alpha_{13} } }
    \nonumber  \\
  & &  \frac{1}{(\ k_2^2 + {\cal M}_2^2\ )^{\alpha_2} }
   \  \frac{1}{ [( k_2-k_3 + p_{23} )^2 + {\cal M}_{23}^2 ]^{\alpha_{23} } } \
    \frac{1}{(\ k_3^2 + {\cal M}_3^2\ )^{\alpha_3} }
  \end{eqnarray}
   with
   \begin{eqnarray}
   {\cal M}_{x_1x_2}^2 \equiv (x_1 - x_2) {\cal M}_{12}^2 + (1-x_1) {\cal M}_1^2 + x_2 {\cal M}_{13}^2
   \end{eqnarray}

    Apply again the usual Feynman parameter method to the remaining
   denominators containing the loop momentum $k_2$. Taking the Feynman parameter
   variables $a_1 = k_2^2 + {\cal M}_2^2 $ and
    $a_2 =  (k_2-k_3 + p_{23} )^2 + {\cal M}_{23}^2 $, and making a momentum shift
   \[ k_{2 \mu} \rightarrow k_{2 \mu} + y_1 (k_{3\mu} - p_{23\mu} )  \]
    we yield
   \begin{eqnarray}
  & &  I^{(3)}_{\alpha_i \alpha_{ij}}  =  \frac{\Gamma (\alpha_1 + \alpha_{12} + \alpha_{13}) }{
  \Gamma(\alpha_1 ) \Gamma(\alpha_{12} ) \Gamma(\alpha_{13}) }
  \frac{\Gamma (\alpha_2 + \alpha_{23}) }{
  \Gamma(\alpha_{2} ) \Gamma(\alpha_{23}) } \int d^4 k_3 \int d^4 k_2 \int d^4 k_1
   \int_0^1 dx_1 \int_0^{x_1} dx_2 \nonumber \\
   & &  \frac{(1-x_1)^{\alpha_1 - 1} (x_1 - x_2 )^{\alpha_{12} - 1}
   x_2^{\alpha_{13} -1 }}{[\ k_1^2 + (x_1-x_2)(1-x_1 + x_2) (k_2 -p_{12})^2
   + x_2(1-x_2) (k_3 - p_{13})^2 + {\cal M}_{x_1x_2}^2\ ] ^{\alpha_1 + \alpha_{12} + \alpha_{13}} }
   \nonumber \\
   & &  \int_0^1 dy_1 \frac{(1-y_1)^{\alpha_2 -1} y_1^{\alpha_{23} - 1}}{[\ k_2^2
   + y_1(1-y_1) (k_3 - p_{13})^2 + {\cal M}_{y_1}^2 \ ]^{\alpha_2 + \alpha_{23} } }
   \frac{1}{(k_3^2 + {\cal M}_3^2 )^{\alpha_3} }
   \end{eqnarray}
   with
   \begin{eqnarray}
   {\cal M}_{y_1}^2 = (1-y_1) {\cal M}_{2}^2 + y_1 {\cal M}_{23}^2
   \end{eqnarray}

   We now adopt as the second step the UV-divergence preserving parameter method. Taking $a =
   k_1^2 + (x_1-x_2)(1-x_1 + x_2) (k_2 -p_{12})^2
   + x_2(1-x_2) (k_3 - p_{13})^2 + {\cal M}_{x_1x_2}^2$ and $b =
   k_2^2 + y_1(1-y_1) (k_3 - p_{13})^2 + {\cal M}_{y_1}^2 $, the integral becomes
   \begin{eqnarray}
  & &  I^{(3)}_{\alpha_i \alpha_{ij}}  =  \frac{\Gamma (\alpha_1 + \alpha_{12} + \alpha_{13}
 + \alpha_2 + \alpha_{23} ) }{ \Gamma(\alpha_1) \Gamma(\alpha_{12} ) \Gamma(\alpha_{13})
  \Gamma(\alpha_{2} ) \Gamma(\alpha_{23}) } \int d^4 k_3 \int d^4 k_2 \int d^4 k_1
   \int_0^1 dx_1 \int_0^{x_1} dx_2 \int_0^1 dy_1 \int_0^{\infty} du_1  \nonumber \\
   & & \frac{(1-x_1)^{\alpha_1 - 1}
   (x_1 - x_2 )^{\alpha_{12} - 1}
   x_2^{\alpha_{13} -1 } (1-y_1)^{\alpha_2 -1} y_1^{\alpha_{23} - 1}
   u_1^{\alpha_2 + \alpha_{23} - 1}}{ \{\ k_1^2 + [u_1 + (x_1-x_2)(1-x_1 + x_2)] k_2^2
   + \mu_{k_3}^2 + u_1 {\cal M}_{y_1}^2 +
    {\cal M}_{x_1x_2}^2\  \}^{\alpha_1 + \alpha_{12} + \alpha_{13} + \alpha_2 + \alpha_{23} }
    }\   \frac{1}{(k_3^2 + {\cal M}_3^2 )^{\alpha_3} } \nonumber
   \end{eqnarray}
   with
   \begin{eqnarray}
   \mu_{k_3}^2 & = & (x_1-x_2)(1-x_1 + x_2)
   [\ 1- \frac{(x_1-x_2)(1-x_1 + x_2)}{ u_1 + (x_1-x_2)(1-x_1 + x_2) } \ ]
   [\ p_{12} - y_1 (k_3 - p_{23} )\ ]^2 \nonumber \\
   & + & [x_2(1-x_2) + y_1(1-y_1) u_1] (k_3 - p_{13})^2
   \end{eqnarray}
   where we have made the momentum shift once more
   \[ k_{2 \mu} \rightarrow k_{2 \mu} +
   \frac{(x_1-x_2)(1-x_1 + x_2)}{ u_1 + (x_1-x_2)(1-x_1 + x_2) } \ [ p_{12\mu}
   -  y_1 (k_{3\mu} - p_{23\mu}) ] \]

   In general, after performing the UV-divergence preserving parameter method, one can safely
   carry out the integration over the momentum $k_1$ as
   $\alpha_1 + \alpha_{12} + \alpha_{13} + \alpha_2 + \alpha_{23} > 2$ for a non-trivial
   overlapping integral. After integration over $k_1$, the integral gets the following form
    \begin{eqnarray}
  & &  I^{(3)}_{\alpha_i \alpha_{ij}}  = \pi^2 \frac{\Gamma (\alpha_1 + \alpha_{12} + \alpha_{13}
 + \alpha_2 + \alpha_{23} -2 ) }{ \Gamma(\alpha_1) \Gamma(\alpha_{12} ) \Gamma(\alpha_{13})
  \Gamma(\alpha_{2} ) \Gamma(\alpha_{23}) } \int d^4 k_3 \int d^4 k_2
   \int_0^1 dx_1 \int_0^{x_1} dx_2 \int_0^1 dy_1  \nonumber \\
   & & \int_0^{\infty} du_1  \frac{u_1^{\alpha_2 + \alpha_{23} - 1} }{
   [u_1 + (x_1-x_2)(1-x_1 + x_2)]^{(\alpha_1 + \alpha_{12} + \alpha_{13}-2)
    + (\alpha_2 + \alpha_{23}) } } \  \frac{1}{(k_3^2 + {\cal M}_3^2 )^{\alpha_3} }   \\
    & &  \frac{(1-x_1)^{\alpha_1 - 1} (x_1 - x_2 )^{\alpha_{12} - 1}
   x_2^{\alpha_{13} -1 } (1-y_1)^{\alpha_2 -1} y_1^{\alpha_{23} - 1}}{
    \{ \  k_2^2 +  [ y_1 (1-y_1) + T_{1 xy} ] (k_3 - p_{23})^2
    + T_{u_1x} [ p_{12} - y_1 (k_3 - p_{23}) ]^2
     + \mu_{u_1}^2  \  \}^{\alpha_1 + \alpha_{12} + \alpha_{13} + \alpha_2 + \alpha_{23} -2 }
     } \nonumber
   \end{eqnarray}
   with
   \begin{eqnarray}
   & & T_{u_1x} = \frac{(x_1-x_2)(1-x_1 + x_2)}{ u_1 + (x_1-x_2)(1-x_1 + x_2) } [\
   1 - \frac{(x_1-x_2)(1-x_1 + x_2)}{ u_1 + (x_1-x_2)(1-x_1 + x_2) } \ ] \\
   & & T_{u_1xy} = \frac{x_2(1-x_2) - (x_1-x_2)(1-x_1 + x_2)y_1(1-y_1) }{
   u_1 + (x_1-x_2)(1-x_1 + x_2) } \\
   & & \mu_{u_1}^2 = \frac{u_1}{ u_1 + (x_1-x_2)(1-x_1 + x_2) } {\cal M}_{y_1}^2 +
    \frac{1}{ u_1 + (x_1-x_2)(1-x_1 + x_2) }{\cal M}_{x_1x_2}^2
   \end{eqnarray}
   where we have factored out the UV-divergence preserving integration over $u_1$.

   It is of interest to notice that the sub-integral over $k_2$ in eq.(7.10) becomes solely quadratic.
  We can then repeatedly adopt the UV-divergence preserving parameter method. Taking $a = k_2^2 +
  [ y_1 (1-y_1) + T_{1 xy} ] (k_3 - p_{23})^2   + T_{u_1x} [ p_{12} - y_1 (k_3 - p_{23}) ]^2
     + \mu_{u_1}^2 $ and  $b = k_3^2 + {\cal M}_3^2 $, we have
  \begin{eqnarray}
  & &  I^{(3)}_{\alpha_i \alpha_{ij}}  = \pi^2 \frac{\Gamma (\alpha_1 + \alpha_{12} + \alpha_{13}
 + \alpha_2 + \alpha_{23} +\alpha_3 -2 ) }{ \Gamma(\alpha_1) \Gamma(\alpha_{12} ) \Gamma(\alpha_{13})
  \Gamma(\alpha_{2} ) \Gamma(\alpha_{23}) \Gamma(\alpha_{3}) } \int d^4 k_3 \int d^4 k_2
   \int_0^1 dx_1 \int_0^{x_1} dx_2 \int_0^1 dy_1  \nonumber \\
   & & \int_0^{\infty} du_1  \frac{u_1^{\alpha_2 + \alpha_{23} - 1} }{
   [u_1 + (x_1-x_2)(1-x_1 + x_2)]^{ (\alpha_1 + \alpha_{12} + \alpha_{13} - 2)
    + (\alpha_2 + \alpha_{23}) } } \  \int_0^{\infty} du_2  \\
    & &  \frac{(1-x_1)^{\alpha_1 - 1} (x_1 - x_2 )^{\alpha_{12} - 1}
   x_2^{\alpha_{13} -1 } (1-y_1)^{\alpha_2 -1} y_1^{\alpha_{23} - 1} u_2^{\alpha_3 -1} }{
    \{ \  k_2^2 +   [ u_2 + y_1 (1-y_1)  + \rho_{u_1} ]\ k_3^2 + u_2 {\cal M}_3^2
    + \mu_{u_1}^2  + p_{u_1}^2 \  \}^{\alpha_1 + \alpha_{12} + \alpha_{13} + \alpha_2 + \alpha_{23} +\alpha_3 -2 }
     } \nonumber
   \end{eqnarray}
   with
   \begin{eqnarray}
   & & p_{u_1}^2 =  [ y_1 (1-y_1)  + \rho_{u_1} ]\  p_{23}^2 +
    T_{u_1x}\ [p_{12}^2 - 2y_1 p_{12}\cdot p_{23} ] \nonumber \\
    & & \rho_{u_1} =  T_{u_1x} y_1^2 + T_{u_1 xy}
   \end{eqnarray}
 where we have made the momentum shift
 \begin{eqnarray}
  k_{3 \mu} \rightarrow k_{3 \mu} + \frac{ [\ y_1 (1-y_1) -  T_{u_1x} y_1^2 +
  T_{u_1xy} ]\ p_{23 \mu} + T_{u_1x} y_1 p_{12 \mu}
    }{u_2 + y_1 (1-y_1)  + \rho_{u_1} }
 \end{eqnarray}

  In general, as $\alpha_1 + \alpha_{12} + \alpha_{13} + \alpha_2 + \alpha_{23} +\alpha_3
  -2 > 2$, we can safely perform the integration over $k_2$ and obtain
  the desired three-fold ILIs
   \begin{eqnarray}
  & &  I^{(3)}_{\alpha_i \alpha_{ij}}  = \pi^4 \frac{\Gamma (\alpha_1 + \alpha_{12} + \alpha_{13}
 + \alpha_2 + \alpha_{23} +\alpha_3 -4 ) }{ \Gamma(\alpha_1) \Gamma(\alpha_{12} ) \Gamma(\alpha_{13})
  \Gamma(\alpha_{2} ) \Gamma(\alpha_{23}) \Gamma(\alpha_{3}) }
  \nonumber \\
  & &  \int_0^1 dx_1 \int_0^{x_1} dx_2  (1-x_1)^{\alpha_1 - 1} (x_1 - x_2 )^{\alpha_{12} - 1}
   x_2^{\alpha_{13} -1 }  \int_0^1 dy_1 (1-y_1)^{\alpha_2 -1} y_1^{\alpha_{23} - 1}  \nonumber \\
   & & \int_0^{\infty} du_1   \frac{u_1^{\alpha_2 + \alpha_{23} - 1} }{
   [u_1 + (x_1-x_2)(1-x_1 + x_2)]^{ (\alpha_1 + \alpha_{12} + \alpha_{13} - 2)
    + (\alpha_2 + \alpha_{23}) } }    \nonumber  \\
    & & \int_0^{\infty} du_2 \frac{u_2^{\alpha_3 -1} }{ [ u_2 + y_1 (1-y_1)
     + \rho_{u_1} ]^{
    (\alpha_1 + \alpha_{12} + \alpha_{13} -2) + (\alpha_2 + \alpha_{23} -2) +\alpha_3 } }
     \  I^{(1)}_{\alpha_i \alpha_{ij}} ( {\cal M}_3^2 + \mu_{u_1 u_2 }^2 )
   \end{eqnarray}
   where $I^{(1)}_{\alpha_i \alpha_{ij}} ( {\cal M}_3^2 + \mu_{u_1 u_2 }^2 )$ defines the
   overall one-fold ILIs
   \begin{eqnarray}
   I^{(1)}_{\alpha_i \alpha_{ij}} ( {\cal M}_3^2 + \mu_{u_1 u_2 }^2 ) =
   \int d^4 k_3 \frac{ 1}{ \{ \  k_3^2 +  {\cal M}_3^2 + \mu_{u_1 u_2 }^2
      \  \}^{(\alpha_1 + \alpha_{12} + \alpha_{13} -2) + (\alpha_2 + \alpha_{23} -2) + \alpha_3  } }
   \end{eqnarray}
   with
   \begin{eqnarray}
  \mu_{u_1 u_2 }^2 & = & \frac{\mu_{u_1}^2 +  p_{u_1}^2
  - [ y_1 (1-y_1)  + \rho_{u_1} ]\   {\cal M}_3^2
    }{u_2 + y_1 (1-y_1)  + \rho_{u_1} } \nonumber \\
 & = & \frac{\mu_{u_1}^2 - [ y_1 (1-y_1)  + \rho_{u_1} ]\
 ( {\cal M}_3^2 - p_{23}^2 )
   +  T_{u_1x}\ [p_{12}^2 - 2y_1 p_{12}\cdot p_{23} ]
    }{u_2 + y_1 (1-y_1)  + \rho_{u_1} }
   \end{eqnarray}
  Similar to the two-fold ILIs of two loop graphs discussed in section V, it is useful to introduce
  the momentum-like integrating variables and measures
  \begin{eqnarray}
   \hat{k}_{i}^2 = u_i q_o^2 \ , \qquad
   \int_0^{\infty} du_i = \int \frac{d^4 \hat{k}_i}{\hat{k}_i^2}
   \frac{1}{q_o^2\pi^2} \quad (i=1,2)
  \end{eqnarray}
  Correspondingly, one needs to make the replacement for the relevant quantities
  \begin{eqnarray}
  & & T_{u_1x} \rightarrow T_{\hat{k}_1^2x}\ , \qquad T_{u_1xy} \rightarrow T_{\hat{k}_1^2xy}
  \nonumber \\
  & & \rho_{u_1}  \rightarrow \rho_{\hat{k}_1^2} \ , \qquad
  p_{u_1}^2 \rightarrow p_{\hat{k}_1^2}^2 \nonumber \\
  & & \mu_{u_1}^2 \rightarrow \mu_{\hat{k}_1^2}^2 \ ,
  \qquad  \mu_{u_1u_2}^2 \rightarrow \mu_{\hat{k}_1^2\hat{k}_2^2}^2
  \end{eqnarray}

   It is noted that in the infinite limit of $\hat{k}_1^2$ (or $u_1$) ,
   the quantities $T_{\hat{k}_1^2x}$,
   $T_{\hat{k}_1^2xy} $ , $\mu^2_{\hat{k}_1^2}$  and $ p_{\hat{k}_1^2}^2$ approach to zero. Consequently,
   we have
   \begin{eqnarray}
  & &  \rho_{\hat{k}_1^2}\ \  ( \rho_{u_1} ) \rightarrow 0 \qquad \mbox{at}
  \quad \hat{k}_1^2 \ \ (  u_1) \rightarrow \infty \\
  & &   \mu^2_{\hat{k}_1^2\hat{k}_2^2}\ \ (  \mu^2_{u_1 u_2} ) \rightarrow 0\ \quad \mbox{at}
  \qquad \hat{k}_1^2\ , \hat{k}_2^2 \  \  (  u_1\ , u_2 ) \rightarrow \infty
   \end{eqnarray}
 This  ensures the factorization property for overlapping divergences in three loop graphs.
 Here the sub-integrals over the momentum-like variables $\hat{k}_1$ and  $\hat{k}_2$
 characterize the UV divergent properties of one-loop and two-loop sub-diagrams respectively,
 and the integral over the loop momentum $k_3$ describes the overall divergent property of three
 loop diagrams.

  From the above demonstration on the evaluation of ILIs for three loop graphs, we can now
  straightforwardly generalize it to arbitrary loop graphs.
 The general prescription on the evaluation of ILIs may be summarized as follows:

  (1). As the first step, one shall repeatedly apply the usual Feynman parameter method to
 the denominators containing the same loop momentum according to the given order
 say $k_1$, $k_2$, $\cdots$, so that one arrives at the integrals in which n-loop Feynman
 integral contains correspondingly n's denominators.
 By making appropriate momentum shifts, each denominator is governed
 by the quadratic of momentum, say $k_1^2$, $k_2^2$, $\cdots$.
 With this procedure, it is not difficult to see that the momentum integral corresponding
 to the first denominator say $k_1$ becomes purely quadratic. In general,
 for n-loop overlapping integrals, one needs to repeatedly use the Feynman parameter method
 by $n-1$ times.

 (2). After the step (1),  one then adopts the UV-divergence preserving parameter method
 to the first two denominators, i.e., the ones involving $k_1^2$ and $k_2^2$. after that,
  one can safely carry out the integration over $k_1$ as the resulting sub-integral
  in such a way becomes convergent. Note that the sub-integral over $k_1$ is quadratic
  and its integration is easily performed.

 (3). One first makes momentum shift again for the momentum involved in the second denominator,
 i.e., $k_2$, so that the integral over $k_2$ becomes purely quadratic.
 Factoring out the coefficient of the momentum $k_2^2$, one arrives at the integral
 which contains UV-divergence preserving
 momentum-like sub-integral over $\hat{k}_1$ instead of the one over the loop momentum $k_1$,
 and the remaining integral has the same form as the one resulting
 from the first step (1) but with the reduced integrals over the loop momenta $k_2$, $k_3$, $\cdots$.

 (4). Repeating the steps (2) and (3) to the remaining loop momentum integrals over
 $k_2$, $k_3$, $\cdots$, but keeping the integral over the last loop momentum,
 we then arrive at the desired ILIs. In general, for n-loop overlapping integrals, the resulting
 ILIs are the n-fold ones with (n-1) sub-integrals over the momentum-like variables and
 one sub-integral over the loop momentum (say $k_n$).

   According to this procedure and prescription, the resulting
  ILIs for any loop graphs possess all the properties
  observed in the ILIs of two loop graphs (see section V).
  In here for the ILIs of more loop graphs, one only needs to notice the essential property that
  all the sub-integrals over the momentum-like variables $\hat{k}_1$, $\hat{k}_2$,
  $\cdots$ characterize the divergences of one-loop, two-loop, $\cdots$ sub-diagrams,
  respectively, and the single sub-integral over the initial loop momentum characterizes the overall
  divergences of the loop graph. One can check that this procedure also ensures
  that all the UV divergences are characterized by the integrals over
  the momentum-like variables $\hat{k}_1$, $\hat{k}_2$,
  $\cdots$, the usual Feynman parameter integrations contain no
   UV divergences and only IR divergences could be hidden in the Feynman parameter
   integrations. For the regularized ILIs in the new
   regularization scheme, all the Feynman parameter integrations
   become convergent. It is then obvious that all the theorems deduced in section V and
  the prescriptions for the regularization and renormalization described in section VI
  hold for ILIs of arbitrary loop graphs.

 It is of interest to observe that Feynman integrals of arbitrary loop graphs can always be
 evaluated into the corresponding ILIs which contains only a single one-fold
 ILI over the initial loop momentum which characterizes the overall
  divergences of the loop graph. Similarly, one can show that the subtracting the divergences from
  all the sub-integrals over the momentum-like variables $\hat{k}_1$, $\hat{k}_2$,
  $\cdots$, the overall divergence becomes harmless.

 \section{ Conformal scaling symmetry breaking and \\ The mass gap/quark confinement}

     It is particularly noted that
    the new regularization method may cause the conformal scaling symmetry
    to be broken down due to the existence of two intrinsic mass scales $M_c$ and $\mu_s$.
     For instance, because of the quadratic `cut-off' terms of the CES $M_c$,
     the conformal scaling symmetry can be broken down in a class of theories
     with scalar interactions.  In general, as long as taking the SES $\mu_s$ to be
     finite ($\mu_s\neq 0$), the conformal scaling
     symmetry will be broken down even in gauge theories.
     Of particular, once a mass gap $\mu_s = \mu_c$ is generated
     via dynamical reasons of strong interactions, it becomes natural to set the SES
     $\mu_s$ to be at the mass gap $\mu_s = \mu_c$.

     An interesting example is for QCD in which an energy scale around $1 $ GeV
     has often be introduced to characterize the low energy dynamics of QCD. This is because
     when the SES $\mu_s$ runs down to such a low energy scale, $\mu_s \sim 1$ GeV,
     the interactions become much stronger. This feature may easily be understood
     from the behavior of asymptotic freedom of QCD\cite{GW,HDP}.
     By evaluating the relevant one loop Feynman diagrams, it is not difficult to
     check that the renormalized gauge coupling constant $g_R$ from the new regularization
     method is consistent with the one from the dimensional regularization
     at $D\to 4$ and $M_c \to \infty$
     \begin{eqnarray}
     g_R(\mu_s) = g(M_c) [\ 1 + \frac{g^2(M_c)}{8\pi^2} \left( \frac{11}{12}C_1 - \frac{N_f}{3}C_2 \right)
     \left( \ln \frac{M_c^2}{\mu_s^2} - \gamma_w + y_0 (\frac{\mu_s^2}{M_c^2} ) \right) \ ]
     \end{eqnarray}
     where only the polynomial of the finite terms $\mu_s^2/M_c^2$ differ from the
     one in the dimensional regularization. In obtaining the above
     result, the masses of quarks have been ignored.

     On the other hand, for a finite SES $\mu_s$, the conformal scaling symmetry
     is explicitly broken down due to higher dimensional effective interaction
     terms, i.e., their momentum power counting dimensions are larger than four. In
     general, the UV convergent but IR divergent Feynman diagrams will result in such higher
     dimensional interaction terms. From the dimensional analysis, these higher dimensional
     interactions can be written down in terms of the expansion of the inverse power of the SES $\mu_s$
     \begin{eqnarray}
    & &  \frac{1}{\mu_s} \bar{q}\sigma^{\mu\nu}G_{\mu\nu}q \ , \quad
      \frac{1}{\mu_s^2} G_{\mu\rho} G^{\rho}_{\nu} G^{\mu\nu} \ , \quad
     \frac{1}{\mu_s^2} D^{\mu}G_{\nu\rho} D^{\mu}G^{\nu\rho} \ , \nonumber \\
    & & \frac{1}{\mu_s^2} (\bar{q}_Lq_R)( \bar{q}_R q_L ) \ , \quad
     \frac{1}{\mu_s^4} \left(G_{\mu\nu}G^{\mu\nu} \right)^2
     \quad \cdots
     \end{eqnarray}
     which implies that the existence of a non-trivial solution for the SES $\mu_s$
     is transmuted to the one of nonzero vacuum expectation values of
     $\langle \bar{q} q \rangle$ and $\langle \alpha_s G^a_{\mu\nu}G^{a \mu\nu}
     \rangle $. Indeed, it has been turned out that at a low energy scale
     $\mu_s \leq m_c \simeq 1.3 $ GeV ($m_c$ is the charm quark mass),
     the light quarks and gluons were found to have nonzero condensates \cite{SVZ}
     due to strong interactions
     \begin{eqnarray}
    & &  \langle \bar{q} q \rangle \simeq (-230 \mbox{MeV} )^3\ , \quad
     \bar{q}\sigma^{\mu\nu}G_{\mu\nu}q \simeq - (0.9 \mbox{GeV})^2 \langle \bar{q} q \rangle \ ,
     \nonumber \\
     & & \frac{1}{4\pi} \langle \alpha_s G^a_{\mu\nu}G^{a \mu\nu}
     \rangle \simeq (238 \mbox{MeV} )^4
     \end{eqnarray}
     which implies that both conformal scaling symmetry and
     chiral symmetry are dynamically broken down at such a low energy scale.
     This strongly indicates that
     a mass gap with $\mu_s =\mu_c \leq m_c $ must be generated and
     the quarks are going to be confined around such a low energy scale.
     Obviously, the conformal symmetry breaking scale or
     the mass gap $\mu_s =\mu_c$ must be related to the
     dynamically chiral symmetry breaking scale $\Lambda_{\chi}$
     which has been known to be around $ 1$ GeV.
     The order of magnitude for the mass gap $\mu_c$ may be estimated
     from the quark and gluon condensates. From simple dimensional and large $N_c$ considerations,
     we have
     \begin{eqnarray}
     \mu_c \sim \frac{\langle \alpha_s G^a_{\mu\nu}G^{a \mu\nu}\rangle}{
     3 \langle \bar{q} q \rangle } \simeq 1.1 \mbox{GeV}
     \end{eqnarray}
     which holds in the large $N_c$ limit with $\alpha_s N_c$ being held fixed.
     As expected, the mass gap $\mu_c$ and the chiral symmetry breaking scale
     $\Lambda_{\chi}$ are at the same order of magnitude, namely:
     \begin{eqnarray}
     \mu_s =\mu_c \sim \Lambda_{\chi}  \sim 1.1 \mbox{GeV}.
     \end{eqnarray}
     Correspondingly, the critical temperature for the quark deconfinement
     is estimated to be at the order of magnitude
     \begin{eqnarray}
     T_c \sim \frac{\mu_c}{2\pi} \sim \frac{\Lambda_{\chi}}{2\pi}
     \sim  175 \mbox{MeV}.
     \end{eqnarray}

     Note that to obtain more definitive results, a detailed quantitative calculation is
     needed. Only the perturbative evaluation of QCD is not enough to determine the numerical value for the mass gap.
     It is necessary to develop low energy dynamics of QCD for a nonperturbation evaluation of QCD.
     One may construct some effective perturbation theories to describe the low energy QCD.
     The crucial issue concerns a consistent matching between the perturbative and nonperturbative QCD.
     In general, the IR cut-off in the perturbative QCD is related to the UV cut-off in the
     nonperturbative QCD. As a good example, it has been shown in ref.\cite{YLWU} that the UV cut-off in the
     chiral perturbation theory of mesons does match to the IR cut-off in perturbative QCD via some appropriate
     matching conditions.  Therefore, it is of interest to further investigate
     the IR and UV correspondence/matching of the short and long distance dynamics of QCD.
     As the new regularization is applicable to both underlying perturbative QCD and
     effective QFTs characterizing nonperturbative QCD at low energy,
     it may provide us a practical tool to understand well
     the genesis of mass gap and the quark confinement via the IR and UV correspondence/matching of
     the short and long distance dynamics of QCD.

  \section{Conclusions}

    We have explicitly presented a proof for the existence of a
    symmetry principle preserving  and infinity free regularization
    and renormalization method.  The main point for the new method is to regularize the
     whole Feynman integrals of loop graphs rather than just the propagators of
     fields. The other important point is to analyze an infinitely large number
     of regulators (or regularization quanta) rather than only a limited few regulators.
     Specifically, the concepts of irreducible loop integrals (ILIs) and UV-divergence
     preserving parameter method as well as intrinsic mass scales (i.e., the characteristic
     energy scale (CES) $M_c$ and the sliding energy scale (SES) $\mu_s$ )  have been found to
     play an essential role. Consequently, the regularized QFTs become well defined.
     The method has been developed to be practically applicable to
     both underlying renormalizable QFTs and effective QFTs.
     Indeed, it provides an explicit demonstration
    that a quantum theory of fields can be applied to describe the laws of nature
     only when the considered energy scale is sufficiently lower than the CES $M_c$.
     Thus the underlying renormalizable QFTs, like Yang-Mills gauge theories,
     become particularly interesting as the CES $M_c$ in such theories is in principle
     allowed to be infinitely large $M_c \to \infty$ through a suitable renormalization
     for the gauge coupling and fields, which is equivalent to the dimensional regularization
     with $D\rightarrow 4$.

     In general, the scales $M_c$ and $\mu_s$ at a particularly interesting energy scale
     $\mu_s = \mu_c$ actually set the UV and IR cut-off energy scales
     respectively in the regularized QFTs, but they distinguish from the naive cutoff momentum scales
    imposed kinematically to the upper and lower bounds of the internal loop momenta,
    the CES $M_c$ and the SES $\mu_s$ are more likely to behave as the dynamical mass
    scales, which remains to be understood from a deeper reason. It is intriguing to see that
     the new regularization method provides a consistent way alternative to string theories
     for making QFTs to be well defined finite theories. Though the explicit
     demonstration for it has been carried out only at one and two loop
     level, its proof to more loops has been shown to be a straightforward
     generalization. In this sense, the divergence problem of QFTs may no
     longer be a strong reason for extending the underlying theory to go
     beyond the quantum theory of fields /elementary particles.

    First of all, it must be very helpful to
    further study the possible interesting phenomena caused
    by the conformal scaling symmetry breaking. In particular, the IR and UV
    correspondence between the short distance physics and long distance physics may
    play an important role to explore the nonperturbative effects and to understand well
    the matching between the short and long distance physics.
    We believe that the conformal scaling symmetry and its breaking mechanism may eventually
    guide us to reveal some longstanding puzzles,
    such as: genesis of mass gap, unseen quarks, missing symmetries,
    origin of quark and lepton masses and mixing angles,
    and also small but nonzero cosmological constant. Finally, we hope that the new method
    described in this paper can widen the applications of QFTs.
 \\
 \\
 {\bf Acknowledgement}

  The author is grateful to Prof. David J. Gross for valuable discussions during
  the string conference at ITP, Beijing, especially for his important comments involving
  the intrinsic mass scales and conformal scaling symmetry breaking
  in the new regularization method. He is also benefited from discussions
  with Prof. L.D. Faddeev concerning the infrared cutoff energy
  scale and the mass gap in Yang-Mills gauge theories. The author wants to express his thanks
  to Profs. K.C. Chou, Y.B. Dai, H.Y. Guo, Z.X. He, X.D. Ji, L.F. Li,
  K.F. Liu, H.W. Peng, M. Yu, Z.X. Zhang, Z.Y. Zhu and many other
  colleagues for stimulating discussions and conversations, and C.J. Zhu for
  carefully reading the manuscript with useful comments.
  He is also grateful to Dr. Y.A. Yan for the numerical check on the conjectures for
  three type of functional limits appearing in the new regularization method.
  Especially, he would like to thank Profs. B.A. Li, W. Palmer and L. Wolfenstein
  for their kind hospitality to visit Kentucky University, Ohio-State University and
  Carnegie-Mellon University during his sabbatical on leave,
  where he was able to continue thinking the subject
  over. This work was supported in part by the key projects
  of Chinese Academy of Sciences and National Science Foundation of China.

 \newpage

 \appendix

 \section{Vacuum polarization functions of gauge fields}

  In this appendix, we present a detailed evaluation of the gauge field vacuum polarization function
  from various non-vanishing diagrams. In the evaluation, we shall often use the Feynman parameter
  formula
  \begin{equation}
    \frac{1}{a_1^{\alpha}a_2^{\beta}} = \frac{\Gamma(\alpha + \beta)}{\Gamma(\alpha)\Gamma(\beta)}
    \int_{0}^{1}\ dx \frac{(1-x)^{\alpha - 1} x^{\beta -1} }{[a_1(1-x) + a_2 x]^{\alpha +\beta } }
    \end{equation}
    and its generalized form
   \begin{eqnarray}
   \frac{1}{a_1^{\alpha_1}a_2^{\alpha_2} \cdots a_n^{\alpha_n}} & = &
   \frac{\Gamma(\alpha_1 + \cdots + \alpha_n)}{\Gamma(\alpha_1) \cdots \Gamma(\alpha_n)}
   \int_0^1 dx_1 \int_0^{x_1} dx_2 \cdots \int_0^{x_{n-2}} dx_{n-1} \nonumber \\
   & & \frac{(1-x_1)^{\alpha_1 -1} (x_1 - x_2)^{\alpha_2 - 1} \cdots x_{n-1}^{\alpha_n - 1} }{
   [\ a_1(1-x_1) + a_2 (x_1 -x_2) + \cdots + a_n x_{n-1}\ ]^{\alpha_1 + \cdots + \alpha_n} }
   \end{eqnarray}
  to simplify the integrals.

   There are in general four Feynman diagrams (see Figures (1)-(4)).
   Consider first the diagram (1) which comes from the Yang-Mills trilinear interaction term.
    With the gauge boson propagator and the Yang-Mills trilinear vertex of the Feynman rules, one has
    \begin{eqnarray}
    \Pi_{\mu\nu}^{(1) ab} = -\frac{1}{2}g^2 f_{acd}f_{bdc} \int \frac{d^4 k}{(2\pi)^4}\
    \frac{ P_{\mu\nu} (p, k) }{((p+k)^2 + i\varepsilon )(k^2 + i\varepsilon) }
    \end{eqnarray}
 with
 \begin{eqnarray}
 P_{\mu\nu} & = & 10 k_{\mu}k_{\nu} + 5 (p_{\mu}k_{\nu} + p_{\nu}k_{\mu}) - 2p_{\mu}p_{\nu} +
 (5p^2 + 2p\cdot k + 2k^2) g_{\mu\nu} \nonumber \\
 & + & \frac{\lambda}{k^2} [\ (k^2+2p\cdot k - p^2 )k_{\mu}k_{\nu} + (k^2 + 3p\cdot k) (p_{\mu}k_{\nu}
 + p_{\nu}k_{\mu}) - k^2 p_{\mu}p_{\nu} - (k^2 + 2p\cdot k )^2 g_{\mu\nu} ] \nonumber \\
 & + & \frac{\lambda}{(p+k)^2} [\ (k^2 -2p^2)k_{\mu}k_{\nu} + (p\cdot k)
 (p_{\mu}k_{\nu} + p_{\nu}k_{\mu}) + (p^2 - 2k^2) p_{\mu}p_{\nu} - (k^2 - p^2)^2 g_{\mu\nu} ]
 \nonumber \\
 & + & \frac{\lambda^2}{(p+k)^2 k^2} [\ p^4 k_{\mu}k_{\nu} + (p\cdot k)^2 p_{\mu}p_{\nu}
 - (p \cdot k) p^2 (p_{\mu}k_{\nu} + p_{\nu}k_{\mu})
 \end{eqnarray}
  After using the Feynman parameter method with  $a_1 = k^2$ and $a_2 = (p+k)^2$, and
 replacing $k_{\mu}$ by $(k_{\mu} - xp_{\mu})$, as well as
 performing some algebra, the gauge field vacuum polarization function
 $\Pi_{\mu\nu}^{(1) ab}$ can be expressed into the following general form in terms of
 the ILIs
 \begin{eqnarray}
 \Pi_{\mu\nu}^{(1) ab}& = & g^2 C_1 \delta_{ab} \{ \ \int_{0}^{1} dx\ [ \
 I_2 g_{\mu\nu} + 5 I_{2 \mu\nu} + \left( \ (5/2 - 2x(1-x) p^2 g_{\mu\nu}
 - (1+5x(1-x)) p_{\mu}p_{\nu}\right) I_0 \  ] \nonumber \\
 & + & \lambda \Gamma(3) \int_{0}^{1} dx\ x\ [ \ - I_2 g_{\mu\nu} + I_{2 \mu\nu} +
 \left(\ 2(1-x)(1+2x) p^2 g_{\mu\nu} + (x^2 -2) p_{\mu}p_{\nu} \right)\ I_{0} \nonumber \\
 & - & \left(\ (x + 2(1-x^2)\ ) p^2 g_{\mu}^{\rho}g_{\nu}^{\sigma} + 4x^2 p^{\rho}p^{\sigma}
 g_{\mu\nu} - (2x^2 + 1) p^{\rho}\ ( g_{\mu}^{\sigma} p_{\nu}
 + g_{\nu}^{\sigma} p_{\mu} ) \right) I_{0 \rho\sigma} \nonumber \\
 & - & \left(\ (1-x)^2(1+2x)^2 p^4 g_{\mu\nu}
 - ( 1 + x(1-2x)(2-x^2)\ ) p^2 p_{\mu}p_{\nu} \right)\ I_{-2}\ ] \nonumber  \\
 & + & \lambda^2\ \Gamma(4)/2 \int_{0}^{1} dx\ x(1-x)\ [ \
 p^4 g_{\mu}^{\rho}g_{\nu}^{\sigma} +  p_{\mu}p_{\nu}p^{\rho}p^{\sigma}
 - p^2 p^{\rho}\ ( g_{\mu}^{\sigma} p_{\nu} + g_{\nu}^{\sigma} p_{\mu} )\ I_{-2\rho\sigma}\ ] \ \}
 \end{eqnarray}
  where $C_1$ and $\lambda$ are defined as
  \begin{equation}
  f_{acd}f_{bcd} = C_1 \delta_{ab}, \qquad \lambda = 1-\xi
 \end{equation}
  and the mass term in all the ILIs is given by
  \begin{equation}
  {\cal M}^2 = - x(1-x)p^2
  \end{equation}

 The diagram (2) is the so-called tadpole diagram  with no momentum
 flowing into the loop and it arises from the Yang-Mills quadrilinear
 interaction terms. This diagram is quadratically divergent from power counting
 and vanishes in dimensional regularization, which has been thought to be the crucial
 point for preserving the gauge invariance in dimension regularization. As a general
 evaluation before applying for any regularization, it is better to treat
 it in the equal foot without worrying about quadratic divergences
 \begin{eqnarray}
    \Pi_{\mu\nu}^{(2) ab} & = & g^2 f_{acd}f_{bdc} \int \frac{d^4 k}{(2\pi)^4}\
    \frac{ 3g_{\mu\nu} - \lambda (g_{\mu\nu} - k_{\mu}k_{\nu}/k^2)
     }{(k^2 + i\varepsilon) } \nonumber \\
     & = & -g^2 C_1 \delta_{ab} \int \frac{d^4 k}{(2\pi)^4}\
    \frac{ (p+k)^2 [\ 3g_{\mu\nu} - \lambda (g_{\mu\nu} - k_{\mu}k_{\nu}/k^2)\ ]
     }{((p+k)^2 + i\varepsilon)(k^2 + i\varepsilon) }
 \end{eqnarray}
 In order to be able to express this loop integral by using the same ILIs
  as those appearing in diagram (1), we have inserted
 the momentum factor $(p+k)^2/(p+k)^2$. Adopting the same Feynman parametrization and
 performing some algebra, we yield
 \begin{eqnarray}
    \Pi_{\mu\nu}^{(2) ab}  & = & -g^2 C_1 \delta_{ab}\ \{\  3\int_{0}^{1} dx\ [\
    I_2 + (1-x)(1-2x) p^2 I_0 \ ] g_{\mu\nu} \nonumber \\
    & + & \lambda \Gamma(3) \int_{0}^{1} dx\ [\ -\frac{1}{2} I_2\ g_{\mu\nu} +
    (1-x) I_{2\mu\nu} \ ]\nonumber \\
    & + & \lambda \Gamma(3) \int_{0}^{1} dx\ [\ \left(\
    -\frac{1}{2}(1-x)(1-2x) p^2 g_{\mu\nu} +
    (1-x)x^2 p_{\mu}p_{\nu} \ \right)\ I_0 \nonumber \\
    & + & \left( \ (1-x)^2(1-2x)p^2 g_{\mu}^{\rho}g_{\nu}^{\sigma}
    - 2(1-x)^2x p^{\rho}\ ( g_{\mu}^{\sigma} p_{\nu} + g_{\nu}^{\sigma} p_{\mu} )\ \right)\
    I_{0\rho\sigma}\ ] \nonumber \\
     & + & \lambda \Gamma(3) \int_{0}^{1} dx\ (1-x)^2(1-2x)x^2 p_{\mu}p_{\nu} p^2
     \ I_{-2} \  \}
  \end{eqnarray}
  which shows that without imposing any regularization, the tadpole graphs of Yang-Mills fields
  are in general no vanishing and has actually all divergent structures given by the
  ILIs.

 We now consider the diagram (3) which arises from the ghost-gauge field interaction term. Using
 the Faddeev-Popov ghost propagator and ghost-gauge field vertex with including a minus sign for
 the closed loop of Grassmann fields, one has
 \begin{eqnarray}
    \Pi_{\mu\nu}^{(3) ab} & = & g^2 f_{acd}f_{bdc} \int \frac{d^4 k}{(2\pi)^4}\
    \frac{ (p+k)_{\mu}k_{\nu} }{((p+k)^2 + i\varepsilon )(k^2 + i\varepsilon) } \nonumber \\
    & = & -g^2C_1 \delta_{ab}\  \int_{0}^{1} dx\ [\ I_{2\mu\nu} - x(1-x) p_{\mu}p_{\nu} I_0 \ ]
 \end{eqnarray}
 In obtaining the second equality, we have used the same
 Feynman parametrization as the previous ones.

 Consider finally the diagram (4) due to fermion-gauge field interactions. Using the fermion
 propagator and fermion-gauge field vertex with including a minus sign for the
 closed fermion loop, one reads
  \begin{eqnarray}
    \Pi_{\mu\nu}^{(f) ab} & = & -g^2 4N_f tr T_a T_b \int \frac{d^4 k}{(2\pi)^4}\
    \frac{(p+k)_{\mu}k_{\nu} + (p+k)_{\nu}k_{\mu} + (m^2 -k^2-p\cdot k)g_{\mu\nu}
     }{((p+k)^2 + i\varepsilon )(k^2 + i\varepsilon) } \nonumber \\
     & = & -g^2 4N_f C_2  \delta_{ab} \  \int_{0}^{1} dx\ [\ 2 I_{2\mu\nu} (m)
     - I_2(m) g_{\mu\nu} + 2x(1-x) (p^2 g_{\mu\nu} - p_{\mu}p_{\nu} ) I_0(m) \ ]
     \end{eqnarray}
   where we have used the definition for the group theory factor
  \begin{equation}
  tr T_a T_b = C_2 \delta_{ab}
  \end{equation}
  Note also that the mass factor in the ILIs is modified by the fermion
  mass $m$ as indicated in the notation $I_{f}(m)$
  \begin{equation}
  {\cal M}^2 =  m^2 - x(1-x)p^2
  \end{equation}

  To simplify the integrals, the following identities are found to be helpful
  \begin{eqnarray}
  & & \int_{0}^{1} dx\ x^2\ I[x(1-x)] = \int_{0}^{1} dx\ x(1-x)\ I[x(1-x)] + \frac{1}{2}\int_{0}^{1} dx\ I[x(1-x)] \nonumber \\
  & & \int_{0}^{1} dx\ x^3\ I[x(1-x)] =
  \frac{3}{2}\int_{0}^{1} dx\ x(1-x)\ I[x(1-x)] + \frac{1}{2}\int_{0}^{1} dx\ I[x(1-x)] \nonumber \\
  & & \int_{0}^{1} dx\ x^4\ I[x(1-x)] = \int_{0}^{1} dx\ x^2(1-x)^2\
  I[x(1-x)] \nonumber \\
  & & \qquad + 2\int_{0}^{1} dx\ x(1-x)\ I[x(1-x)] + \frac{1}{2} \int_{0}^{1} dx\ I[x(1-x)]  \\
  & & \int_{0}^{1} dx\ x^5\ I[x(1-x)] = \frac{5}{2}\int_{0}^{1} dx\ x^2(1-x)^2\
  I[x(1-x)] \nonumber \\
   & & \qquad + \frac{5}{2}\int_{0}^{1} dx\ x(1-x)\ I[x(1-x)] + \frac{1}{2} \int_{0}^{1} dx\ I[x(1-x)] \nonumber
  \end{eqnarray}
  \\

  \section{Some Regularized one-fold ILIs in Cut-off and dimensional regularizations }

    We present here some regularized one-fold ILIs in the cut-off and dimensional regualrizations.
    The relevant one-fold ILIs in the text are
     \begin{eqnarray}
     & & I_2  = \int \frac{d^4 k}{(2\pi)^4}\ \frac{1}{k^2 - {\cal M}^2} \qquad
     I_0 = \int \frac{d^4 k}{(2\pi)^4}\  \frac{1}{(k^2 - {\cal M}^2)^2}  \\
     & & I_{2\ \mu\nu}  =  \int \frac{d^4 k}{(2\pi)^4}\
     \frac{k_{\mu}k_{\nu}}{(k^2 - {\cal M}^2)^2} \qquad
     I_{0\ \mu\nu} = \int \frac{d^4 k}{(2\pi)^4}\
       \frac{k_{\mu}k_{\nu}}{(k^2 - {\cal M}^2)^3}
    \end{eqnarray}

    Consider first the cut-off regularization, the ILIs can be easily
    evaluated in four dimensional Euclidean space of momentum by performing a Wick rotation
   \begin{eqnarray}
     & & I_2^R  = i\int^{\Lambda^2} \frac{d^4 k}{(2\pi)^4}\ \frac{1}{-k^2 -{\cal M}^2}
     = \frac{-i}{16\pi^2} \ [\ \Lambda^2 - {\cal M}^2\
     \ln \frac{\Lambda^2 + {\cal M}^2}{{\cal M}^2} \ ] \\
    & & I_0^R = i\int^{\Lambda^2} \frac{d^4 k}{(2\pi)^4}\  \frac{1}{(-k^2 - {\cal M}^2)^2} =
      \frac{i}{16\pi^2} \ [\ \ln \frac{\Lambda^2 + {\cal M}^2}{{\cal M}^2}  +
      \frac{{\cal M}^2}{\Lambda^2 + {\cal M}^2} -1 \ ]
    \end{eqnarray}
    and
    \begin{eqnarray}
     I_{2\ \mu\nu}^R & = & i\int^{\Lambda^2} \frac{d^4 k}{(2\pi)^4}\
     \frac{-k_{\mu}k_{\nu}}{(-k^2 - {\cal M}^2)^2}
     =  \frac{1}{2}g_{\mu\nu}\ \frac{-i}{16\pi^2}\ [\ \frac{1}{2} \Lambda^2
     + \frac{1}{4} {\cal M}^2
     - {\cal M}^2\ \ln \frac{\Lambda^2 + {\cal M}^2}{{\cal M}^2} \ ]  \\
    I_{0\ \mu\nu}^R  & = & i\int^{\Lambda^2} \frac{d^4 k}{(2\pi)^4}\
       \frac{-k_{\mu}k_{\nu}}{(-k^2 - {\cal M}^2)^3}
       =  \frac{1}{4}g_{\mu\nu}\ \frac{i}{16\pi^2}
     \ [\ \ln \frac{\Lambda^2 + {\cal M}^2}{{\cal M}^2}  + \frac{2{\cal M}^2}{\Lambda^2 + {\cal M}^2}
     - \frac{{\cal M}^4}{2(\Lambda^2 + {\cal M}^2)^2} -  \frac{3}{2} \ ]
    \end{eqnarray}
 where $\Lambda^2$ is the cut-off momentum. Note that the above integrals
  are carried out in the Euclidean space and
 the final results are given by rotating back to the Minkowski space,
 all integrations over momentum will be performed in this way except with a specific mention.

  Consider next the dimensional regularization in which the space-time dimension $D$ is
  taken $D = 4 -\varepsilon$
 \begin{eqnarray}
     & & I_2^R  = i\int \frac{d^D k}{(2\pi)^D}\ \frac{1}{-k^2 - {\cal M}^2}
     = \frac{-i}{(4\pi)^{D/2}}\ \left({\cal M}^2\right)^{D/2-1} \  \Gamma(1-\frac{D}{2}) \\
     & & I_0^R = i\int \frac{d^D k}{(2\pi)^D}\  \frac{1}{(-k^2 - {\cal M}^2)^2} =
      \frac{i}{(4\pi)^{D/2}}\ \left({\cal M}^2\right)^{D/2-2} \  \Gamma(2-\frac{D}{2})
    \end{eqnarray}
    and
    \begin{eqnarray}
     I_{2\ \mu\nu}^R & = &i\int \frac{d^D k}{(2\pi)^D}\
     \frac{-k_{\mu}k_{\nu}}{(-k^2 - {\cal M}^2)^2}
     =  \frac{1}{D}g_{\mu\nu}\ \frac{-i}{(4\pi)^{D/2}}\ \left({\cal M}^2\right)^{D/2-1} \
     \frac{\Gamma( 1 + \frac{D}{2}) \Gamma(1-\frac{D}{2}) }{ \Gamma( \frac{D}{2}) }  \\
     I_{0\ \mu\nu}^R & = & i\int \frac{d^D k}{(2\pi)^D}\
     \frac{-k_{\mu}k_{\nu}}{(-k^2 - {\cal M}^2)^3} =
      \frac{1}{D}g_{\mu\nu}\ \frac{i}{(4\pi)^{D/2}}\
      \left({\cal M}^2\right)^{D/2-2} \
      \frac{\Gamma( 1 + \frac{D}{2}) \Gamma(2-\frac{D}{2}) }{ \Gamma( \frac{D}{2}) \Gamma(3)}
    \end{eqnarray}

\section{Mathematically interesting functions and \\ limiting numbers involved in the new method}

     We first present in the following table some numerical results for the
     functions $h_w(N)$, $\gamma_w(N)$ and $L_n(N)$ when the regulator number
     $N$ is taken to be typically large.
     \\
     \\
     Table 1. The quantities $h_w(N)$, $\gamma_w(N)$ and $L_n(N)$ ($n=1,2,3,4,5,6$) as
     functions of the regulator number $N$.
     \\

     \begin{tabular}{|c|c|c|c|c|c|c|c|c|c|c|c|c|}
    \hline
     N   &  2 & 4  &  6 & 10 & 100 & 500 & 5000 & 10000 & 20000
     \\  \hline  \hline
    $ h_w(N)$ & 1.0407 & 1.0614 & 1.0676 & 1.0715 & 1.0667 & 1.0586 & 1.0487 & 1.0462 & 1.0439
    \\ \hline
    $\gamma_w(N)$ & 1.0198 & 0.7645 & 0.7059 & 0.6640 & 0.6037 & 0.5927 & 0.5860 & 0.5848 & 0.5839
    \\  \hline
    $L_1(N)$ & 2.079  & 1.416 &   1.281 & 1.187 & 1.0560 & 1.0325 & 1.0182& 1.0158 & 1.0138
    \\  \hline
    $L_2 (N)$ & 6.726  & 2.662 &  2.048 & 1.664 & 1.1828 & 1.1040 & 1.0574 & 1.0495 & 1.0431
    \\  \hline
    $ L_3(N)$ & 29.97  & 6.37 &   4.00 & 2.74 & 1.4124 & 1.2266 & 1.1220  & 1.1047 & 1.0909
    \\  \hline
     $L_4(N)$ & 171.74 & 18.79 &  9.42  & 5.27 &  1.8072 & 1.4218 & 1.2191 & 1.1868 & 1.1613
    \\  \hline
     $L_5(N)$ & 1209.62  & 66.59&  26.24 &  11.74 & 2.4888 & 1.7265  & 1.3599 & 1.3040 & 1.2606
    \\  \hline
     $L_6(N)$ & 10141.19  & 277.45 &  85.08 & 30.08 & 3.7152 & 2.2060 & 1.5614  & 1.4690 & 1.3981
    \\  \hline
  \end{tabular}
\\
\\

   As the functions behave well. They are expected to approach some finite values at $N\to \infty$,
   one may use a functional fitting method to extract their values at $N\to \infty$
   with the required precision. Here we choose the following fitting function
   \begin{eqnarray}
   F_a (N) = F_a (\infty) + \sum_{n=1} C_a^{(n)} \left(\frac{1}{\ln N} \right)^n
   \end{eqnarray}
  Here $C_a^{(n)}$ are the fitting coefficients. In general, one can choose, instead of $1/\ln N $,
  other functions with vanishing limit at $N\to \infty$.
    Taking $F_1 (N) = h_w(N)$, $F_2(N) = \gamma_w(N)$
  and $F_{2+n} (N) = L_n (N)$ ($n=1,2 \cdots$), one then has $F_1 (\infty) = h_w$, $F_2(\infty) =
\gamma_w$ and $F_{2+n} (\infty) = L_n $
 ($n=1,2 \cdots$). Practically, by computing a set of numerical values of the functions
 $h_w(N)$, $\gamma_w(N)$ and $L_n(N)$ ($n=1,2 \cdots$) for some sufficiently large values of $N$,  and
  solving the corresponding linear equations, one is then able to obtain $F_a(\infty)$
  with the required precision. Here we only present the solutions at the precision $10^{-3}$
 \begin{eqnarray}
 & & h_w = F_1 (\infty) \simeq 1.000 \\
 & & \gamma_w = F_2(\infty) \simeq 5.772\times 10^{-1} \\
 & & L_1 =  F_{3} (\infty) \simeq 1.000 \\
 & & L_2 =  F_{4} (\infty) \simeq 1.000 \\
 & & L_3 =  F_{5} (\infty) \simeq 1.000 \\
 & & L_4 =  F_{6} (\infty) \simeq 1.000 \\
 & & L_5 =  F_{7} (\infty) \simeq 1.000 \\
 & & L_6 =  F_{8} (\infty) \simeq 1.000
 \end{eqnarray}
 They are resulted from a set of values for the functions  $h_w(N)$, $\gamma_w(N)$ and $L_n(N)$ ($n=1,2
 \cdots$) obtained for $N$ from $N=300$ to $N=10000$.
  For a consistent check, such solutions are found to provide, with the same precision, a prediction
  for the numerical values of the functions $h_w(N)$, $\gamma_w(N)$ and $L_n(N)$ ($n=1,2 \cdots$)
  at $N= 20000$ (see table).

 The above numerical analysis provides a reasonable check on the three conjectures
 made in section IV.

\section{Some useful regularized one-fold ILIs \\ in the new method}

     Using the definitions and prescriptions of the new regularization, we present the
     following useful regularized ILIs

  \begin{eqnarray}
     & & \int [d^4 k]_l \frac{1}{k^2 + M_l^2 + {\cal M}^2 } = \pi^2 \{ M_c^2
     - \mu^2 [\ \ln \frac{M_c^2}{\mu^2} - \gamma_w + 1 + y_2(\frac{\mu^2}{M_c^2}) \ ] \} \\
     & & \int [d^4 k]_l \frac{1}{ (k^2 + M_l^2 + {\cal M}^2 )^2} = \pi^2 \{
     \ \ln \frac{M_c^2}{\mu^2} - \gamma_w + y_0(\frac{\mu^2}{M_c^2}) \  \} \\
     & & \int [d^4 k]_l \frac{1}{ (k^2 + M_l^2 + {\cal M}^2)^3} = \frac{1}{2\mu^2} \pi^2  \{
     \ 1- y_{-2}(\frac{\mu^2}{M_c^2}) \  \} \\
     & & \int [d^4 k]_l \frac{1}{ (k^2 + M_l^2 + {\cal M}^2 )^{\alpha}} = \pi^2
     \frac{\Gamma(\alpha -2)}{\Gamma(\alpha) } \frac{1}{( \mu^2 )^{\alpha -2} } \{
     \ 1 -  y_{-2(\alpha -2)} (\frac{\mu^2}{M_c^2}) \  \} \qquad \alpha > 3 \\
     & & \int [d^4 k]_l \frac{k{\mu}k_{\nu}}{ (k^2 + M_l^2 + {\cal M}^2 )^2 } =
     \frac{1}{2} \delta_{\mu\nu} \pi^2 \{ M_c^2
     - \mu^2 [\ \ln \frac{M_c^2}{\mu^2} - \gamma_w + 1 + y_2(\frac{\mu^2}{M_c^2}) \ ] \} \\
     & & \int [d^4 k]_l \frac{k{\mu}k_{\nu}}{ ( k^2 + M_l^2 + {\cal M}^2 )^3 } =
     \frac{1}{4} \delta_{\mu\nu} \pi^2 \{
     \ \ln \frac{M_c^2}{\mu^2} - \gamma_w + y_0(\frac{\mu^2}{M_c^2}) \  \} \\
      & & \int [d^4 k]_l \frac{k{\mu}k_{\nu}}{ ( k^2 + M_l^2 + {\cal M}^2 )^4 } =
      \frac{1}{4}\delta_{\mu\nu}  \frac{1}{3\mu^2} \pi^2  \{
     \ 1- y_{-2}(\frac{\mu^2}{M_c^2}) \  \} \\
      & & \int [d^4 k]_l \frac{k{\mu}k_{\nu}}{ ( k^2 + M_l^2 + {\cal M}^2 )^{\alpha + 1} } =
      \frac{1}{2} \delta_{\mu\nu} \pi^2 \frac{\Gamma(\alpha -2)}{\Gamma(\alpha + 1) }
      \frac{1}{( \mu^2 )^{\alpha -2} } \{
     \ 1 -  y_{-2(\alpha -2)} (\frac{\mu^2}{M_c^2}) \  \}  \\
     & & \int [d^4 k]_l \frac{k{\mu}k_{\nu}k_{\rho}k_{\sigma}}{ (k^2 + M_l^2 + {\cal M}^2 )^3 } =
     \frac{1}{8} \delta_{\{ \mu\nu\rho\sigma\} }
      \pi^2 \{ M_c^2
     - \mu^2 [\ \ln \frac{M_c^2}{\mu^2} - \gamma_w + 1 + y_2(\frac{\mu^2}{M_c^2}) \ ] \} \\
     & & \int [d^4 k]_l \frac{k{\mu}k_{\nu}k_{\rho}k_{\sigma} }{ ( k^2 + M_l^2 + {\cal M}^2 )^4 } =
     \frac{1}{24} \delta_{\{ \mu\nu\rho\sigma\} }  \pi^2 \{
     \ \ln \frac{M_c^2}{\mu^2} - \gamma_w + y_0(\frac{\mu^2}{M_c^2}) \  \} \\
     & & \int [d^4 k]_l \frac{k{\mu}k_{\nu} k_{\rho}k_{\sigma} }{ (k^2 + M_l^2 + {\cal M}^2 )^5 } =
       \frac{1}{24} \delta_{\{ \mu\nu\rho\sigma\} }  \frac{1}{4\mu^2} \pi^2  \{
     \ 1- y_{-2}(\frac{\mu^2}{M_c^2}) \  \} \\
     & & \int [d^4 k]_l \frac{k{\mu}k_{\nu}  k_{\rho}k_{\sigma} }{ ( k^2 + M_l^2 + {\cal M}^2 )^{\alpha +
     2} }
     =  \frac{1}{4} \delta_{\{ \mu\nu\rho\sigma\} } \pi^2 \frac{\Gamma(\alpha -2)}{\Gamma(\alpha + 2) }
      \frac{1}{( \mu^2 )^{\alpha -2} } \{
     \ 1 -  y_{-2(\alpha -2)} (\frac{\mu^2}{M_c^2}) \  \}  \\
    \end{eqnarray}
    with
    \begin{eqnarray}
      \int [d^4 k]_l \equiv   \lim_{N\to \infty} \sum_{l=0}^{N} c_l^N \int d^4 k
      =   \lim_{N\to \infty} \sum_{l=0}^{N} (-1)^l \frac{N!}{(N-l)!\ l!}  \int d^4 k
    \end{eqnarray}
  where
  \begin{eqnarray}
   & & M_l^2 = \mu_s^2 + l M_R^2 \ , \quad M_R^2 = M_c^2 h_w(N) \ln N \\
 & &   \mu^2 = \mu_s^2 + {\cal M}^2  \\
 & & \delta_{\{ \mu\nu\rho\sigma\} } \equiv \delta_{\mu\nu} \delta_{\rho\sigma} + \delta_{\mu\rho} \delta_{\nu\sigma}
     + \delta_{\mu\sigma} \delta_{\rho\nu}  \\
 & &  y_{-2(\alpha -2)} (x) = -\lim_N \sum_{l=1}^{N} c_l^N \left( \frac{x/(l h_w(N) \ln N )}{
  1 + x/(l h_w(N) \ln N ) } \right)^{\alpha -2}  \\
  & & \gamma_w = \gamma_E = 0.5772 \cdots \ , \quad h_w(N\to\infty) = 1
  \end{eqnarray}
  The explicit forms for $y_{2}(x)$, $y_0(x)$ and $y_{-2}(x)$ are given in eqs.(4.10-4.13).

\newpage
 \centerline{\large{Figures}}

\newcommand{\PICL}[2]
{
\begin{center}
\begin{picture}(500,170)(0,0)
\put(0,-18){ \epsfxsize=8cm \epsfysize=14cm
\epsffile{#1} } \put(115,90){\makebox(0,0){#2}}
\end{picture}
\end{center}
}

\newcommand{\PICR}[2]
{
\begin{center}
\begin{picture}(300,0)(0,0)
\put(160,10){ \epsfxsize=8cm \epsfysize=14cm
\epsffile{#1} } \put(275,120){\makebox(0,0){#2}}
\end{picture}
\end{center}
}

\newcommand{\PICLN}[2]
{
\begin{center}
\begin{picture}(500,10)(0,0)
\put(-80,-18){ \epsfxsize=13cm \epsfysize=16cm
\epsffile{#1} } \put(115,70){\makebox(0,0){#2}}
\end{picture}
\end{center}
}

\newcommand{\PICRN}[2]
{
\begin{center}
\begin{picture}(300,0)(0,0)
\put(80,10){ \epsfxsize=13cm \epsfysize=16cm
\epsffile{#1} } \put(275,97){\makebox(0,0){#2}}
\end{picture}
\end{center}
}


\small
\mbox{} \vspace{3cm}
 \PICL{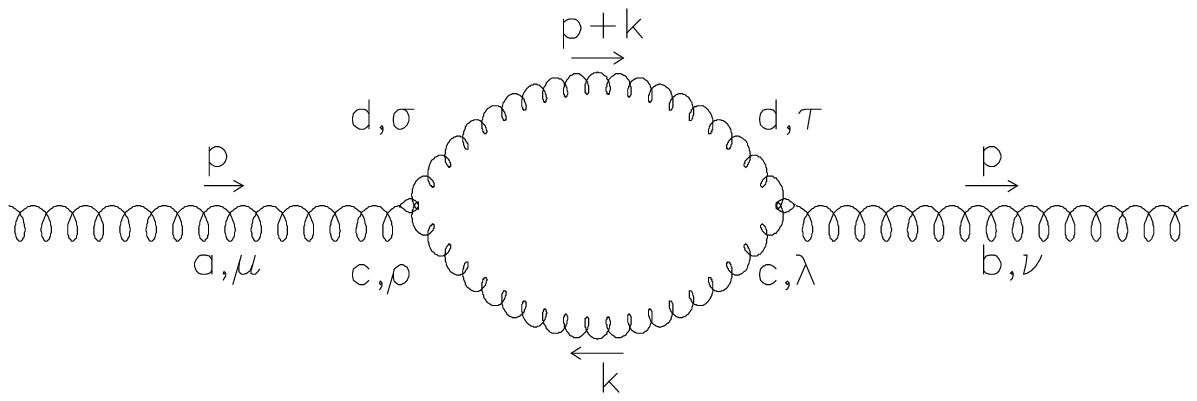}{(1)}
 \PICR{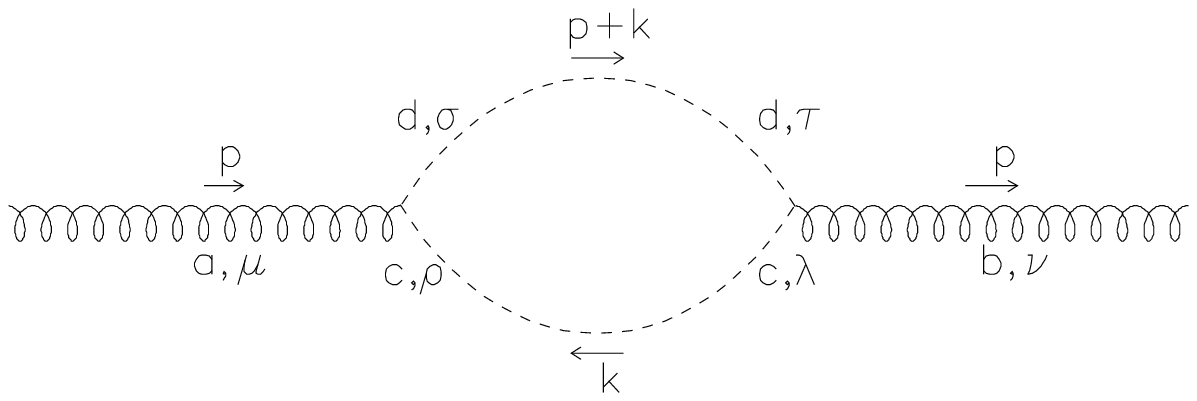}{(3)}
 \vspace{1cm}
 \PICL{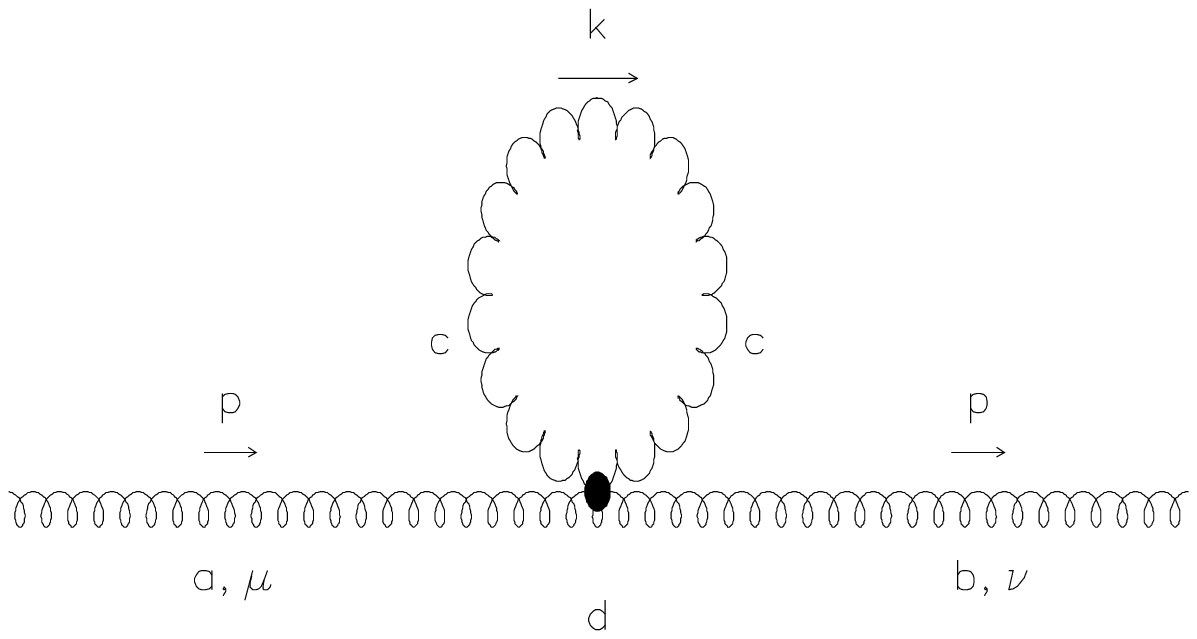}{(2)}
 \PICR{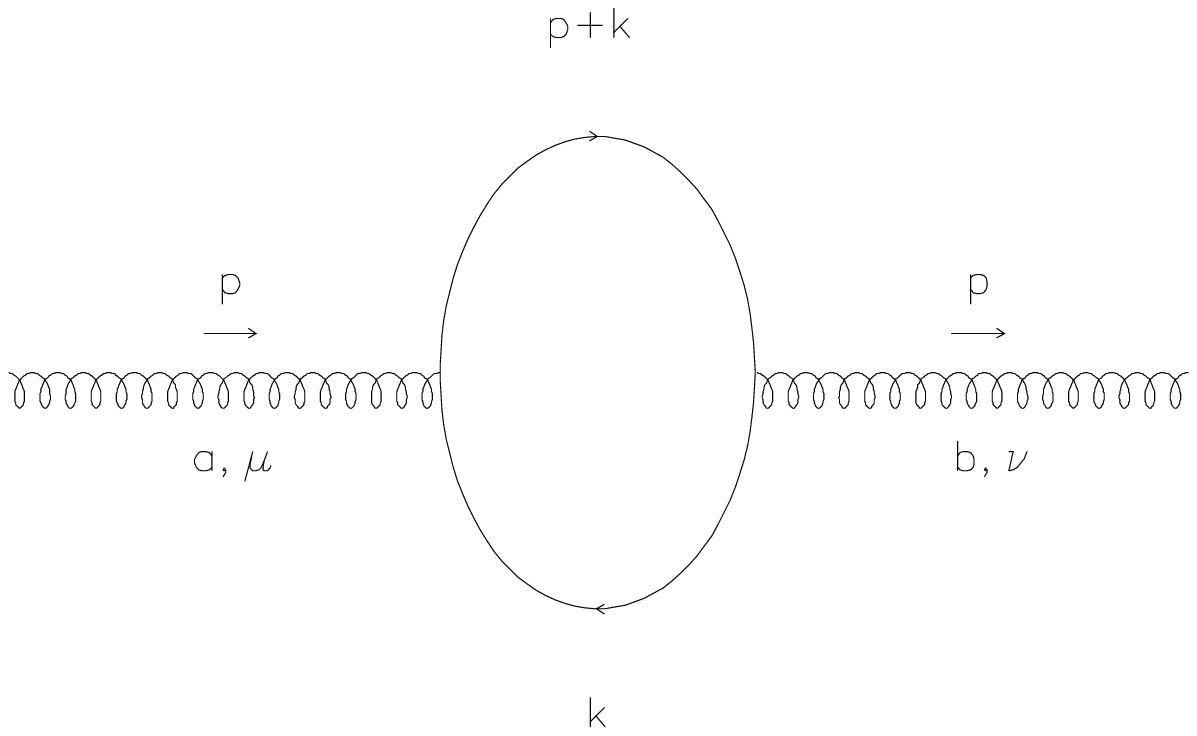}{(4)}
 \vspace{-2.5cm}
   Vacuum polarization diagrams of gauge field:  (1) from trilinear gauge boson interactions ; \\
  (2) from quartic gauge boson interactions;  (3) from
   gauge-ghost boson interactions;   \\ (4) from fermion-gauge
  boson interactions.



\begin{thebibliography}{99}
 \bibitem{TDL}  See e.g.: T.D. Lee, {\it Particle Physics and Introduction to Field Theory},
  (Harwood Academic Publishers GmbH, 1981); {\it Symmetries, Asymmetries, and the World of Particles},
  (University of Washington Press, 1988).
 \bibitem{YM} C.N. Yang and R.L. Mills, Phys. Rev. {\bf 96}, 191 (1954).
 \bibitem{SW1} See e.g.: S. Weinberg,  {\it The Quantum Theory of Fields}-I and II,
 (Cambridge University Press, Cambridge, 1995).
 \bibitem{CL} See also: Ta-Pei Cheng and Ling-Fong Li, {\it Gauge theory of
 elementary particle physics}, (Oxford University Press, Walton Street, Oxford OX2 6DP, 1984;
 reprinted 1985-1996); \\
 Yuan-Ben Dai, {\it Gauge Theory of Interactions}, Scientific publishing, Beijing, 1987; \\
 M. E. Peskin and D.V. Schroeder, {\it An Introduction to Quantum Field Theory}, Addison-Wesley Publishing Co. ,
 1995.
 \bibitem{GTH} See e.g.: G. 't Hooft, {\it Under the Spell of the Gauge Principle},
 ( World Scientific Publishing Co., 1994).
 \bibitem{JZJ} See e.g.: Jean Zinn-Justin, {\it Quantum
 Field Theory and Critical Phenomena}, (Oxford University Press, Walton Street, Oxford OX2 6DP, 1996).
 \bibitem{COR} W. Heisenberg, Ann. d. Phys. {\bf 32}, 20 (1938).
\bibitem{PV} W. Pauli and F. Villars, Rev. Mod. Phys. {\bf 21}, 434 (1949).
\bibitem{PR} J. Schwinger, Phys. Rev. {\bf 82}, 664 (1951).
 \bibitem{BPHZ} N.N. Bogoliubov and O.S. Parasiuk, Acta Math.,
 {\bf 97}, 227 (1957); N.N. Bogoliubov and D.V. Shirkov, {\it
 Introduction to the Theory of Quantized Fields}, Interscience,
 New York, 1959; \\ K. Hepp, Comm. Math. Phys. {\bf 2}, 301
 (1966); \\ W. Zimmermann, Cmmmun. Math. Phys. {\bf 15}, 208
 (1969); ``Lectures on Elementary Particles and Quantum Field
 Theory'', Brandeis Summer Institute 1970, edited by S. Deser, M.
 Grisaru and H.Pendleton, MIT press.
\bibitem{DR} G.'t Hooft and M. Veltman, Nucl. Phys. {\bf B44}, 189 (1972); \\
J. Ashmore, Lett. Nuovo Cimento {\bf 4}, 289 (1972); \\
C.G. Bollini and J.J Giambiaggi, Phys. Lett. {\bf 40B} (1972) 566; Nuovo Cimento {\bf 12B} (1972).
\bibitem{LR} K.G. Wilson, {\it New Phenomena in Subnuclear Physics }, Erice 1975, Edited by
A. Zichichi (Plenum, New York 1977); \\
T. Reisz, Commun. Math. Phys. {\bf 117}, 79, 639 (1988); \\
H.B. Nielsen and M. Ninomiya, Nucl. Phys. {\bf B185}, 20 (1981); \\
T. Banks, L. Susskind and J. Kogut, Phys. Rev. {\bf D13}, 1043 (1976).
\bibitem{DFR} D.Z. Freedman, K. Johnson and J.I. Latorre, Nucl.
 Phys. {\bf B371}, 353 (1992);
  D.Z. Freedman, G. Grignani, K. Johnson and N. Rius, Ann. Phys. {\bf 218}, 75
  (1992); \\
  P.E. Haagense and J.I. Latorre, Phys. Lett. {\bf B283}, 293 (1992);
  P.E. Haagense and J.I. Latorre, Ann. Phys.  {\bf B221}, 77 (1993).
\bibitem{NJL}Y. Nambu and G. Jona-Lasinio, Phys. Rev. {\bf 122}, 345 (1961).
\bibitem{TG} T. Gherghetta, Phys. Rev. {\bf D50}, 5985 (1994).
\bibitem{AAS} A.A. Slavnov, Theor. Math. Phys. {\bf 33}, 977 (1977).
\bibitem{LMR} J.H. Leon, C.P. Martin and F.R. Ruiz, Phys. Lett. {\bf B355}, 531 (1995).
\bibitem{MR}  C.P. Martin and F.R. Ruiz, Nucl. Phys. {\bf B436}, 545 (1995).
\bibitem{SW2} S. Weinberg, Phys. Rev. {\bf 118}, 838 (1959).
 \bibitem{DRED1} W. Siegel, Phys. Lett. {\bf B84}, 193 (1979).
 \bibitem{DRED2} W. Siegel, Phys. Lett. {\bf B94}, 37 (1980).
 \bibitem{YLWU} Y.L. Wu, Phys. Rev. {\bf D64} 016001 (2001).
 \bibitem{CDFR} F. del Aguila, A. Culatti, R. Munoz Tapia and M.
 Perez-Victoria, Phys. Lett. {\bf B419}, 263 (1998); M.
 Perez-Victoria, Phys. Lett. {\bf B442}, 315 (1998).
 \bibitem{SST} See e.g.: M.B. Green, J. H. Schwarz and E. Witten, {\it Superstring Theory},
  (Cambridge University Press, Cambridge, 1987); \\
  D.J. Gross, J.A. Harvey, E. Martinec and R. Rohm, Nucl. Phys. {\bf B256}, 235 (1985); {\bf B267},
  75 (1986); \\
  J. Polchinski, {\it String Theory} (Cambridge University Press, Cambridge, 1998).
 \bibitem{SW3} S. Weinberg, hep-th/9702027, 1997.
\bibitem{KGW} K.G. Wilson, Phys. Rev. {\bf B4}, 3174, 3184 (1971);
Rev. Mod. Phys. {\bf 47}, 773 (1975).
\bibitem{GML} M. Gell-Mann and F.E. Low, Phys. Rev. {\bf 95}, 1300 (1954).
\bibitem{FP} L.D. Faddeev and V.N. Popov, Phys. Lett. {\bf 25B}, 29 (1967).
\bibitem{F1} R.P. Feynman, Acta Phys. Polonica {\bf 24}, 697 (1963).
\bibitem{F2} B.S. De Witt, Phys. Rev. {\bf 162}, 1195, 1239 (1967).
 \bibitem{IR} O.A. Batistel, A.L. Mota and M.C. Nemes, Mod. Phys.
 Lett. {A13}, 1597 (1998); hep-th/0203261 and references therein.
\bibitem{GW} D.J. Gross and F. Wilczek, Phys. Rev. Lett. {\bf 30}, 1343 (1973).
 \bibitem{HDP} H. D. Politzer, Phys. Rev. Lett. {\bf 30}, 1346 (1973).
 \bibitem{SVZ} M.A. Shifman, A.I. Vainstein and V.I. Zakharov, Nucl.
 Phys. {\bf B147}, 385 (1979).
\end{thebibliography}
\end{document}